\documentclass[aps,prd,reprint,floatfix,superscriptaddress,showkeys,nofootinbib]{revtex4-1}

\setcitestyle{authoryear,round}
\usepackage[utf8]{inputenc}
\usepackage[T1]{fontenc}

\usepackage{graphicx}

\usepackage{siunitx}
\usepackage{mathtools}
\usepackage{amsfonts}
\usepackage{bm}
\usepackage{bbm}

\usepackage{verbatim}
\usepackage{rotate}
\usepackage{color}
\usepackage{aas_macros}
\usepackage{ulem}
\DeclareFontFamily{OT1}{pzc}{}
\DeclareFontShape{OT1}{pzc}{m}{it}%
            {<-> s * [1.10] pzcmi7t}{}
\DeclareMathAlphabet{\mathscr}{OT1}{pzc}%
                                {m}{it}

\definecolor{RedWine}{rgb}{0.743,0,0}
\definecolor{green(pigment)}{rgb}{0.0,0.65,0.31}
\definecolor{RoyalBlue}{rgb}{0.25,0.41,0.88}

\newcommand{\be}{\begin{equation}}
\newcommand{\ee}{\end{equation}}
\newcommand{\bea}{\begin{eqnarray}}
\newcommand{\eea}{\end{eqnarray}}
\def\ba#1\ea{\begin{align}#1\end{align}}

\def\({\left(}
\def\){\right)}
\def\<{\left\langle}
\def\>{\right\rangle}

\newcommand{\vs}{\nonumber\\}

\def\vr{{\bm{r}}}

\def\rhat{{\hat{\bm{r}}}}

\def\nbar{{\bar{n}}}

\def\fnl{f_\mathrm{NL}}

\DeclareSIUnit \parsec {pc}
\DeclareSIUnit \h {\text{$h$}}
\DeclareSIUnit \year {yr}
\DeclareSIUnit \solarmass {M_\odot}
\DeclareSIUnit \Mpc {\mega\parsec}

\def\obs{\mathrm{obs}}

\def\true{\mathrm{true}}

\def\min{\mathrm{min}}
\def\max{\mathrm{max}}

\def\sky{\mathrm{sky}}

\def\tr{\mathrm{tr}}

\def\dd{\mathrm{d}}

\def\myapp#1#2{%
  \mathrel{%
    \setbox0=\hbox{$#1\sim$}%
    \setbox2=\hbox{%
      \rlap{\hbox{$#1\propto$}}%
      \lower1.1\ht0\box0%
    }%
    \raise0.25\ht2\box2%
  }%
}

\newcommand{\incgraph}[2][0.49]{\includegraphics[width=#1\textwidth]{#2}}

\usepackage[colorlinks]{hyperref}
\usepackage[capitalise]{cleveref}

\hypersetup{
colorlinks=true,
citecolor=cyan,}

\begin{document}

\title{Validation of Spherical Fourier-Bessel power spectrum analysis with lognormal simulations and eBOSS DR16 LRG EZmocks}

\author{Henry S. \surname{Grasshorn Gebhardt}}
\email{gebhardt@caltech.edu}
\thanks{NASA Postdoctoral Program Fellow}
\affiliation{Jet Propulsion Laboratory, California Institute of Technology, Pasadena, CA 91109, USA}
\affiliation{California Institute of Technology, Pasadena, CA 91125, USA}

\author{Olivier \surname{Dor\'e}}
\affiliation{Jet Propulsion Laboratory, California Institute of Technology, Pasadena, CA 91109, USA}
\affiliation{California Institute of Technology, Pasadena, CA 91125, USA}

\begin{abstract}
  Tuning into the bass notes of the large-scale structure requires careful
  attention to geometrical effects arising from wide angles. The spherical
  Fourier-Bessel (SFB) basis provides a harmonic-space coordinate system that
  fully accounts for all wide-angle effects. To demonstrate the feasibility of
  the SFB power spectrum, in this paper we validate our SFB pipeline by
  applying it to lognormal, and both \emph{complete} and \emph{realistic} EZmock
  simulations that were generated for eBOSS DR16 LRG sample.
  We include redshift space distortions and the local
  average effect (aka integral constraint). The covariance matrix is obtained
  from 1000 EZmock simulations, and inverted using eigenvalue decomposition.
\end{abstract}


\maketitle

\section{Introduction}
Upcoming galaxy surveys such as SPHEREx, Euclid, DESI, PFS, and others will
have significant constraining power on very large scales. This will allow
constraints on local-type primordial non-Gaussianity that is predicted by
inflation.

The typical power spectrum multipole estimator was introduced by
\citet{Yamamoto+:2006PASJ...58...93Y} \citep[for a review and implementation
see][e.g.]{Hand+:2017JCAP...07..002H}. In the Yamamoto estimator, a single
line-of-sight (LOS) is chosen for each pair of galaxies, and for efficient
implementation that LOS is typically chosen to one of the galaxies in each
pair. However, while the Yamamoto estimator is a significant improvement over
the assumption of a flat sky with a fixed LOS, it still suffers from wide-angle
effects that can mimic a non-Gaussianity parameter $f_\mathrm{NL}\sim5$
\citep{Benabou+Sands+:inprep}. Further extensions to the Yamamoto estimator
such as using the midpoint or bisector between the galaxy pair as the LOS
\citep{Philcox+:2021PhRvD.103l3509P} are also possible. However, while
certainly less, those estimators, too, will contain wide-angle effects.

The spherical Fourier-Bessel decomposition fully accounts for all wide-angle
effects, because it is the natural coordinate system for the radial/angular
separation. \citep{Heavens+:1995MNRAS.275..483H, Percival+:2004MNRAS.353.1201P,
Wang+:2020JCAP...10..022W, GrasshornGebhardt+:2021PhRvD.104l3548G}

In this paper we validate the SFB power spectrum estimator developed in
\citet{GrasshornGebhardt+:2021PhRvD.104l3548G} on the realistic eBOSS DR16
luminous red galaxy EZmocks \citep{Zhao+:2021MNRAS.503.1149Z}.
For the covariance matrix we use all 1000 EZmocks.

This paper is organized as follows. In \cref{sec:sfb_basis} we review the SFB
basis, in \cref{sec:sfb_estimator} we review our \emph{SuperFaB} estimator, in
\cref{sec:sfb_theory} we detail the theoretical modeling including window,
weights, shot noise, local average effect, and pixel window. We validate our
analysis pipeline with lognormal mocks in \cref{sec:ln}, and with EZmocks in
\cref{sec:ez}. The likelihood is constructed and run with an adaptive
Metropolis-Hastings sampler in \cref{sec:mcmc}. We conclude in
\cref{sec:conclusion}. Many details such as a derivation of the local average
effect and velocity boundary conditions are delegated to the appendices.

\section{Spherical Fourier-Bessel basis}
\label{sec:sfb_basis}
Similar to the Cartesian basis power spectrum analysis, the spherical
Fourier-Bessel analysis also separates scales according to a wavenumber $k$.
This stems from the Laplacian eigenequation $\nabla^2 f = -k^2 f$ that defines
both bases, with the difference coming mainly from the geometry of the boundary
conditions.

The spherical Fourier-Bessel basis offers several advantages and disadvantages
over the more familiar Cartesian-space analysis. The two chief advantages of SFB
is in the separation of angular and radial modes due to the use of spherical
polar coordinates, and in the complete modeling of wide-angle effects. Many
observational effects are local in origin, and, thus, manifest themselves
purely as angular systematics. Wide-angle effects come from the geometry of the
curved sky, and they have a similar signature as local non-Gaussianity
in the endpoint estimator
\citep{Yamamoto+:2006PASJ...58...93Y, Bianchi+:2015MNRAS.453L..11B,
Scoccimarro:2015PhRvD..92h3532S, Hand+:2017JCAP...07..002H,
Castorina+:2020MNRAS.499..893C, Beutler+:2019JCAP...03..040B}. Due to the use
of a spherical coordinate system and spherical boundaries, SFB is well-suited
for both these types of systematics.

Besides the lesser familiarity of the SFB power spectrum compared to
Cartesian-space power spectrum multipoles, the SFB power spectrum suffers from
an increased computational cost due to the lack of the fast Fourier-transform,
and it suffers from an increase in the number of modes. The large number of
modes contains a wealth of information that may be exploited, such as
distinguishing between redshift space distortions and evolution of galaxy bias
and growth rate with redshift, to name a few. However, the presence of so many
modes also presents challenges in the analysis.

The spherical Fourier-Bessel basis is defined by the eigenfunctions of the
Laplacian in spherical polar coordinates with spherical boundaries. Thus, the
density contrast is expanded in SFB coefficients as
\citep[e.g.][]{Nicola+:2014PhRvD..90f3515N}
\ba
\label{eq:sfb_discrete_fourier_pair_a}
\delta(\vr)
&= \sum_{n\ell m} g_{n\ell}(r)\,Y_{\ell m}(\rhat)\,\delta_{n\ell m}\,,
\\
\label{eq:sfb_discrete_fourier_pair_b}
\delta_{n \ell m}
&= \int\dd^3\vr \,g_{n\ell}(r)\,Y^*_{\ell m}(\rhat) \,\delta(\vr)\,,
\ea
where the second is the inverse of the first, and $Y_{\ell m}(\rhat)$ are
spherical harmonics, and
\ba
\label{eq:gnl}
g_{n\ell}(r) = c_{n\ell} \, j_\ell(k_{n\ell}r) + d_{n\ell} \, y_\ell(k_{n\ell}r)
\ea
are linear combinations of spherical Bessel functions of the first and second
kind, chosen to satisfy the orthonormality relation
\ba
\label{eq:gnl_orthonormality}
\int_{r_\min}^{r_\max}\dd r\,r^2\,g_{n\ell}(r)\,g_{n'\ell}(r)
&=
\delta^K_{nn'}\,.
\ea
The index $n$ denotes the wavenumber $k_{n\ell}$, and $c_{n\ell}$ and
$d_{n\ell}$ are constants \citep{Samushia:2019arXiv190605866S,
GrasshornGebhardt+:2021PhRvD.104l3548G}. For comparison, we include the SFB
transform over an infinite flat universe in
\cref{eq:sfb_fourier_pair_a,eq:sfb_fourier_pair_b}.

In this paper, we assume potential boundaries
\citep{Fisher+:1995MNRAS.272..885F} which ensure that the density contrast is
continuous and smooth on the boundaries. Further, we assume a spherical
boundary at some minimum distance $r_\min$ and some maximum distance $r_\max$
so that the SFB analysis is performed in a thick shell between $r_\min$ and
$r_\max$ \citep{GrasshornGebhardt+:2021PhRvD.104l3548G,
Samushia:2019arXiv190605866S}. We have also verified that velocity boundary
conditions (that ensure vanishing derivative at the boundary) give essentially
the same result, see \cref{sec:ln_boundary}.

\section{\emph{SuperFaB} estimator}
\label{sec:sfb_estimator}
For our SFB estimator we use \emph{SuperFaB}. We refer the interested reader to
\citet{GrasshornGebhardt+:2021PhRvD.104l3548G} for all the details, and give
only a very short overview here. In brief, \emph{SuperFaB} is similar to the
3DEX approach developed in \citet{Leistedt+:2012A&A...540A..60L}.
\emph{SuperFaB} assumes the number density is given by a discrete set of
points,
\ba
\label{eq:number_density}
  n(\vr) &= \sum_p \delta^D(\vr - \vr_p)\,,
\ea
where the sum is over all points (galaxies) in the survey.
To calculate the fluctuation field, we include a weighting
\citep{Feldman+:1994ApJ...426...23F}
\ba
\label{eq:fkp_weights}
w(\vr) &= \frac{1}{1 + \nbar \, W(\vr) \, C_{\ell nn'}}\,,
\ea
and we approximate $C_{\ell nn'}\sim\SI{e4}{\per\h\cubed\mega\parsec\cubed}$,
so that the observed density fluctuation field is
\ba
  \delta^\obs(\vr)
    &= w(\vr)\,\frac{n(\vr) - \alpha\,n_r(\vr)}{\nbar_\max}\,,
  \label{eq:delta_fkp_weighted}
\ea
where $n(\vr)$ is the observed number density and $n_r(\vr)$ is a uniform
random catalog subject to the same window function and systematics.
$\nbar_\max$ is the maximum number density in the data catalog, in our
convention.

To perform the discrete SFB transform \cref{eq:sfb_discrete_fourier_pair_b}, we
first perform the radial integral for each $(n,\ell)$ combination by directly
summing over the galaxies, and pixelizing on the spherical sky using the
HEALPix scheme \citep{Gorski+:2005ApJ...622..759G}. The angular integration is
then performed using \texttt{HEALPix.jl} \citep{Tomasi+:2021ascl.soft09028T}.

This SFB transform is performed for both the data catalog and the random
catalog, and the result is subtracted with appropriate normalization to obtain
the fluctuation field $\hat\delta_{n\ell m}$.

Finally, we construct the pseudo-SFB power spectrum
\ba
\hat C^{wWA}_{\ell nn'}
&= \frac{1}{2\ell+1} \sum_{m} \hat\delta_{n\ell m} \hat\delta^*_{n'\ell m}\,,
\ea
where we attach the suffixes `$wW$' for the weight and window, and the suffix
`$A$' for the local average effect, and the caret ($\hat{\phantom{x}}$)
indicates estimation from data. Window function and other effects will be
forward-modeled, and we turn to that next.

\section{SFB Power Spectrum Model}
\label{sec:sfb_theory}
In this section we detail our approach for calculating a theoretical model for
the estimator of \cref{sec:sfb_estimator}. We start with a full-sky calculation, and
then add the window function, weights, local average effect (LAE), and pixel
window. In the \emph{window function} we generally include survey geometry and
radial selection function.

In the SFB basis, the full-sky linear power spectrum $C_{\ell nn'} \simeq
C_\ell(k_{n\ell},k_{n'\ell})$ on the lightcone in redshift space is expressed
as \citep{Khek+:2022arXiv221205760K}
\ba
C_{\ell nn'}
&=
\int_0^\infty\dd q\,\mathcal{W}_{n\ell}(q)\,\mathcal{W}_{n'\ell}(q)\,\gamma^{-3}\,P_m(\gamma^{-1}q)\,,
\ea
where $P_m(q)$ is the matter power spectrum, $\gamma$ is an
Alcock-Paczynski-like parameter, and
\ba
\label{eq:Wnlq}
\mathcal{W}_{n\ell}(q)
&=
\sqrt{\frac{2}{\pi}}\,q
\int_{r_\min}^{r_\max}\dd r\,r^2\,g_{n\ell}(r)\,w(r)\,\phi(r)\,D(r)
\vs&\quad\times
e^{\frac12\sigma_u^2q^2\partial_{qr}^2}
\left[
  b(r,q)\,j_\ell(qr) - f(r)\,j''_\ell(qr)
\right],
\ea
with the radial basis functions $g_{n\ell}(r)$ defined in \cref{eq:gnl},
where $w(r)$ is a weighting function, $\phi(r)$ is a radial selection function,
$D(r)$ is the linear growth factor, $b(r,q)$ is a scale-dependent linear galaxy
bias, $\sigma_u^2$ is a velocity dispersion, and $f$ is the linear growth rate.

The velocity dispersion term we approximate by expanding the exponential
operator (that acts on the spherical Bessels only) in a Taylor series to get
\ba
e^{\frac12\sigma_u^2q^2\partial_{qr}^2} j^{(d)}_\ell(qr)
&\approx
j^{(d)}_\ell(qr)
- \frac12\sigma_u^2q^2\, j^{(d+2)}_\ell(qr)\,,
\ea
where $j_\ell^{(d)}(qr)$ is the $d$th derivative w.r.t.\ $qr$.\footnote{Another
reasonably well-performing approximation is \citep{Khek+:2022arXiv221205760K}
\ba
e^{\frac12\sigma_u^2q^2\partial_{qr}^2} j^{(d)}_\ell(qr)
&\approx
e^{-\frac12\sigma_u^2q^2} j^{(d)}_\ell(qr)\,,
\ea
which is motivated by two extreme cases: when $q$ is small, the exponential
operator does nothing; if it is large, the convolution implied by the
exponential operator averages over many oscillations, thus leading to a
vanishing value. However, for large $\ell$ this approximation suppresses power
significantly.
}

The velocity dispersion is a combination of Fingers-of-God (FoG) effect and
redshift measurement uncertainty, $\sigma_u^2 = \sigma_{u,\mathrm{FoG}}^2 + \sigma_z^2$.
The FoG effect is modeled to be proportional to the linear growth rate $f(z)$.
The redshift measurement uncertainty is kept constant \SI{1.05}{\per\h\mega\parsec} in our model, taken
from \citet[their Fig.~2]{Ross+:2020MNRAS.498.2354R}.

Local non-Gaussianity is modeled via a scale-dependent bias
\citep{Dalal+:2008PhRvD..77l3514D}
\ba
b(r,q) &= b(r) + \fnl b_\phi(r)\,\frac{3\Omega_m H_0^2}{2 q^2 T(q) \bar D(r)}\,,
\ea
where for this paper we assume the universal bias relation
\citep{Dalal+:2008PhRvD..77l3514D, Slosar+:2008JCAP...08..031S,
Afshordi+:2008PhRvD..78l3507A, Matarrese+:2008ApJ...677L..77M}
\ba
b_\phi(r) &= 2\delta_c\,[b(r) - 1]\,,
\ea
$T(q)$ is the transfer function, and $\bar D(r)$ is the growth factor
normalized to the scale factor $a$ during matter domination,
\ba
\bar D(r) &= \frac{D(r)}{(1 + z_\mathrm{md})\,D(r_\mathrm{md})}\,,
\ea
where the ``$\mathrm{md}$'' suffix indicates a time deep within matter domination.

In its most complete form, the SFB power spectrum contains off-diagonal terms
where $k_{n\ell}\neq k_{n'\ell}$. In a homogeneous universe we would have
$C_\ell(k,k')\,\propto\,\delta^D(k-k')$. However, both redshift space
distortions and growth of structure on the light cone break homogeneity in the
observed galaxy sample. For details we refer the reader to
\citet{Pratten+:2013MNRAS.436.3792P, Pratten+:2014MNRAS.442..759P,
Khek+:2022arXiv221205760K}.

The bulk of the computation is spent on the spherical Bessel functions in
$\mathcal{W}_{n\ell}(q)$ in \cref{eq:Wnlq}. However, those only depend on the
combination $qr$, not on $q$ and $r$ separately. Thus, we choose
discretizations for $q$ and $r$ such that in $q$-$r$ space the ``iso-$qr$''
lines go precisely through grid points. This condition demands that the
discretizations for $q$ and $r$ are logarithmic, i.e.,
\ba
q_i &= \bar q_\min \, R^{i\Delta n} \,,\\
r_j &= \bar r_\min \, R^{j}\,,
\ea
where $\bar q_\min$ and $\bar r_\min$ are the central values of the lowest
bins, $\ln R$ is the spacing in log-space, and $\Delta n\geq1$ is an integer
to allow sparse sampling of $q$.
Then, $qr$ is sampled at only a few discrete $m=i\Delta n + j$,
\ba
(qr)_m &= \bar q_\min \bar r_\min \, R^{i\Delta n + j}\,.
\ea
This transforms the problem of calculating the spherical Bessels from a
2-dimensional to a 1-dimensional problem.

\subsection{Window function convolution}
The window function limits the density contrast to that observed by a survey
$\delta^\obs(\vr) = W(\vr)\,\delta(\vr)$, where we define the window
$W(\vr)=\bar n(\vr) / \nbar_\max$. In SFB space the window effect becomes a
convolution,
\ba
\label{eq:delta_obs}
\delta^\obs_{n\ell m} &= \sum_{NLM} W_{n\ell m}^{NLM} \, \delta_{NLM}\,,
\ea
where the coupling matrix is
\ba
\label{eq:wmix}
W_{n \ell m}^{n'\ell'm'}
&=
\int\dd r\,r^2
\,g_{n\ell}(r)
\,g_{n'\ell'}(r)
\vs&\quad\times
\int\dd^2\rhat
\,Y^*_{\ell m}(\rhat)
\,Y_{\ell'm'}(\rhat)
\,W(r,\rhat)\,.
\ea
As a result, the modes of the SFB power spectrum are coupled,
\ba
C^\obs_{\ell nn'} = \sum_{LNN'} \mathcal{M}_{\ell nn'}^{LNN'} \, C_{LNN'}\,,
\ea
where we calculate the coupling matrix $\mathcal{M}$ by
\ba
\label{eq:cmix}
\mathcal{M}_{\ell nn'}^{LNN'}
&=
\frac{1}{2\ell+1}\sum_{mM}
W_{n\ell m}^{NLM}
W_{n'\ell m}^{N'LM,*}
\,.
\ea
However, in practice \cref{eq:wmix,eq:cmix} can be evaluated more efficienctly
as described in \citet{GrasshornGebhardt+:2021PhRvD.104l3548G}.

For the eBOSS data set, it is sufficient to assume a separable window,
\ba
W(\vr) &= M(\rhat)\,\phi(r)\,,
\ea
where $M(\rhat)$ is an angular window, and $\phi(r)\,\propto\,n(z)$ is the
radial selection function.

\begin{figure*}
  \centering
  \incgraph[0.32]{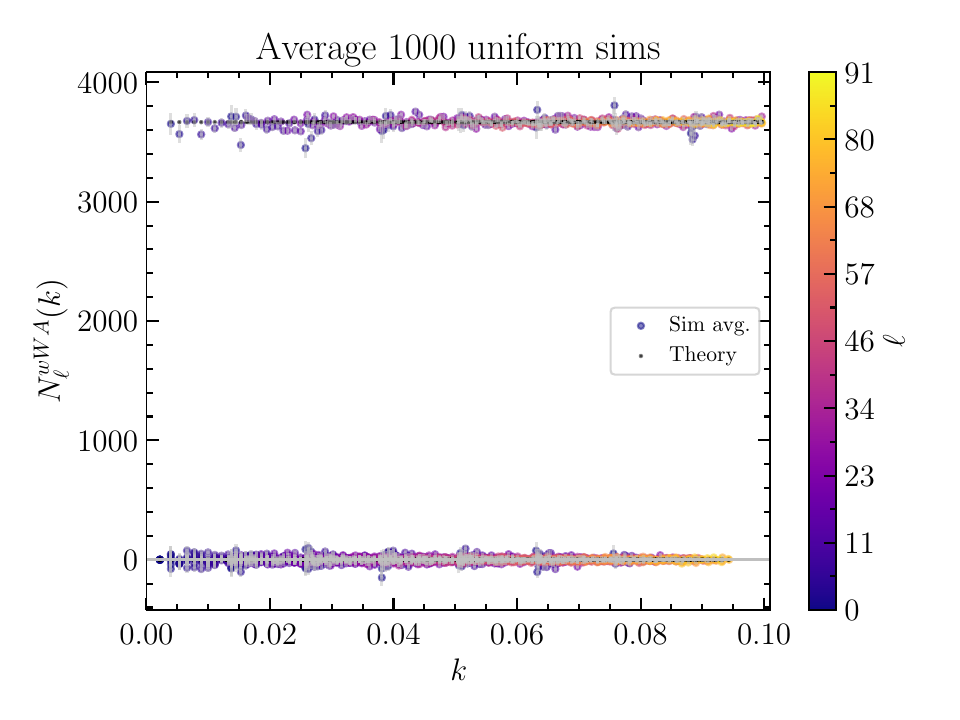}
  \incgraph[0.32]{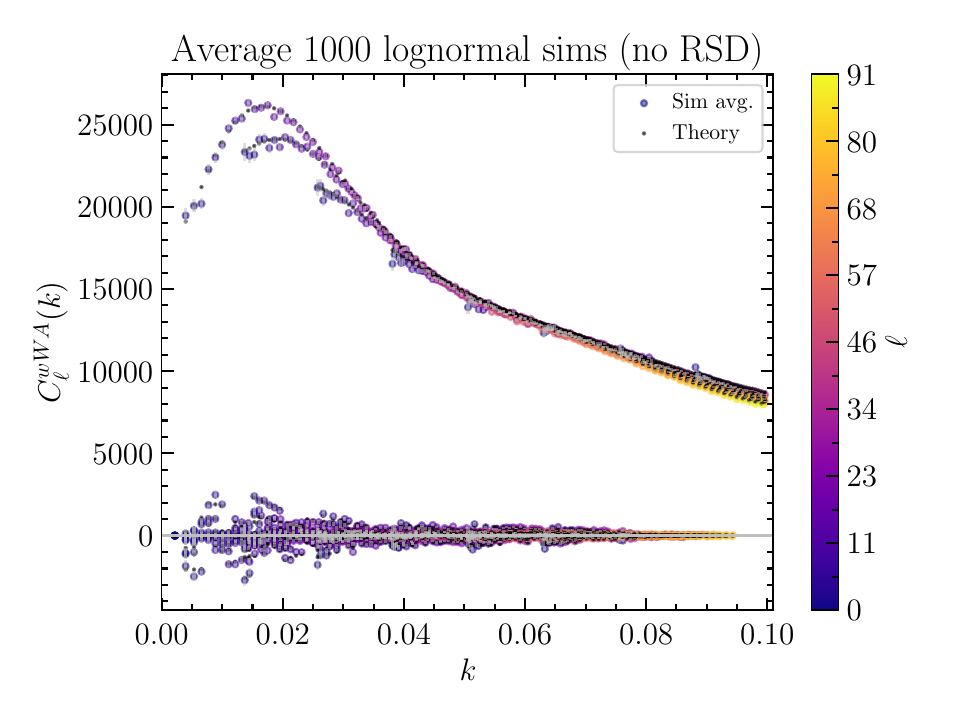}
  \incgraph[0.32]{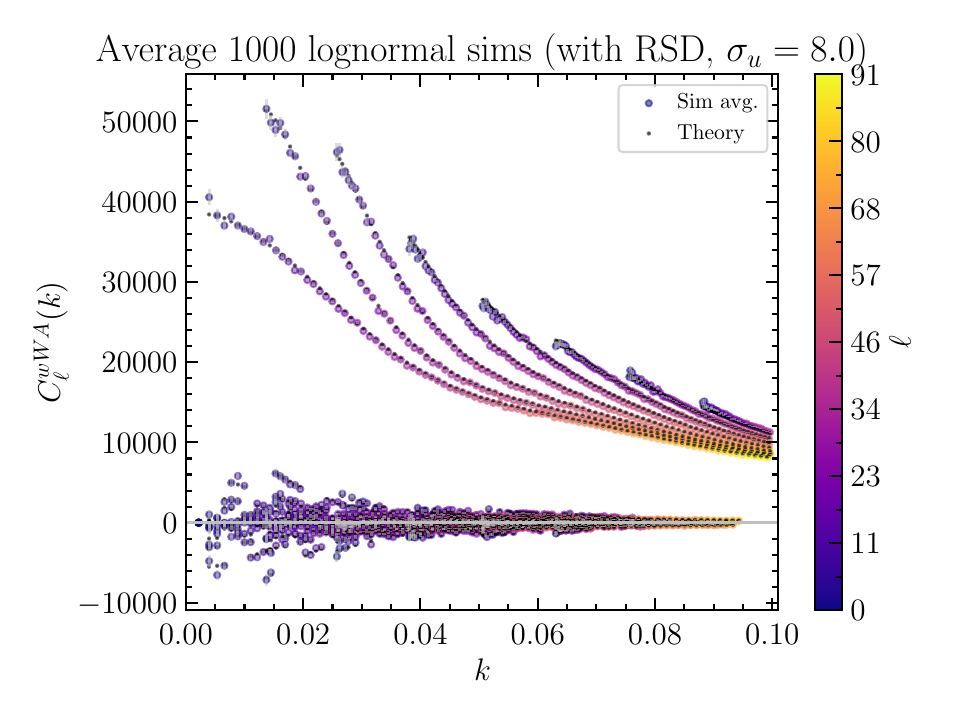}
  \caption{
    Here we show the SFB power spectrum for a full-sky survey with top-hat
    redshift bin \SIrange{750}{1000}{\per\h\mega\parsec}. In each panel the
    colored points show the simulations average with the color indicating the
    $\ell$-mode.
    Small black points indicate the theoretical calculation results.
    Left: Shot noise only;
    Center: Real-space lognormal simulations;
    Right: Redshift-space lognormal simulations.
    In all panels the SFB power spectrum modes separates into on-diagonal
    ($k=k'$) and off-diagonal ($k\neq k'$) modes. The off-diagonal modes
    cluster around zero power.
  }
  \label{fig:ln_fullsky_thinshell}
\end{figure*}

\subsection{FKP weights}
The effect of adding arbitrary weights $w(\vr)$, e.g., for optimizing the
statistical power similar to \citet{Feldman+:1994ApJ...426...23F} as in
\cref{eq:delta_fkp_weighted} is the following.

Since in the limit of infinitely many galaxies both the observed number density
$n(\vr)$ and the random catalog density $n_r(\vr)$ are proportional the window
$W(\vr)$, \cref{eq:delta_fkp_weighted} implies that the observed density
contrast is related to the true underlying theoretical density contrast by
\ba
\delta^\obs = w W \delta\,.
\ea

However, the window $W(\vr)$ and the weights $w(\vr)$ enter the observed power
spectrum differently. In particular, for the clustering signal we simply change
$W \to wW$, e.g., in \cref{eq:delta_obs}. On the other hand, the shot noise
changes from $N\,\propto\,W$ to $N\,\propto\,w^2W$. We leave the detailed
argument to \cref{app:fkp_shotnoise}.

\subsection{Local average effect}
\label{sec:lae}
The local average effect or integral constraint comes from measuring the number
density of galaxies from the survey itself. This implies that the observed
density contrast in the survey is measured relative to the average true density
contrast. That is, $\delta^\obs(\vr) \sim \delta(\vr) - \bar\delta(z)$, where
$\bar\delta(z)$ is the average density contrast in a thin redshift bin. This
removes power from the largest angular scales. As a result the observed
clustering and shot noise power spectra are given by four terms each
\ba
C^{wWA} &= C^1 - C^{23} - C^4\,, \\
N^{wWA} &= N^1 - N^{23} - N^4\,,
\ea
where the terms on the right-hand side are derived in
\cref{sec:sfb_local_average_effect_with_weighting}. In practice, all but the
$C^1$ and $N^1$ terms essentially vanish for $\ell \gtrsim 2/f_\sky$.

\subsection{Pixel window}
With our \emph{SuperFaB} estimator only angular modes are affected by the
HEALPix pixelization scheme \citet{Gorski+:2005ApJ...622..759G}. We include the
HEALPix pixel window in the theoretical modeling after the window and weights
convolution, and before adding the shot noise, since shot noise is not affected
by the pixel window.

\section{Validation with lognormal mocks}
\label{sec:ln}
\begin{figure*}
  \centering
  \incgraph{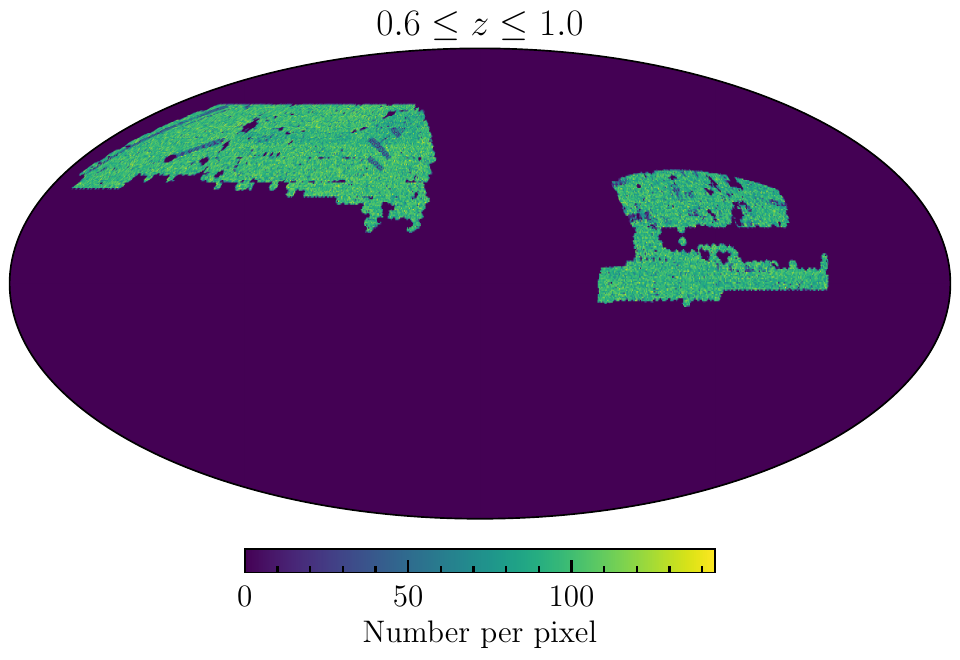}
  \incgraph{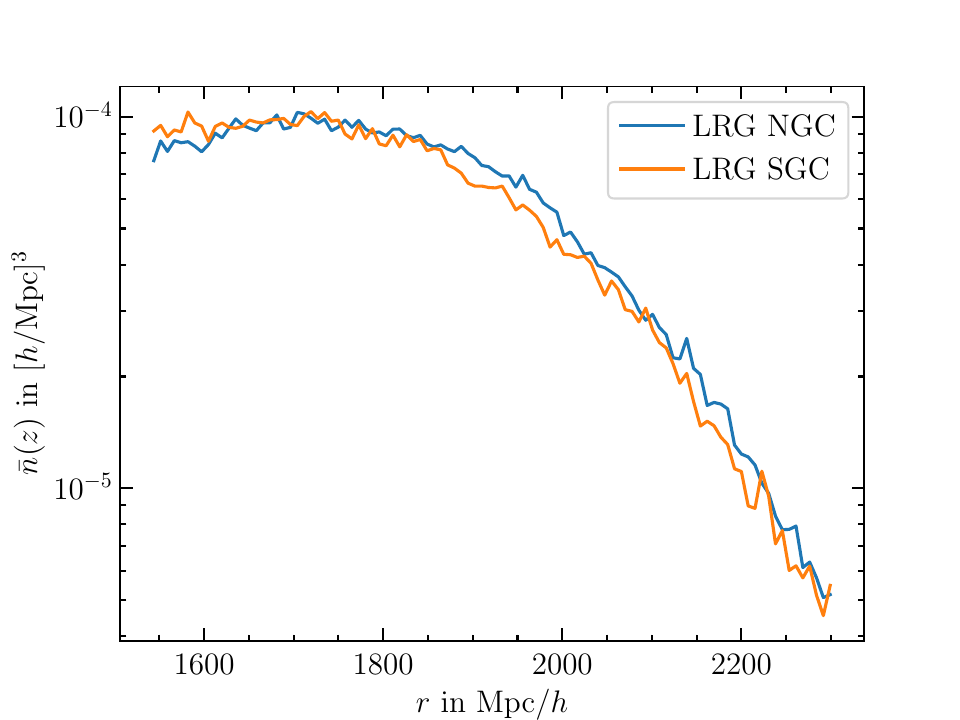}
  \caption{
    Catalog projection eBOSS DR16 LRGs.
    Left: angular projection at Healpix resolution $n_\mathrm{side}=256$.
    Right: radial selection function.
  }
  \label{fig:window}
\end{figure*}
In this section we validate our pipeline using our own implementation of
lognormal mocks \citep{Coles+:1991MNRAS.248....1C,
Xavier+:2016MNRAS.459.3693X}. Similar to \citet{Agrawal+:2017JCAP...10..003A},
our lognormals contain linear redshift space distortions, and optionally can
have a large Gaussian FoG component.

We start with validating our model calculation on the full sky, then our window
convolution and the importance of off-diagonal terms, local average effect, and
boundary conditions.

\subsection{Full-sky lognormal mocks}
In \cref{fig:ln_fullsky_thinshell} we start with a full-sky survey in the radial range
\SIrange{750}{1000}{\per\h\mega\parsec}. The three panels compare our model
described in \cref{sec:sfb_theory} to the average over \num{1000} simulations; the
first panel to the left compares the shot noise calculation with the estimator
run on a uniform random data catalog; the center panel compares the model with
a lognormal simulation in real space; the last panel compares the same in
redshift space.

In each panel of \cref{fig:ln_fullsky_thinshell}, the color indicates the
$\ell$-mode. Further, the modes broadly break into those that are on the
diagonal $k=k'$ and those that are off-diagonal. For the modes shown, the
diagonal terms are all above $C_{\ell
nn}\gtrsim\SI{1000}{\per\h\cubed\mega\parsec\cubed}$, while the off-diagonal
$k\neq k'$ terms are closer (but not necessarily equal) zero. Crucially,
non-zero power is present in the off-diagonal terms even in real space.

\subsection{eBOSS DR16 LRG lognormal mocks}
\begin{figure*}
  \centering
  \incgraph[0.32]{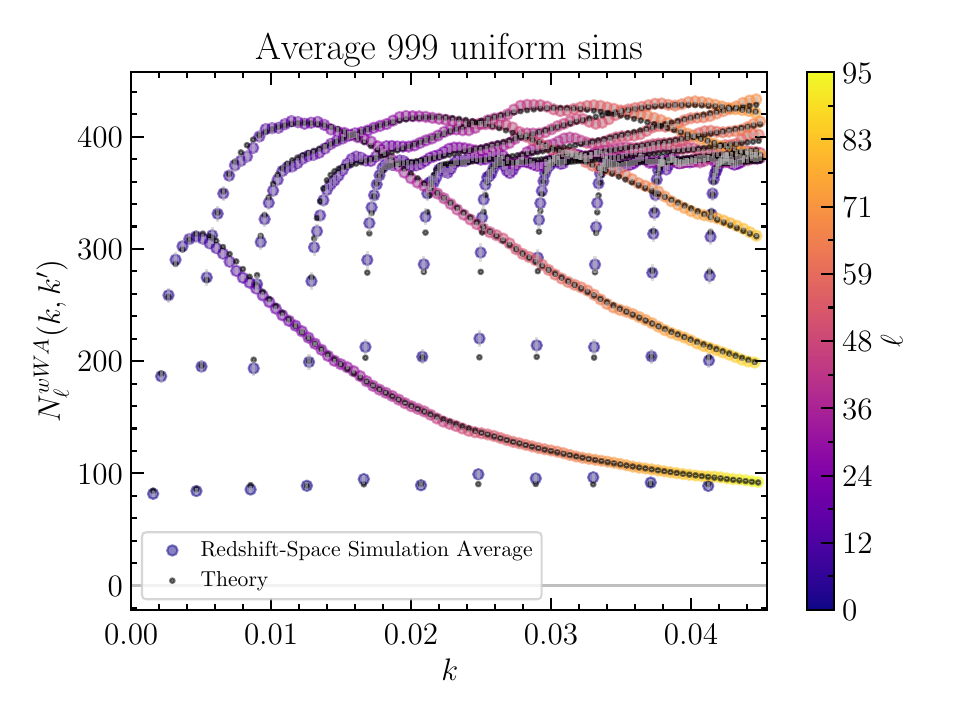}
  \incgraph[0.32]{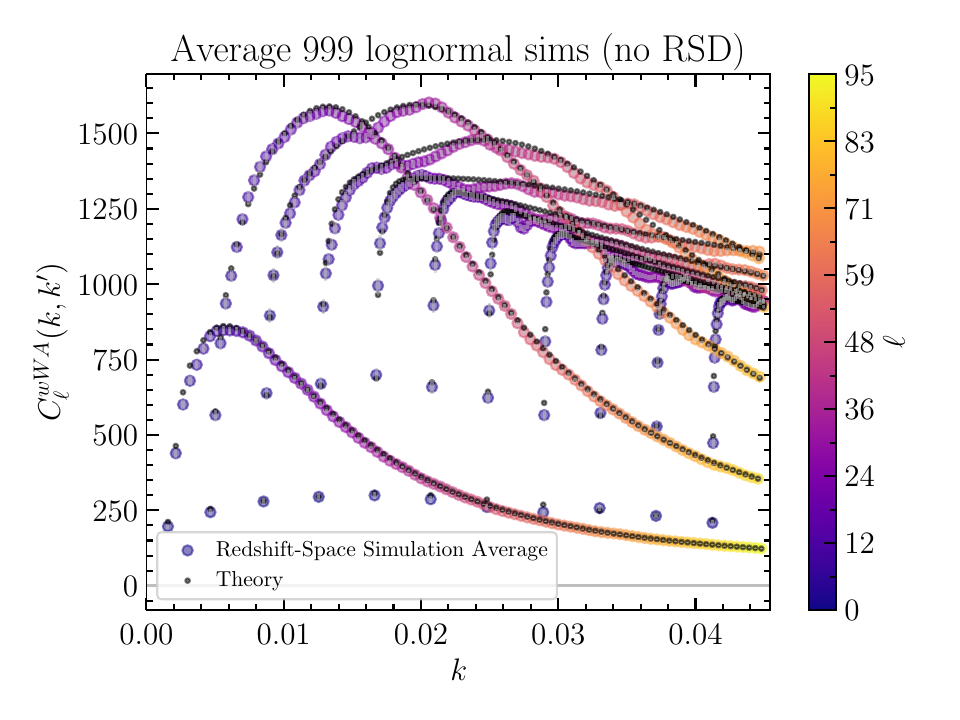}
  \incgraph[0.32]{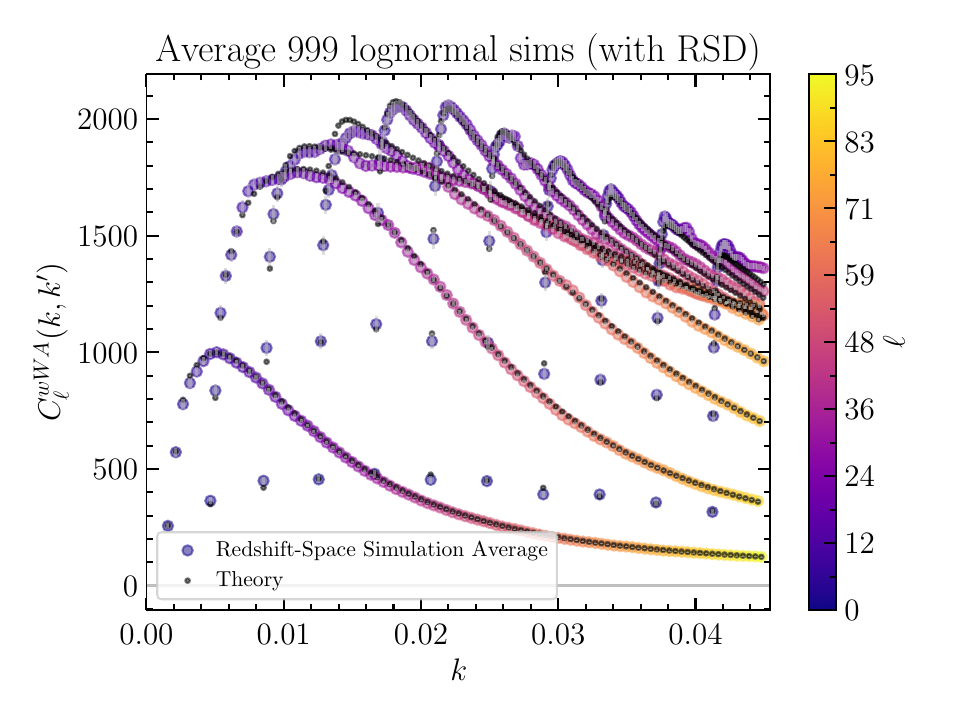}
  \incgraph[0.32]{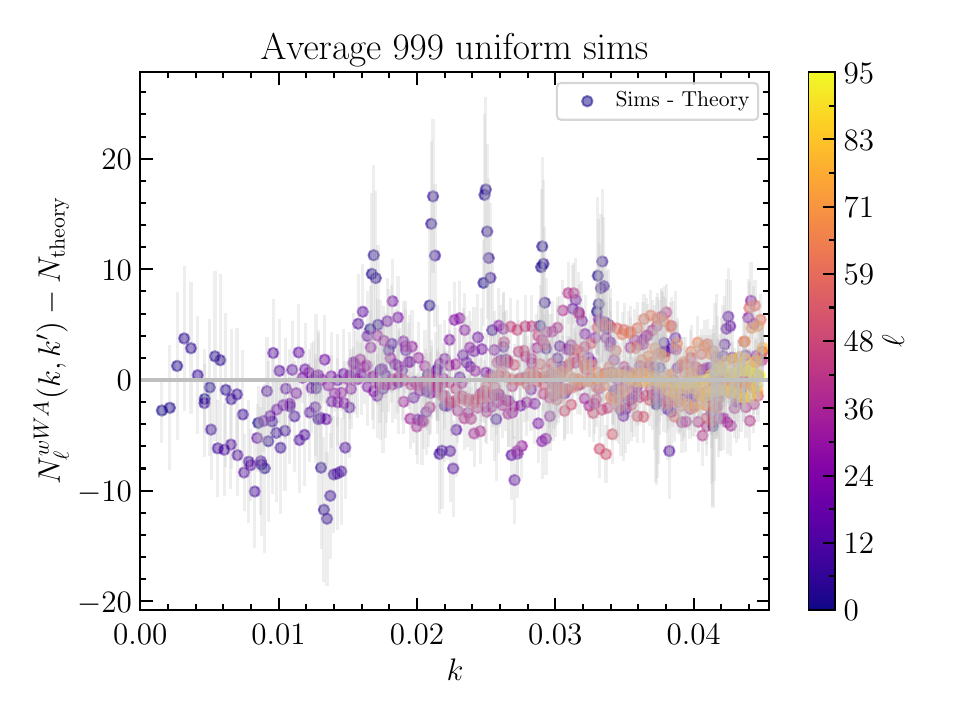}
  \incgraph[0.32]{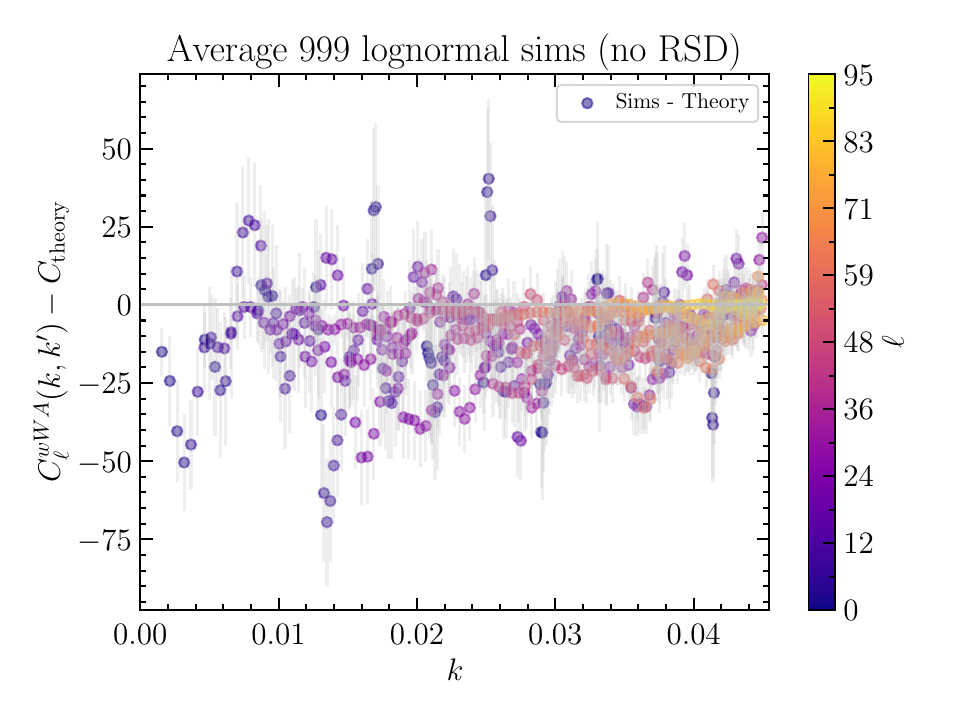}
  \incgraph[0.32]{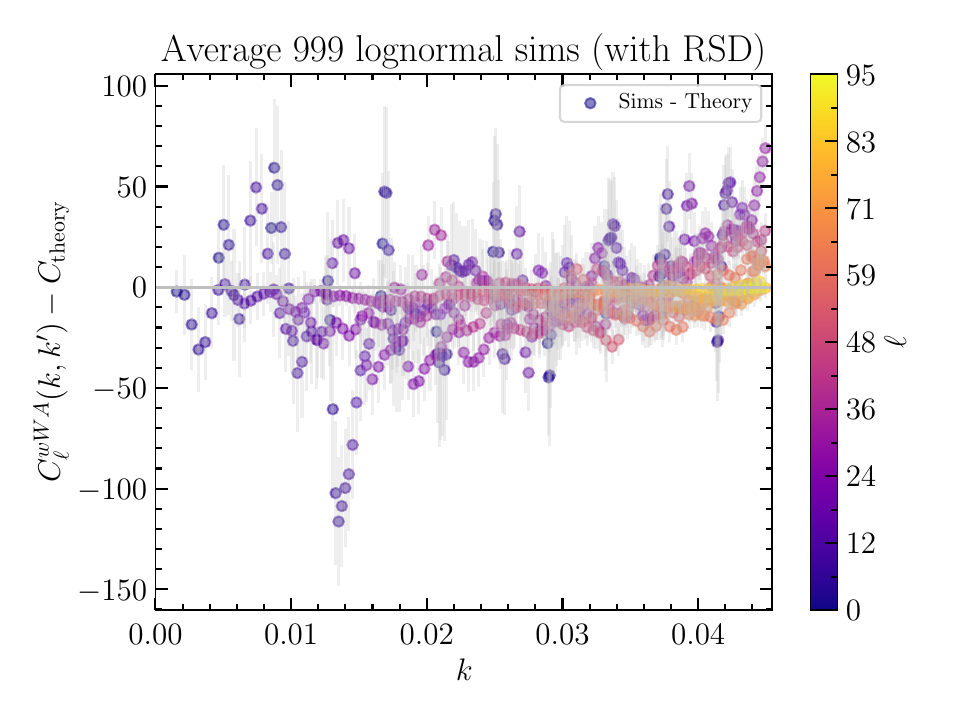}
  \caption{eBOSS-NGC-LRG-sky lognormal mocks.
    We only display $k=k'$ ($\Delta n=0$) modes.
    The top row of panels displays the power spectrum for the three cases of
    uniform random mocks (left), lognormal mocks in real space (center), and
    lognormal mocks in redshift space (right).
    The bottom row displays the corresponding residuals to the theory.
  }
  \label{fig:ln_lrg_ngc}
\end{figure*}
\begin{figure*}
  \centering
  \makebox[0.32\textwidth]{$\Delta n_\max=0$:}
  \makebox[0.32\textwidth]{$\Delta n_\max=2$:}
  \makebox[0.32\textwidth]{$\Delta n_\max=4$:}
  \\
  \incgraph[0.32]{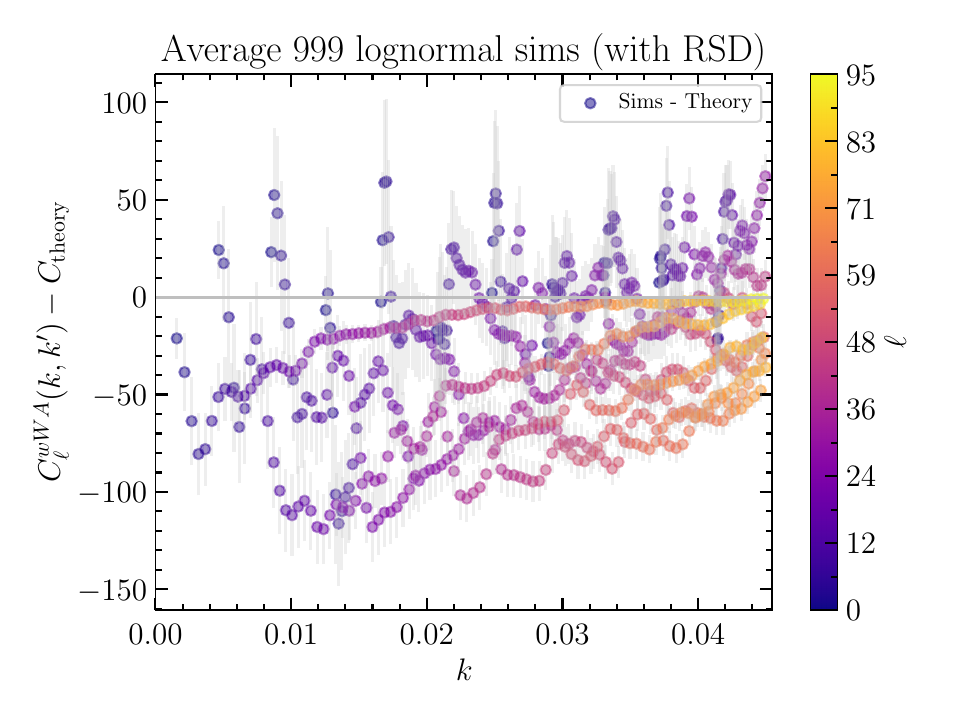}
  \incgraph[0.32]{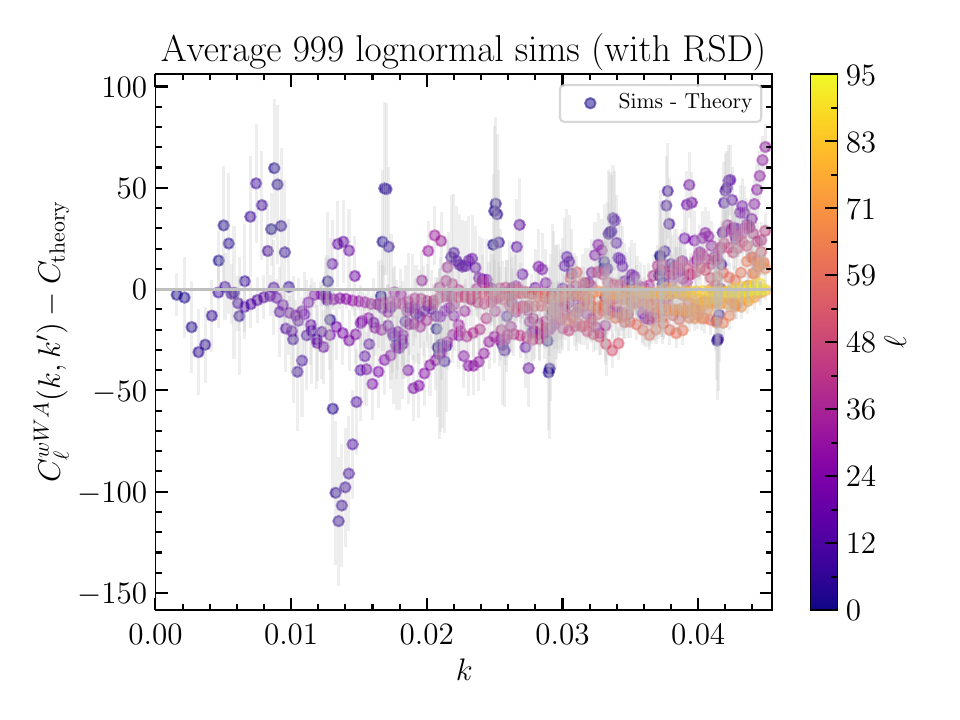}
  \incgraph[0.32]{fig_3f}
  \caption{
    LRG-sky redshift-space lognormals. If the model is calculated only with
    on-diagonal modes ($k=k'$, or $\Delta n=0$), then there is significant
    discrepancy between the simulation average and the theory model, as shown
    in the left panel. Even with just 2 off-diagonals included (center panel)
    the modeling is significantly improved. The right panel shows when modeling
    with up to $\Delta n_\max=4$ off-diagonals.
  }
  \label{fig:ln_lrg_ngc_dnmax}
\end{figure*}
\begin{figure*}
  \centering
  \incgraph[0.49]{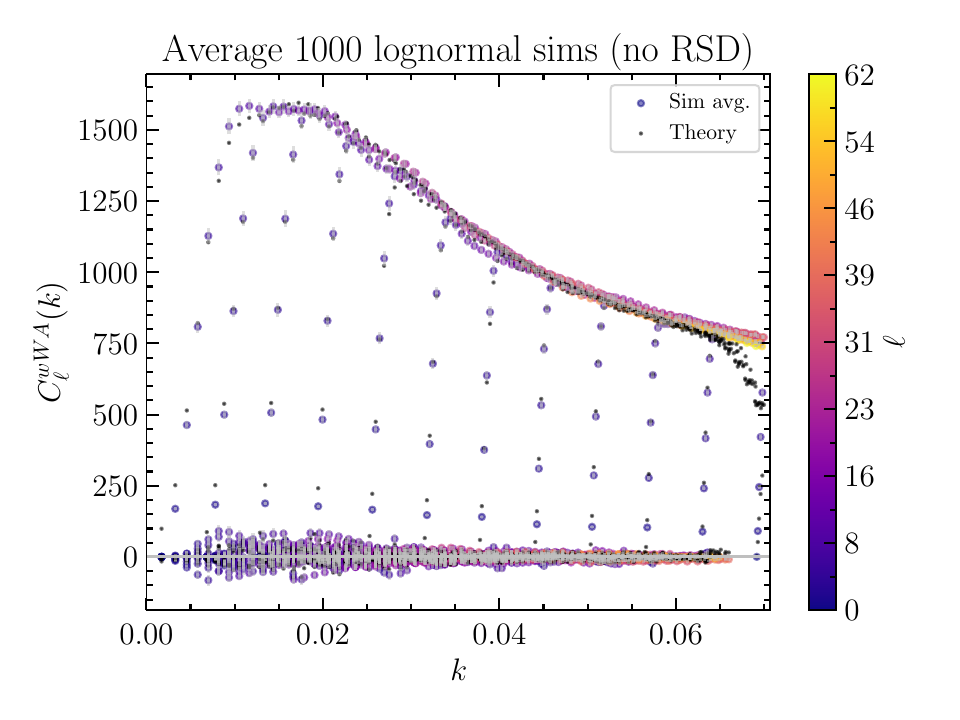}
  \incgraph[0.49]{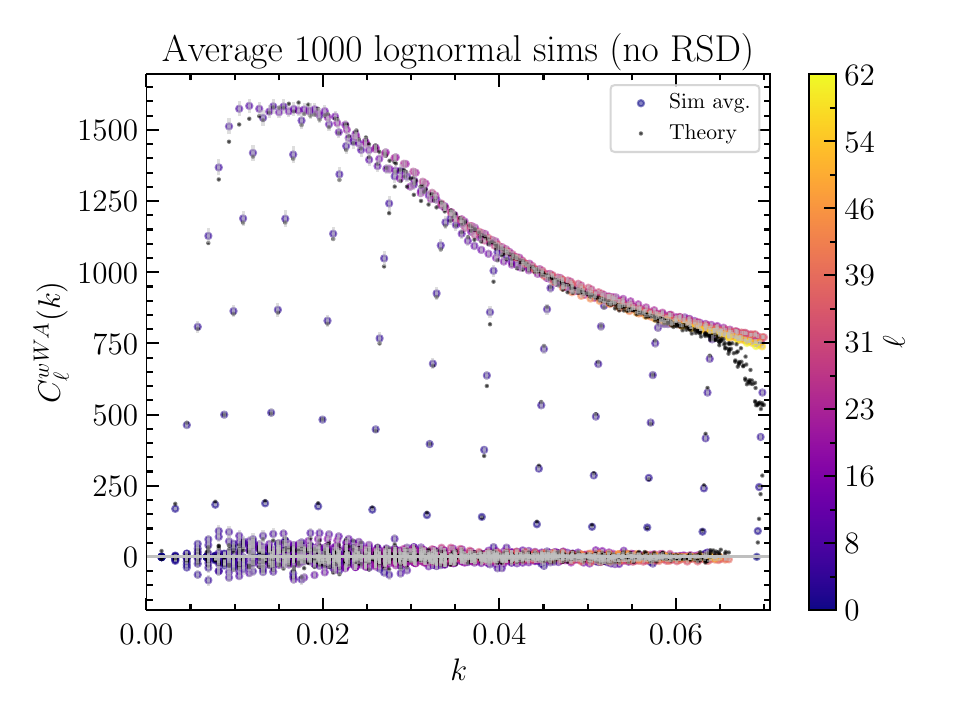}
  \caption{1/16-th sky, constant radial selection between
    \SIrange{500}{1000}{\per\h\mega\parsec}.
    Left: theory local-average effect (LAE) calculation up to $\ell_\max\lesssim16$.
    Right: theory LAE calculation up to $\ell_\max\lesssim32$.
  }
  \label{fig:ln_lae}
\end{figure*}
\begin{figure*}
  \centering
  \incgraph[0.32]{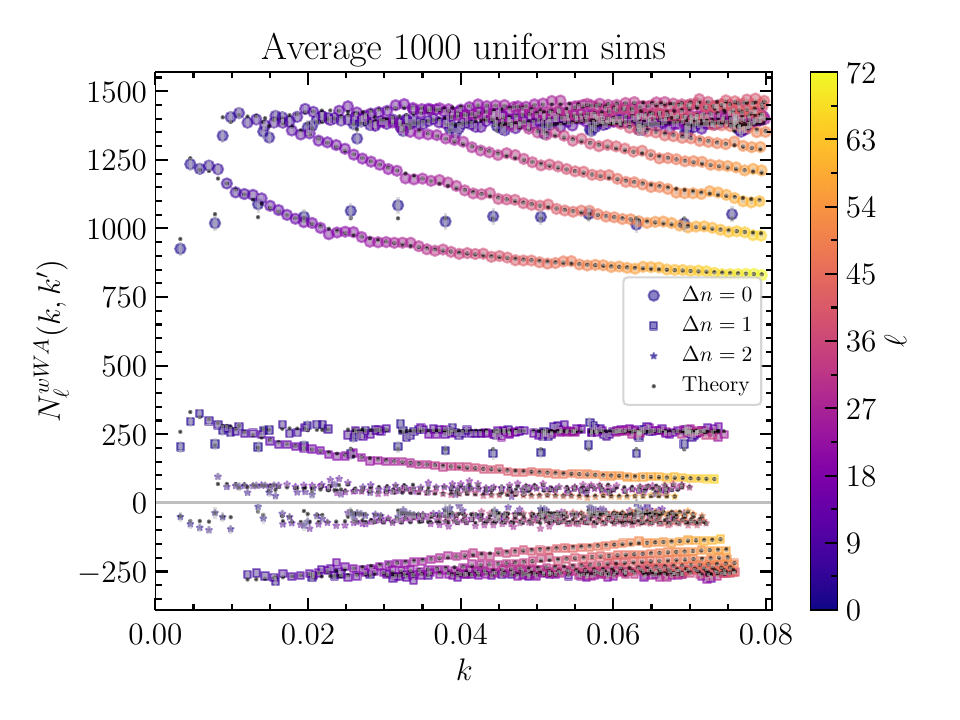}
  \incgraph[0.32]{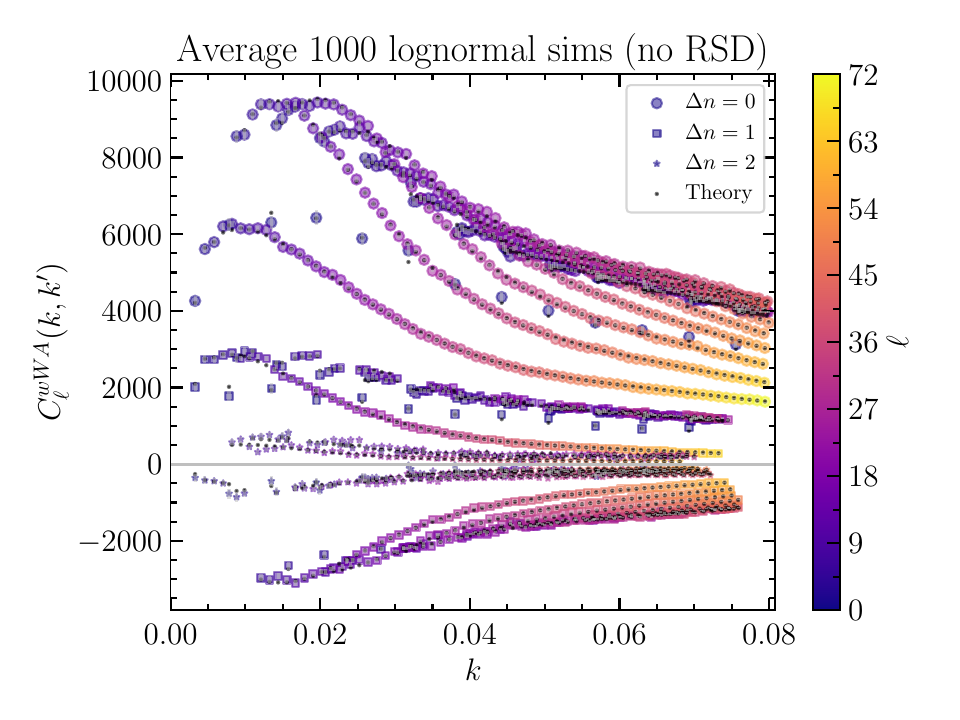}
  \incgraph[0.32]{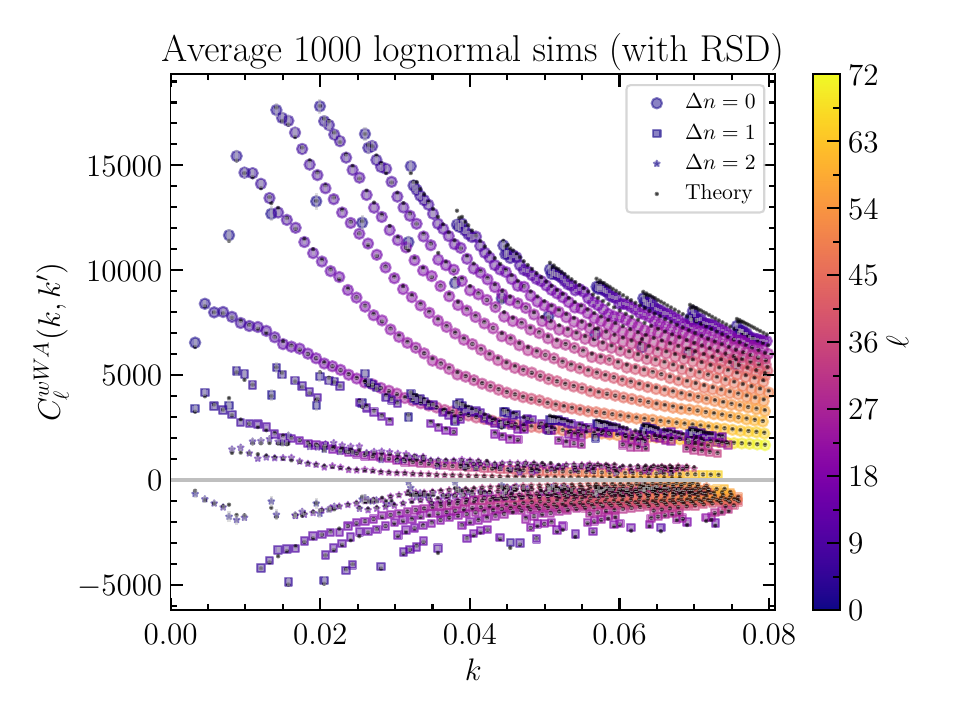}
  \incgraph[0.32]{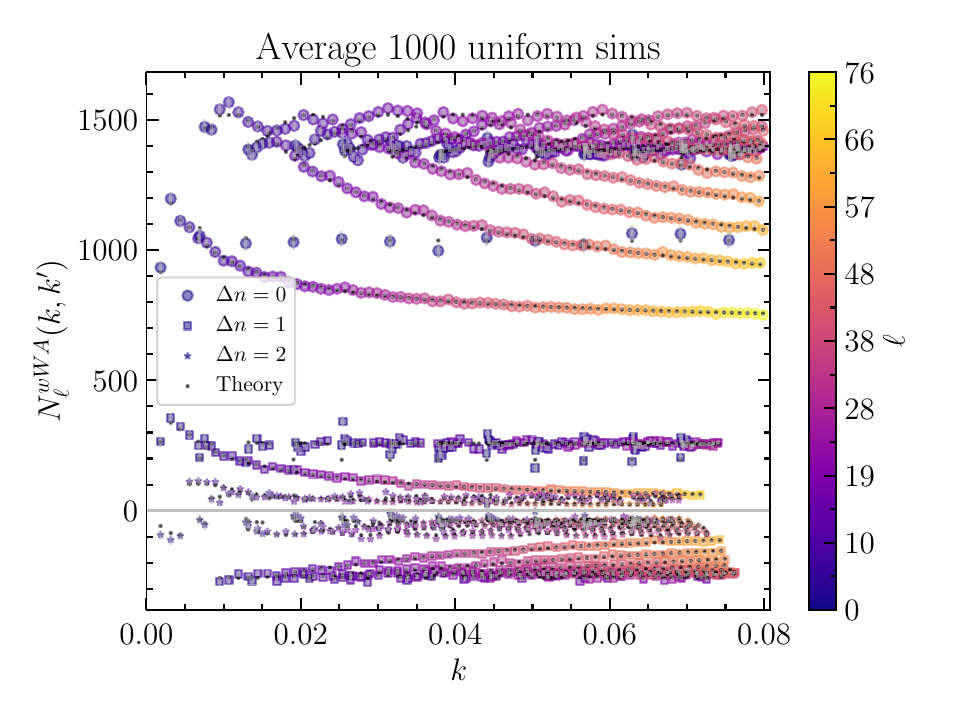}
  \incgraph[0.32]{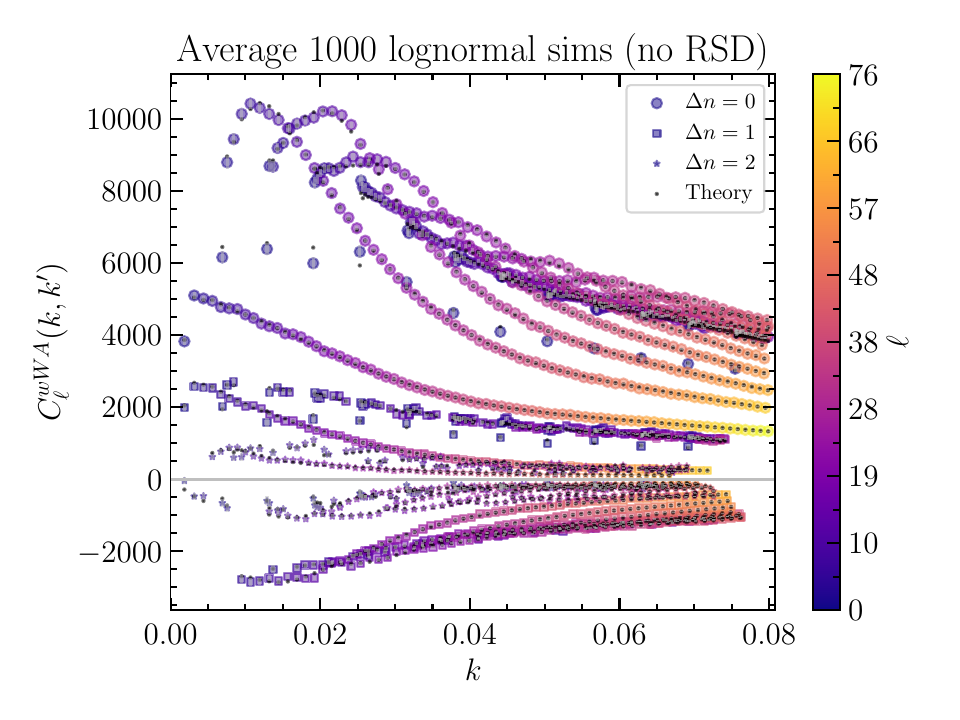}
  \incgraph[0.32]{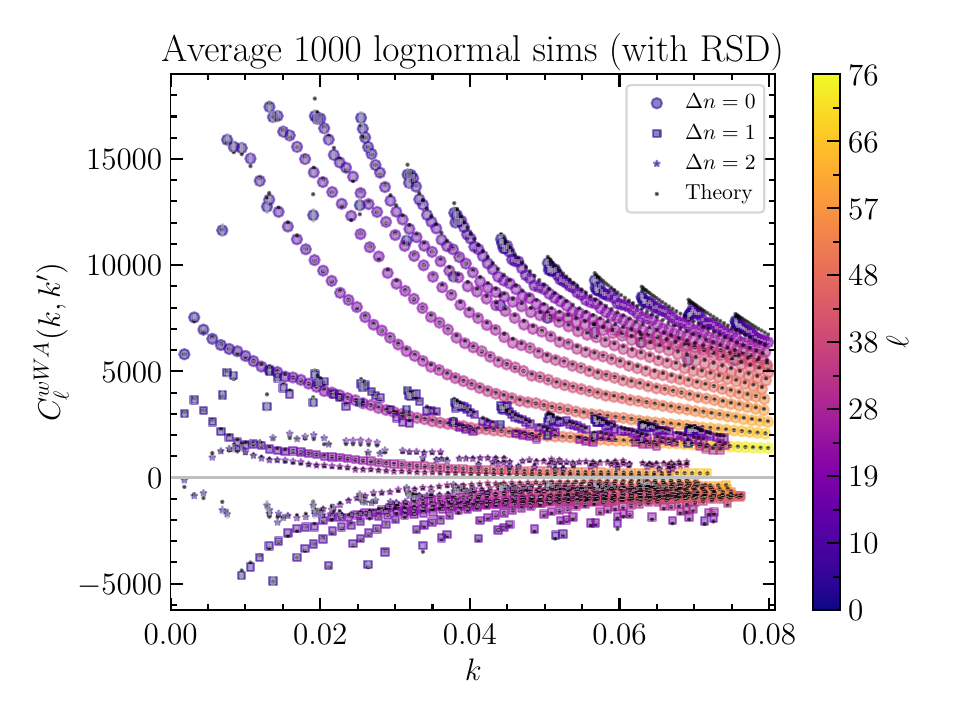}
  \caption{Comparison between potential (top) and velocity (bottom) boundary conditions with
    lognormal simulations for a thin shell covering half the sky.
    Top row shows the modes with potential boundaries that ensure a continuous
    and smooth field at the boundary, bottom row with velocity boundaries that
    set the vanishing velocity at the boundary.
  }
  \label{fig:ln_boundaries}
\end{figure*}
\begin{figure}
  \centering
  \incgraph[0.23]{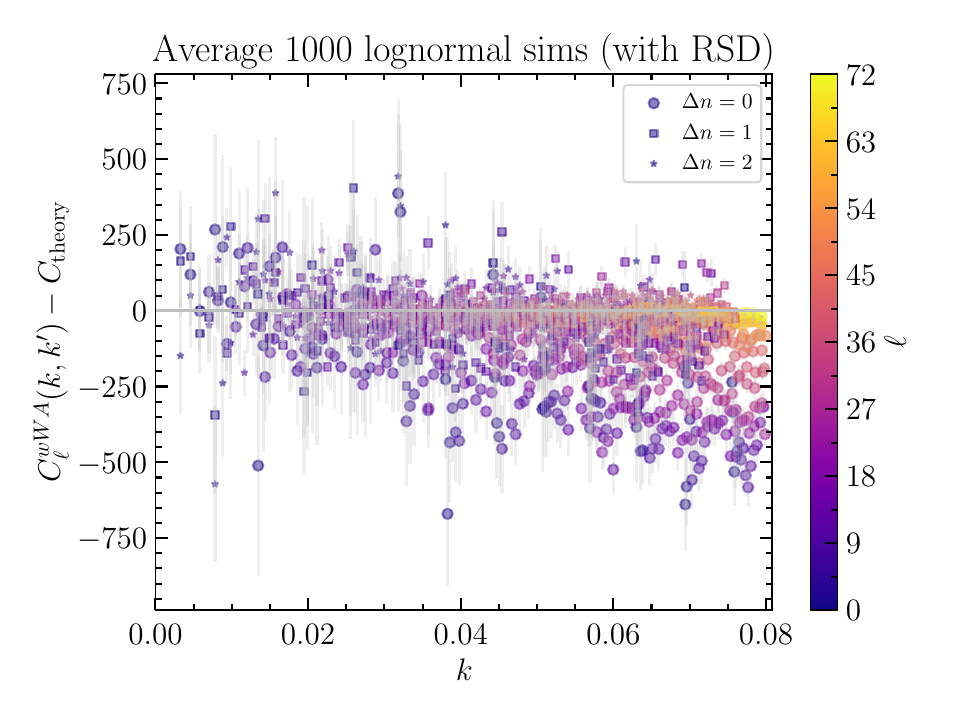}
  \incgraph[0.23]{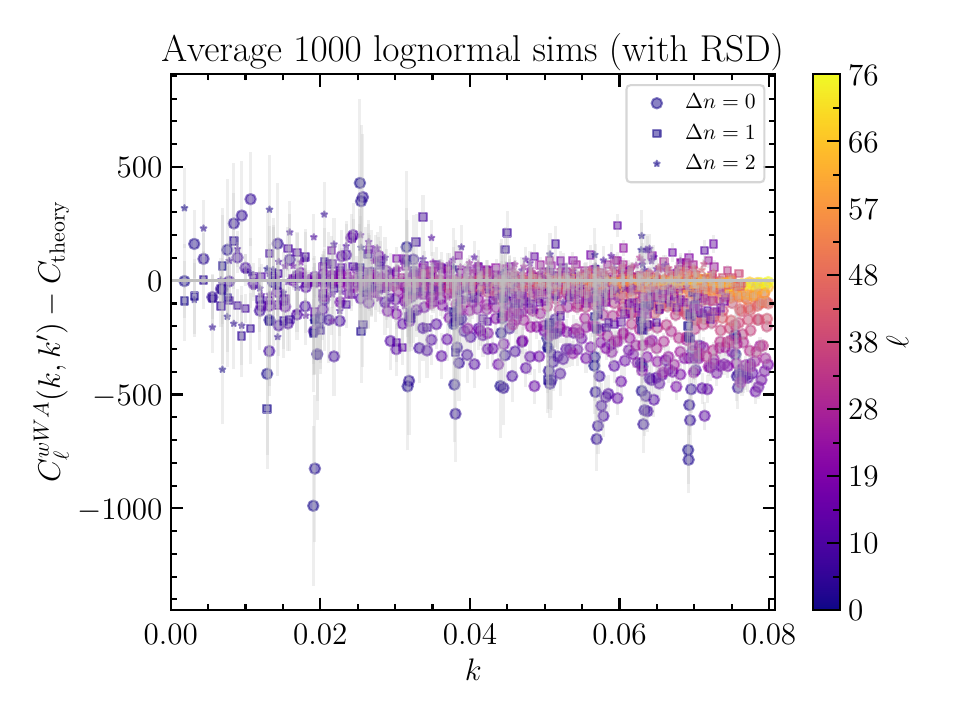}
  \incgraph[0.23]{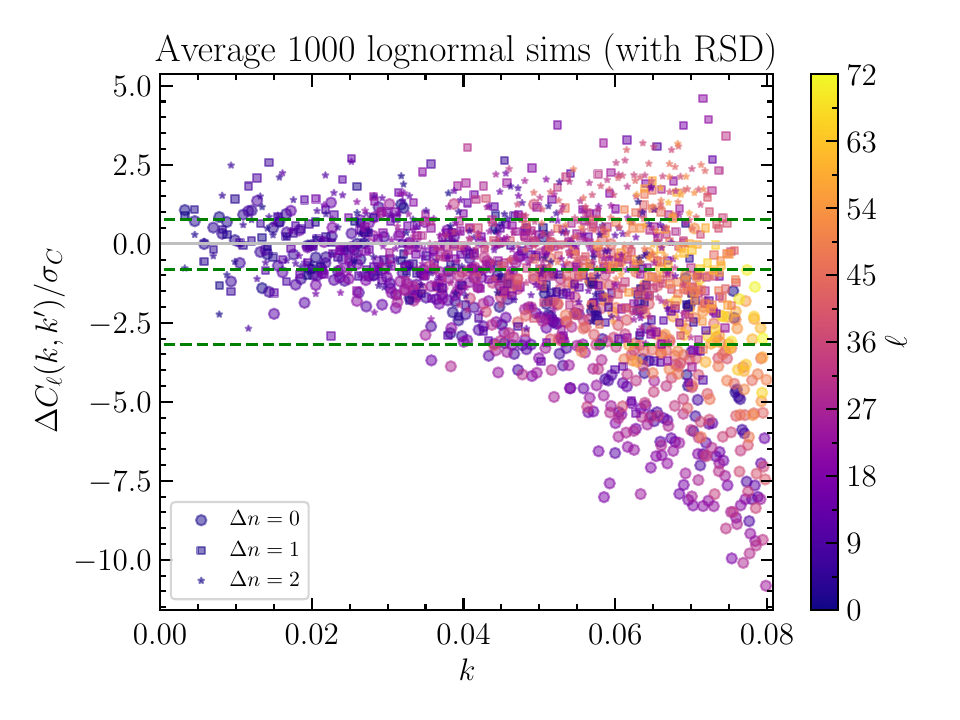}
  \incgraph[0.23]{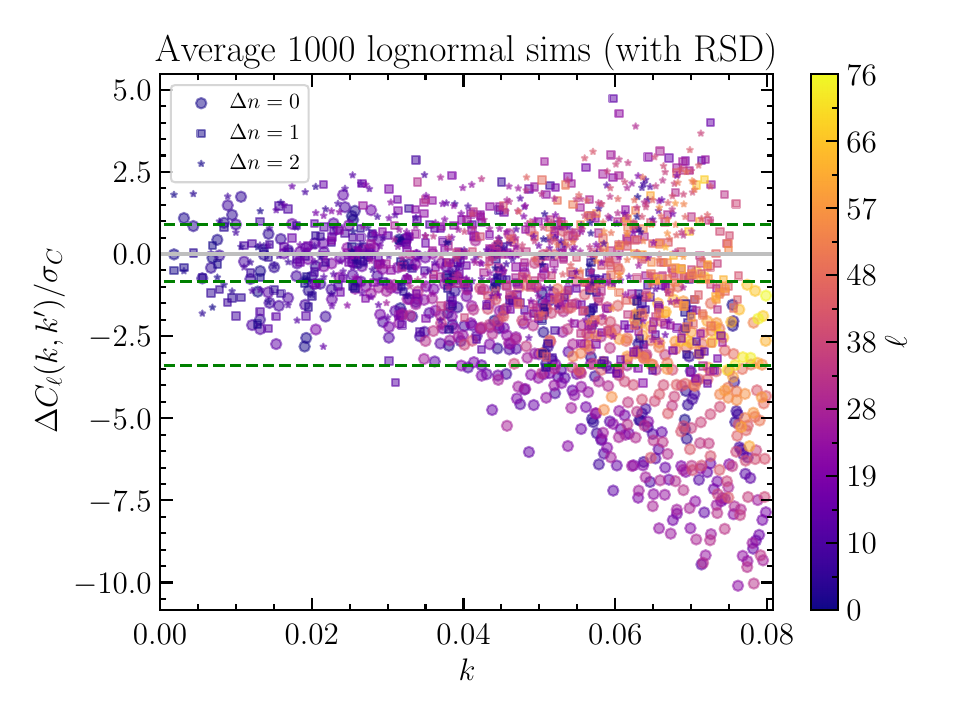}
  \caption{Left column shows the potential boundary modes, right column the
    velocity boundary modes, and the difference to our theory code for
    lognormal simulations with RSD. While the lognormals are afflicted by
    systematics, they are essentially the same whether potential or velocity
    boundaries are used. See right column of \cref{fig:ln_boundaries} for the
    absolute values of the modes.
  }
  \label{fig:ln_boundaries_diffs}
\end{figure}
Next, we validate our pipeline using the eBOSS DR16 LRG mask in the North
Galactic Cap (NGC) and its radial selection function. The radial selection and
angular windows are estimated from 1000 random catalogs provided by the SDSS
collaboration \citep{Zhao+:2021MNRAS.503.1149Z}. Since we use the estimated
window as the exact window for both generating and analysing the lognormal
mocks, the details of this window-estimation procedure are irrelevant here, and
we defer its description to \cref{sec:ez}.

Suffice to say that the resulting angular window encodes the fraction of each
pixel that is observed. On the left of \cref{fig:window} we show the number of
points in each pixel for the random catalog. Both NGC and SGC are shown.
However, in this section we only use the NGC. The radial selection is shown on
the right of \cref{fig:window}, and it is estimated from the data catalog. We
again defer the details to \cref{sec:ez}.

To ensure the presence of the local average effect (\cref{sec:lae}), we
generate uniform random catalogs that match the radial number density
distribution of each ``data'' catalog in the following way. For each redshift
bin we draw exactly the same number of galaxies as in the data catalog. We
adhere to the angular varying-depth window at that redshift by first uniformly
drawing galaxies on the full sky, then rejecting points with probability
proportional to the depth of the window. In this way a random catalog is
created with the exact same redshift distribution as the data catalog (within
the resolution of the analysis), and an angular distribution matching the
angular window at each redshift.

We show our lognormal simulation results in \cref{fig:ln_lrg_ngc}, for uniform
random data catalogs (left panels), real-space lognormals (center panels), and
redshift-space lognormals (right panels). In addition to the observed power
spectrum (convolved with window), we also show the difference to the calculated
theory power spectrum. We get generally good agreement. The discrepancy at $k
\gtrsim \SI{0.04}{\h\per\mega\parsec}$ comes from the incomplete window
convolution that needs to make use of modes above our maximum
$k_\max=\SI{0.05}{\h\per\mega\parsec}$ used in the analysis.

\subsubsection{Importance of off-diagonal terms}
\label{sec:ln_dnmax}
In \cref{fig:ln_lrg_ngc} we used off-diagonal $k_{n+\Delta n,\ell}\neq
k_{n\ell}$ terms up to $\Delta n_\max=4$. As already mentioned for the full-sky
case, these off-diagonal terms are important. We explicitly show this in
\cref{fig:ln_lrg_ngc_dnmax}, where we limit $\Delta n_\max=0$ (left panel),
$\Delta n_\max=2$ (center panel), and $\Delta n_\max=4$ (right panel) for the redshift-space mocks. The figure
shows that the on-diagonal $\Delta n=0$ terms are sufficiently modeled using
$\Delta n_\max=2$. However, to model the off-diagonal terms as well, we
conclude from \cref{fig:ln_lrg_ngc,fig:ln_lrg_ngc_dnmax} that for the eBOSS
DR16 LRG redshift range and radial selection, $\Delta n_\max=4$ is needed and
sufficient.

\subsubsection{Local average effect modeling}
\label{sec:ln_lae}
Since the calculation for the local average effect in
\cref{sec:sfb_local_average_effect_with_weighting} is computationally
expensive, we perform the calculation only up to some $\ell_\max$. In
\cref{fig:ln_lae} we show the result when $\ell_\max=1/f_\sky$ and when it is
$\ell_\max=2/f_\sky$. The residuals only become percent-level with the larger
$\ell_\max$.

\subsubsection{Potential vs. velocity boundaries}
\label{sec:ln_boundary}
While potential boundaries have the property of being able to better model the
density field with fewer modes, they have the unfortunate property that a
sphere in real space does not stay a sphere in redshift space. Thus, our
modeling in \cref{sec:sfb_theory} is technically inconsistent. However, those
boundary effects are vanishingly small in the case of eBOSS DR16 LRGs, which we
show in \cref{fig:ln_boundaries}, where we compare with \emph{velocity}
boundary conditions \citep{Fisher+:1995MNRAS.272..885F} that set the derivative
to vanish on the boundaries, and, thus, a sphere in real space is also a sphere
in redshift space. We show the derivation of the basis functions for velocity
boundaries (including $k=0$ mode) in \cref{app:gnl_velocity}.

To better assess if there is a relevant difference between the boundary
conditions, \cref{fig:ln_boundaries_diffs} shows the difference between them.
Since we use lognormal simulations that are inaccurate especially on small
scales, we don't expect the theory and measured average to agree perfectly.
However, both potential and velocity boundaries show the same systematics.
Thus, we conclude that the difference is insignificant.

\section{Validation with EZmocks}
\label{sec:ez}
To capture the full complexity of the eBOSS data, we use the EZmocks generated
by \citet{Zhao+:2021MNRAS.503.1149Z}. This allows us to test several potential
sources of systematics that were not included in the lognormal mocks. First, we
estimate the window from the 1000 random catalogs provided, and so the randoms
are afflicted by a shot noise component. However, we find that this is
negligible. Second, the \emph{complete} EZmocks include redshift-evolution in
the growth factor, galaxy bias, and radial dispersion (i.e., FoG and redshift
uncertainty) that is lacking in the lognormal mocks. Third, the
\emph{realistic} EZmocks build on the \emph{complete} EZmocks adding
photometric systematics, fiber collisions, and redshift failures
\citep{Zhao+:2021MNRAS.503.1149Z, Ross+:2020MNRAS.498.2354R,
  Raichoor+:2021MNRAS.500.3254R, Gil-Marin+:2020MNRAS.498.2492G,
Neveux+:2020MNRAS.499..210N}.

We match the cosmology to that used to create the EZmocks: a flat $\Lambda$CDM
cosmology with $H_0=\SI{67.77}{\kilo\meter\per\second\per\mega\parsec}$ and
$\Omega_m=0.307115$. In our fiducial cosmology the LRG redshift range $0.6 \leq
z \leq 1.0$ corresponds to radial distances
$\num{1540}<\frac{r}{\si{\per\h\mega\parsec}}<\num{2467}$.

Our procedure for estimating the angular mask/window is as follows. We average
over 1000 random catalogs provided by the SDSS collaboration
\citep{Zhao+:2021MNRAS.503.1149Z}. At HEALPix resolution $n_\mathrm{side}=256$
and with 554 comoving distance bins, the number of random catalog points is
added up for each voxel\footnote{We use the word \emph{voxel} in 3 dimensions
in analogy to \emph{pixel} in 2 dimensions.}, to then estimate the number
density in each voxel. The angular window is estimated by taking the mean over
all redshift bins, and then normalizing.
The angular footprint on sky is shown on the left of \cref{fig:window}.

The radial selection function is estimated from the data in each cap
separately. First, the data catalog is binned into 40 radial bins of width
$\Delta z=0.01$, and then cubically splined.
The radial selection function is shown on the right of \cref{fig:window}.

\subsection{\emph{Complete} EZmocks}
\label{sec:ez_complete}
\begin{figure*}
  \centering
  Slice 1: $z=0.6-0.645$\\
  \incgraph[0.28]{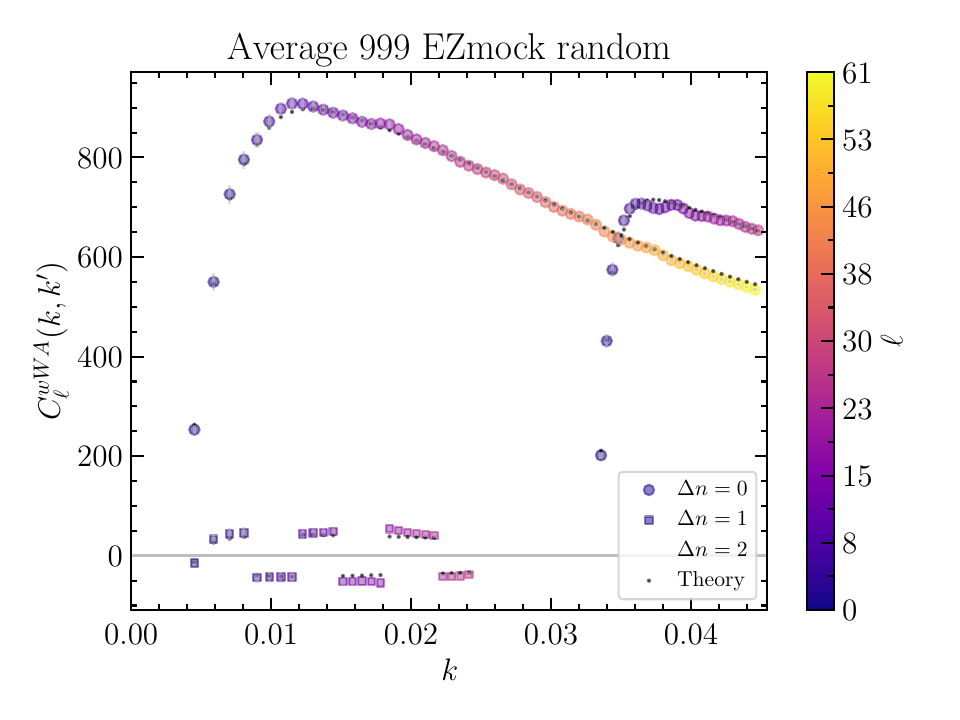}
  \incgraph[0.28]{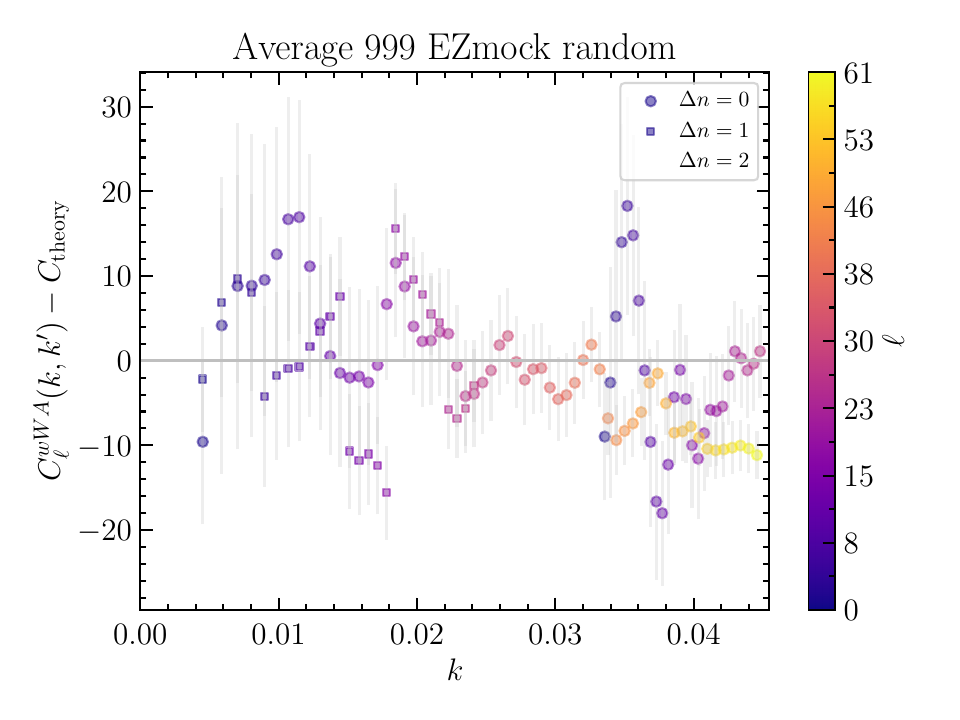}
  \incgraph[0.28]{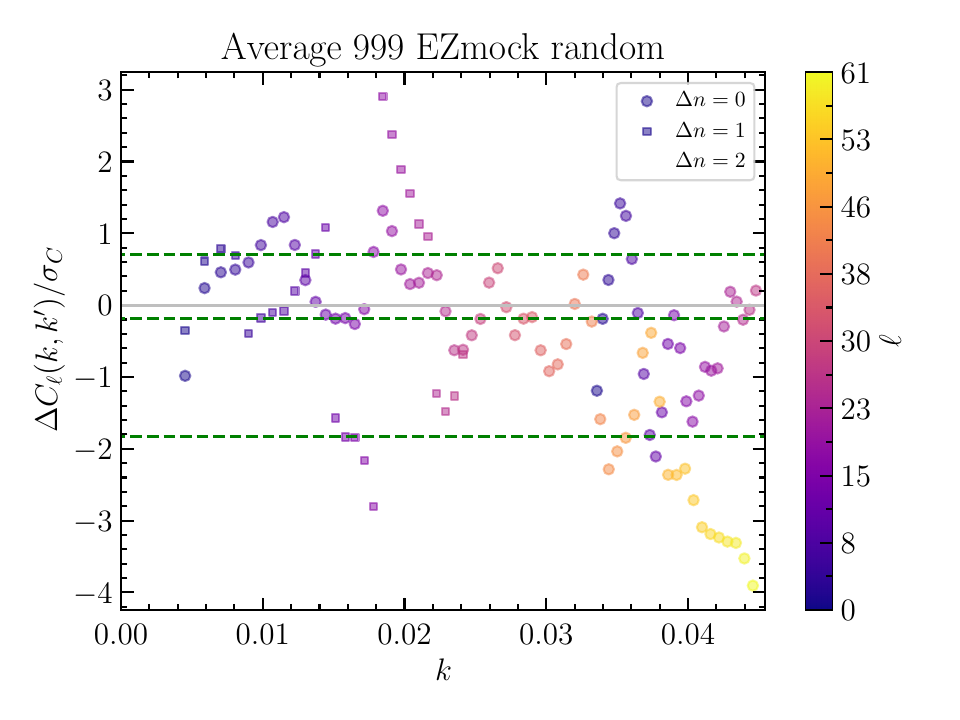}
  \incgraph[0.14]{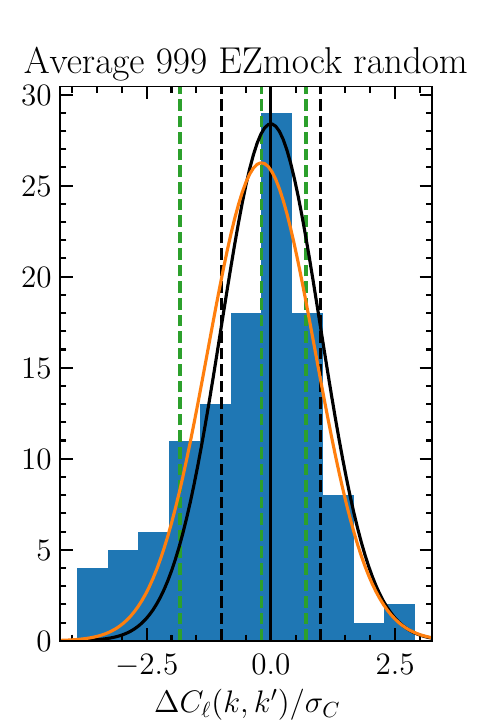}
  Slice 2: $z=0.655-0.695$\\
  \incgraph[0.28]{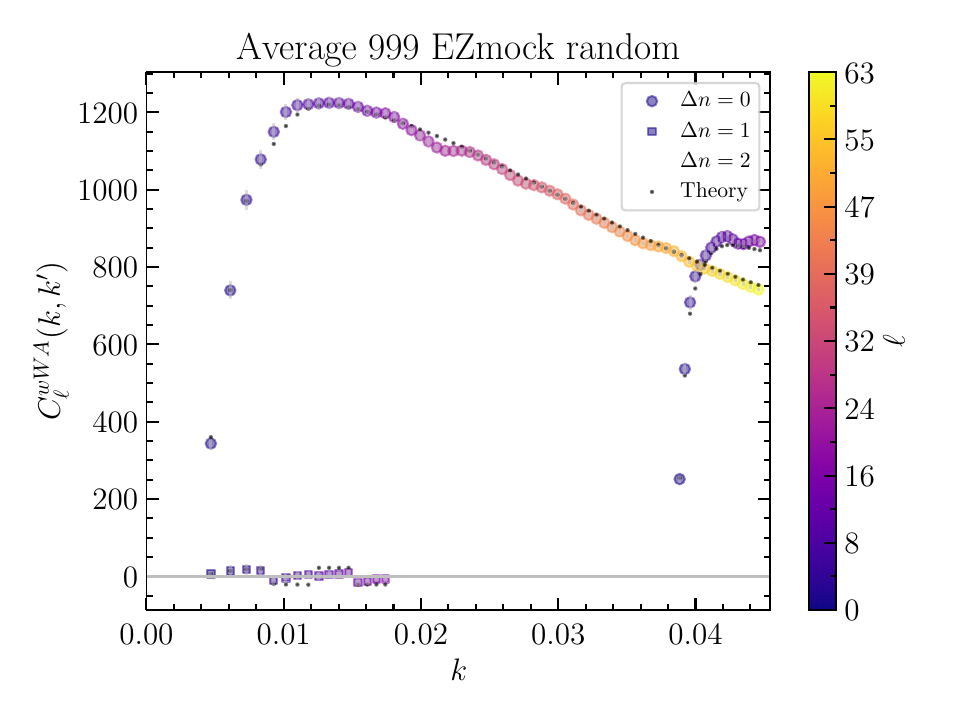}
  \incgraph[0.28]{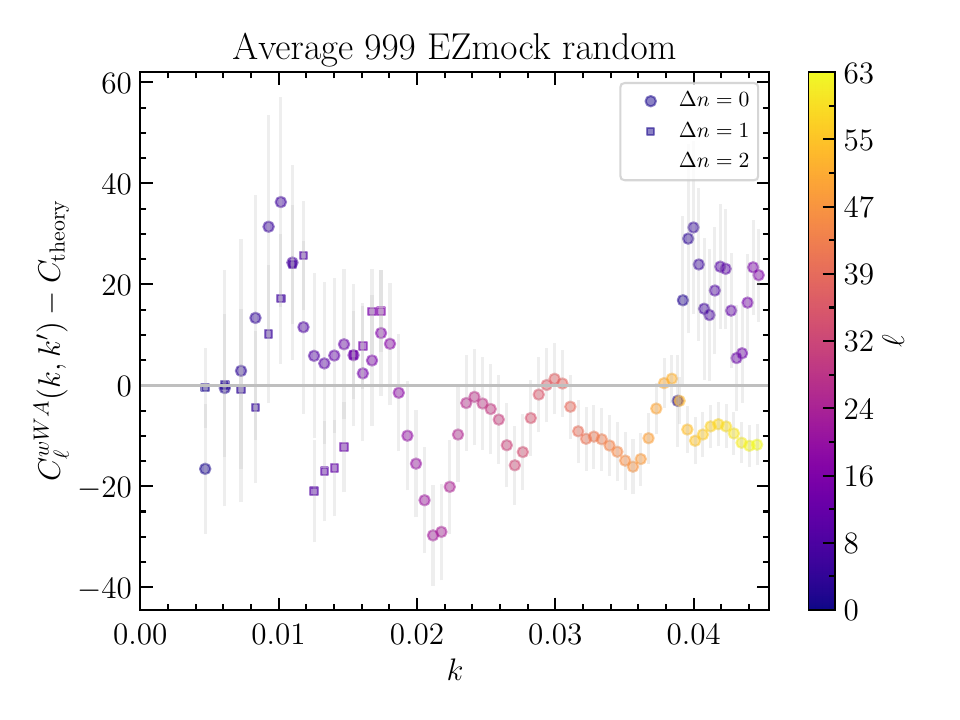}
  \incgraph[0.28]{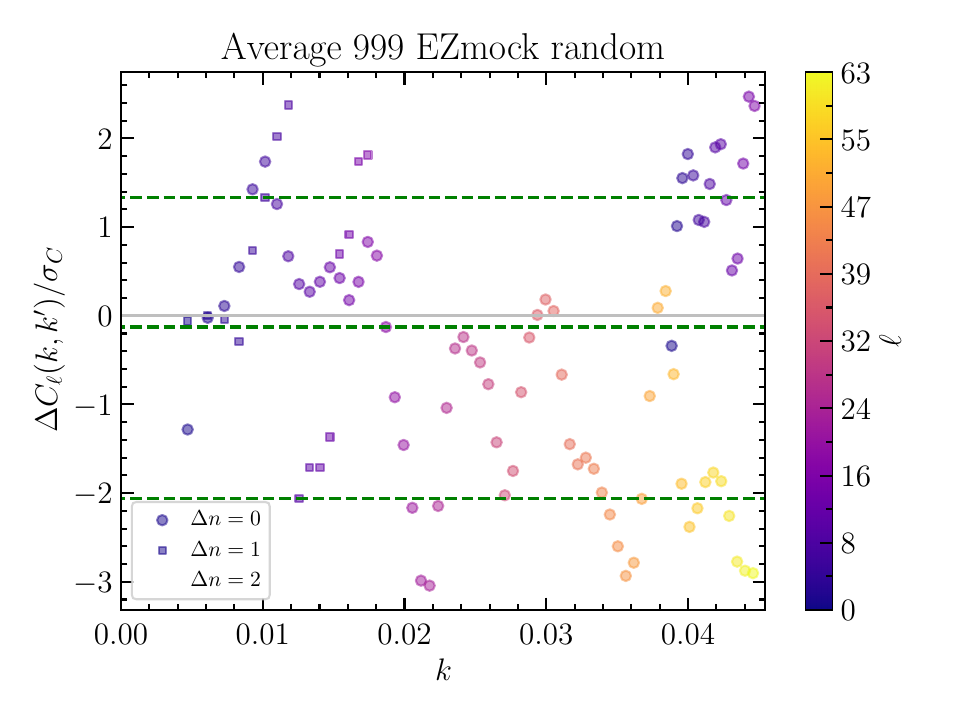}
  \incgraph[0.14]{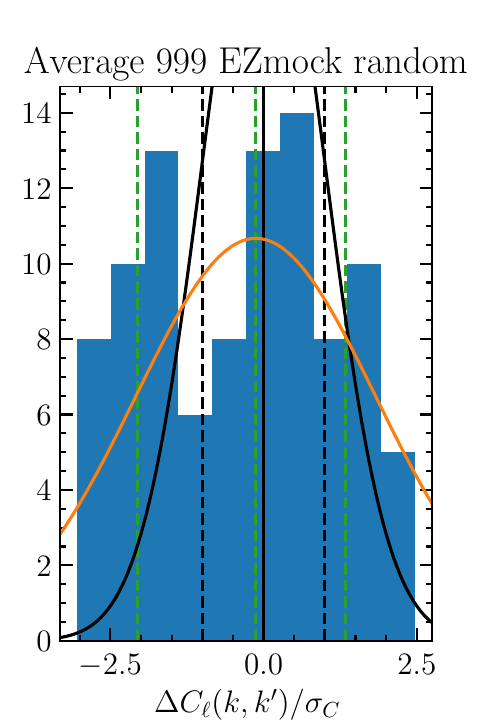}
  Slice 3: $z=0.705-0.795$\\
  \incgraph[0.28]{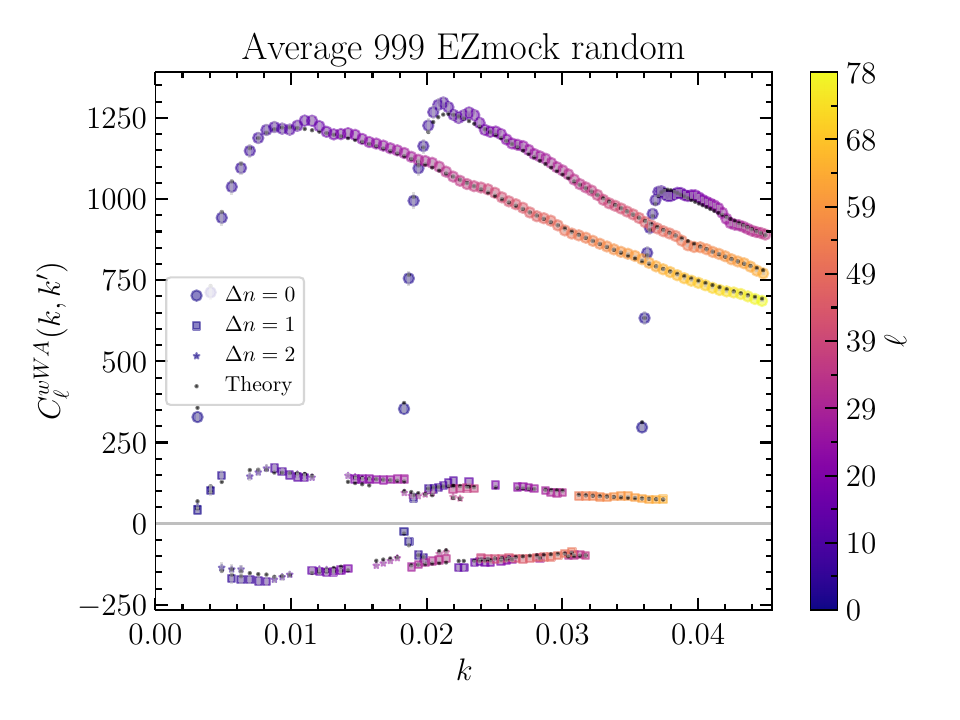}
  \incgraph[0.28]{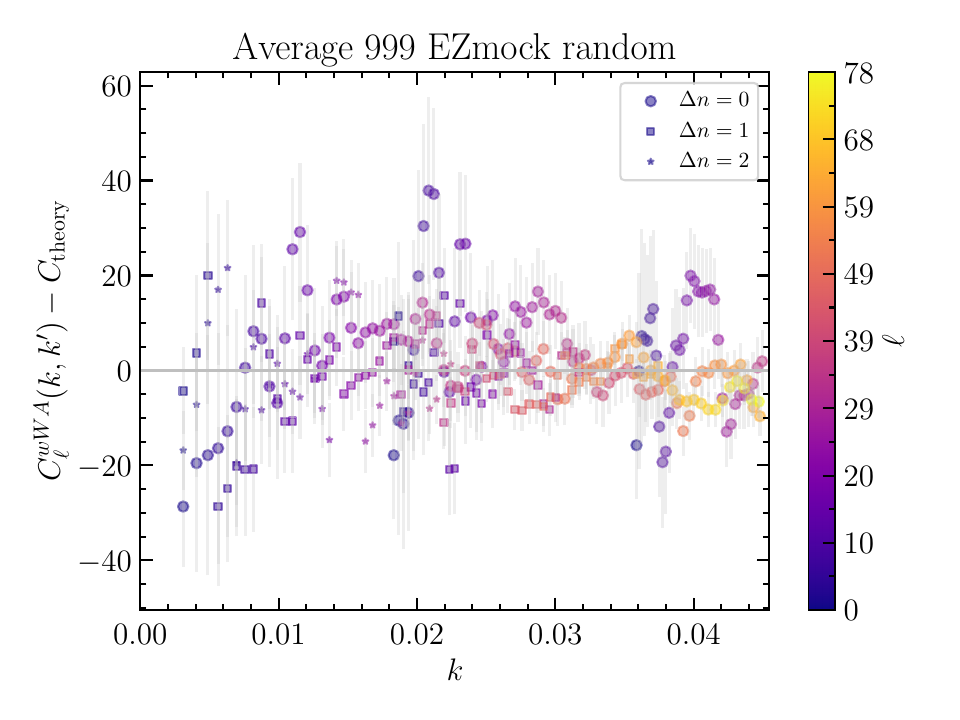}
  \incgraph[0.28]{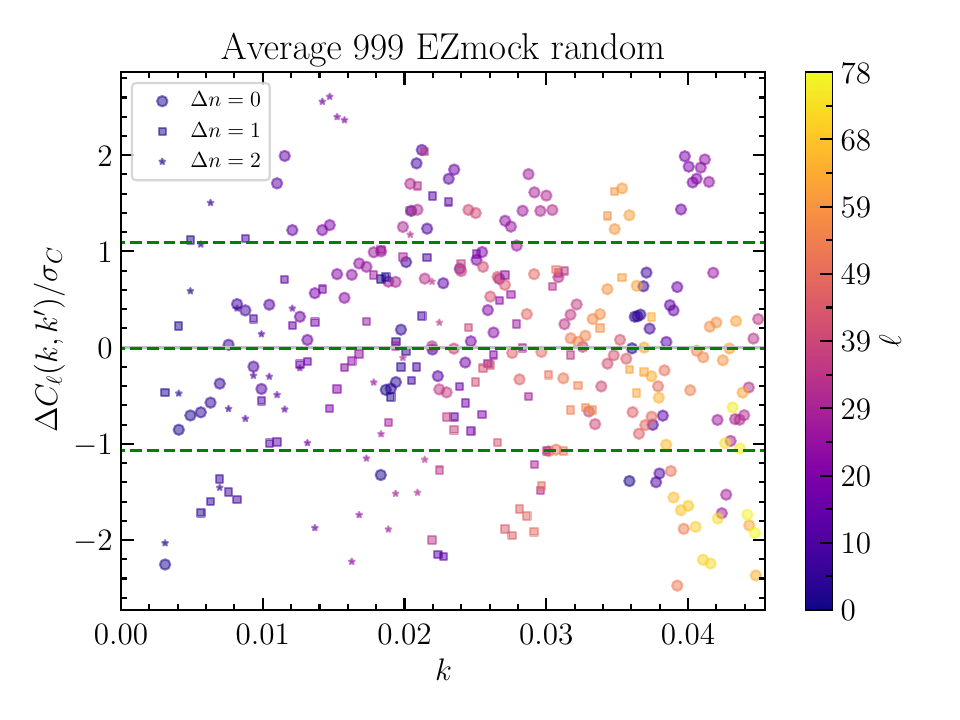}
  \incgraph[0.14]{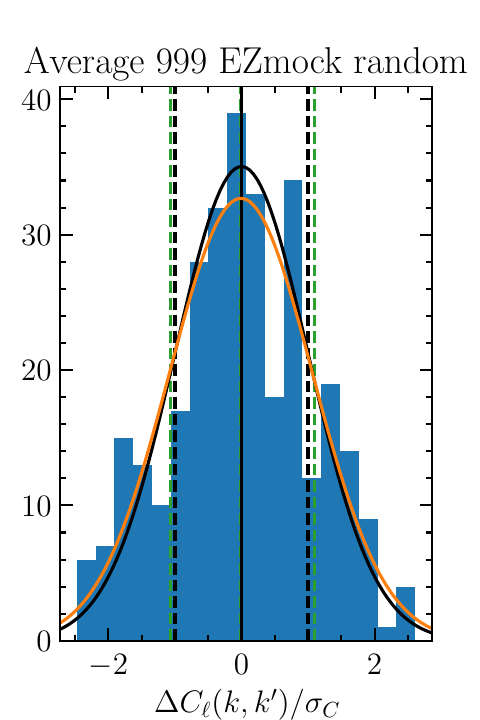}
  Slice 4: $z=0.805-0.895$\\
  \incgraph[0.28]{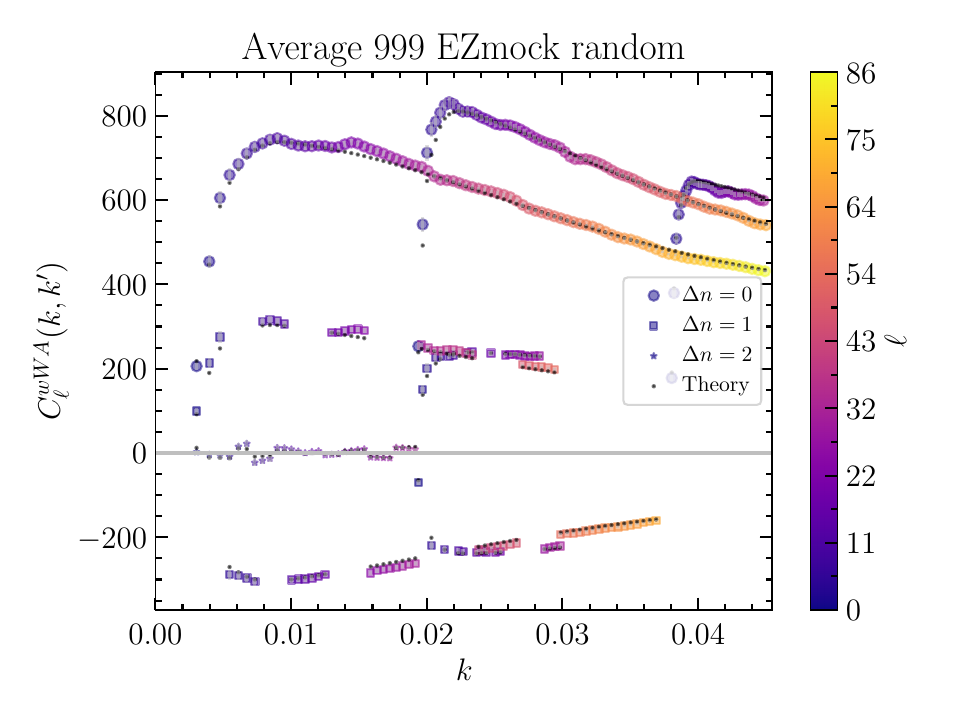}
  \incgraph[0.28]{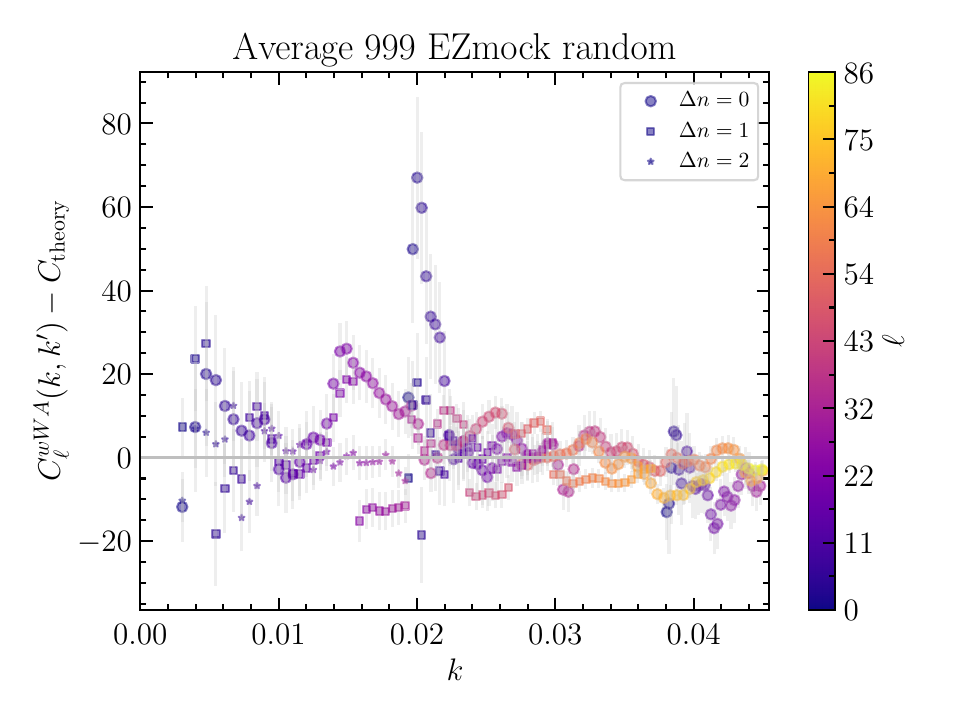}
  \incgraph[0.28]{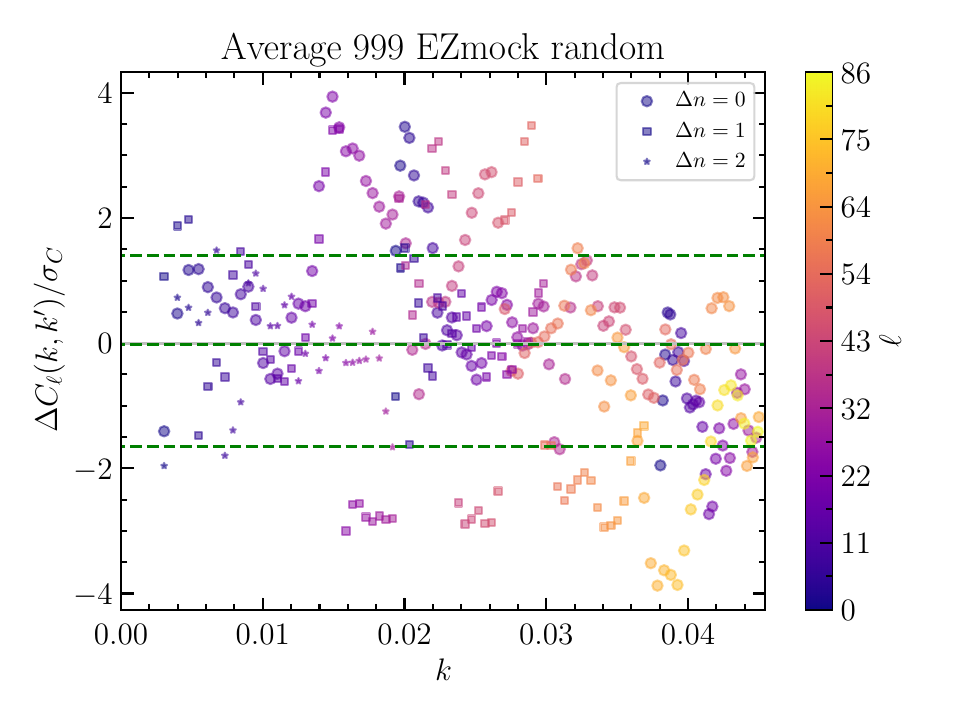}
  \incgraph[0.14]{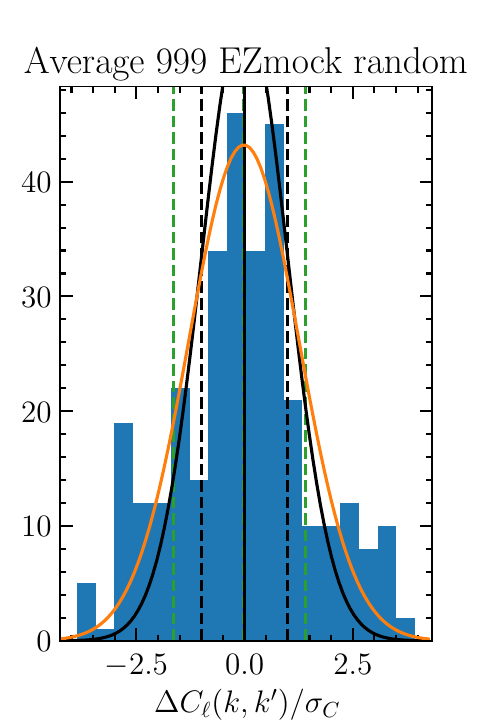}
  Slice 5: $z=0.905-1.0$\\
  \incgraph[0.28]{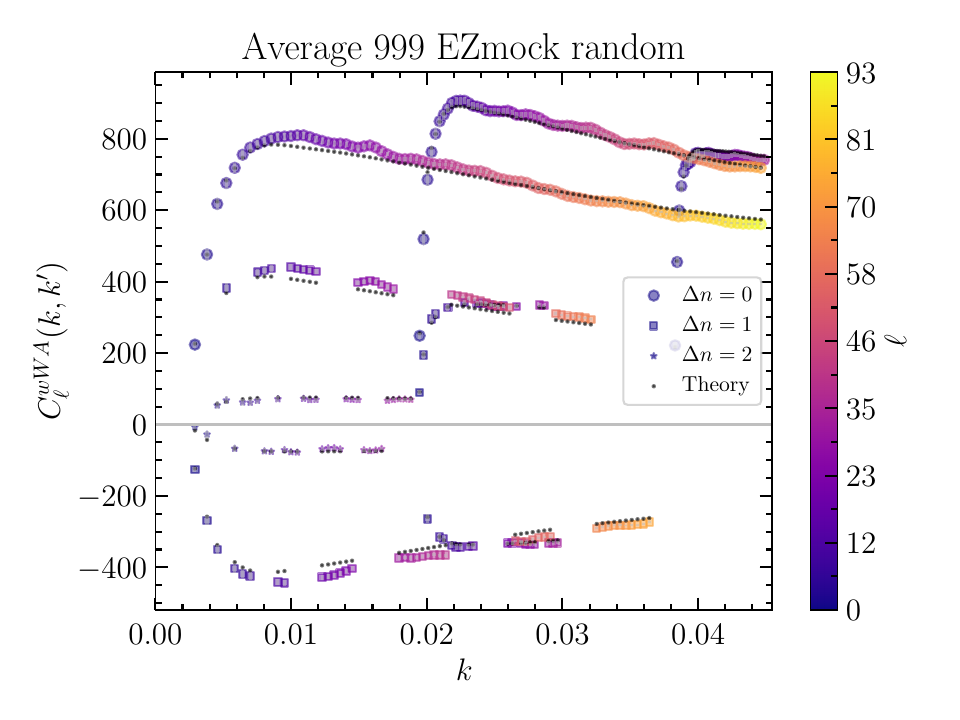}
  \incgraph[0.28]{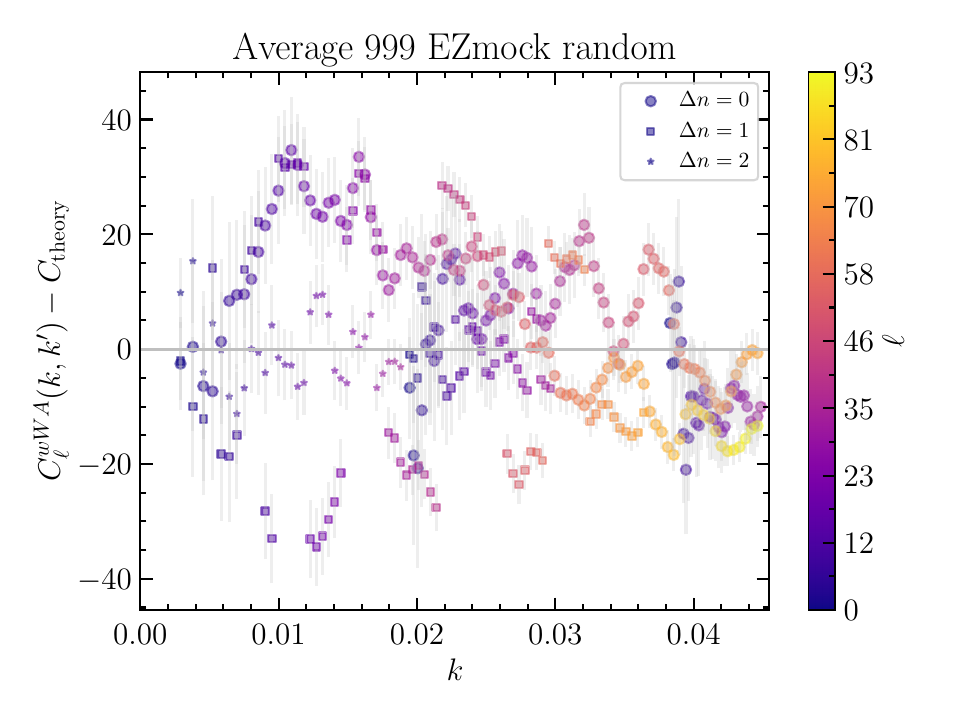}
  \incgraph[0.28]{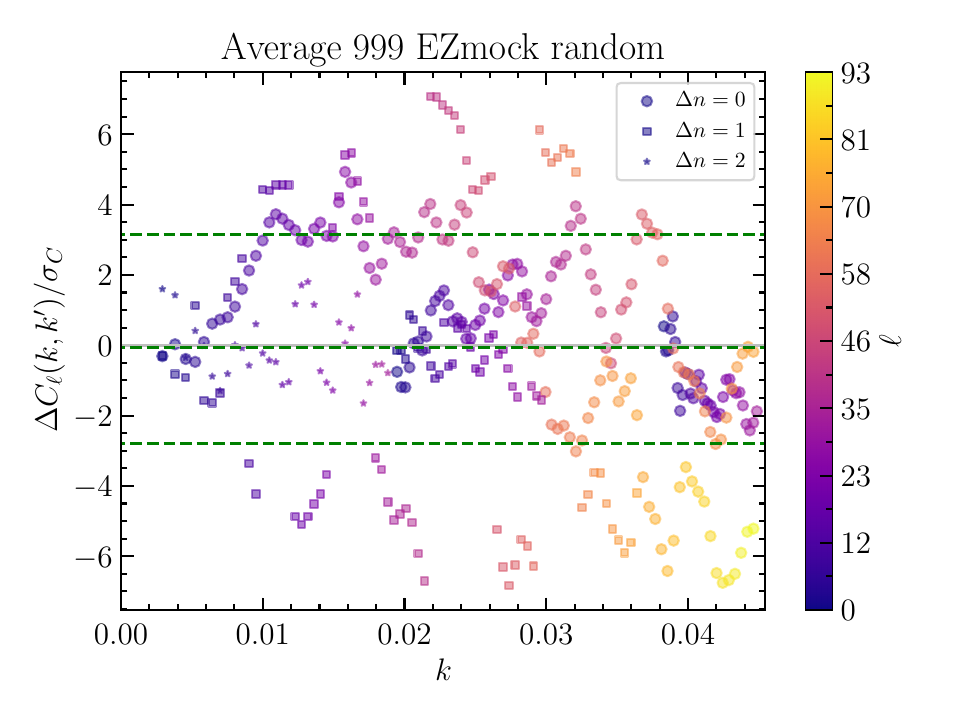}
  \incgraph[0.14]{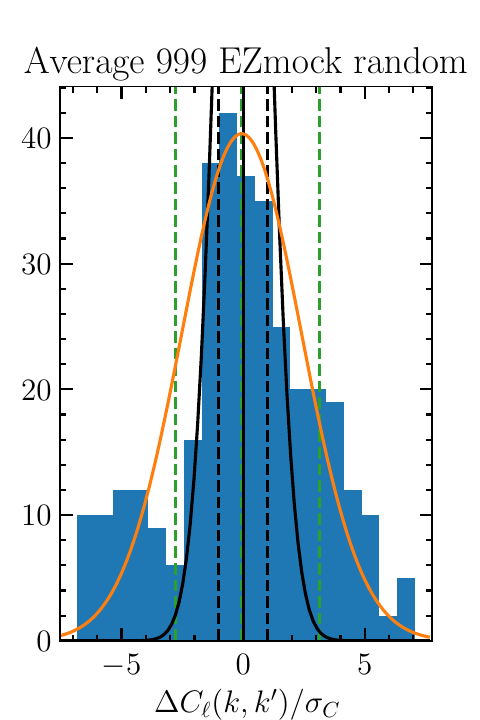}
  \caption{
    Each row shows the SFB power spectrum measured in each of the redshift
    slices used to generate the LRG EZmocks in the NGC. We show the full SFB
    power spectrum in the slice in the left panels, the residual relative to
    the theory in the center left panels, the same residual rescaled by the
    standard deviation as measured from the simulations in the center right
    panels, and a histogram of those standard-deviation-rescaled residuals in
    the left panels. In the two right-most columns the green dashed lines are
    the 16, 50, and 84 percentiles. In the right-most column we also show in
    orange a Gaussian fit, in black solid a Gaussian of unity width with black
    dashed lines as the corresponding 16, 50, and 84 percentiles.
    We adjusted the bias parameters to fit each slice.
    We added a buffer of $\Delta z=0.05$ to each slice in order to avoid
    leakage due to RSD from neighboring slices.
  }
  \label{fig:ez_complete_zslices}
\end{figure*}
\begin{figure}
  \centering
  \incgraph[0.49]{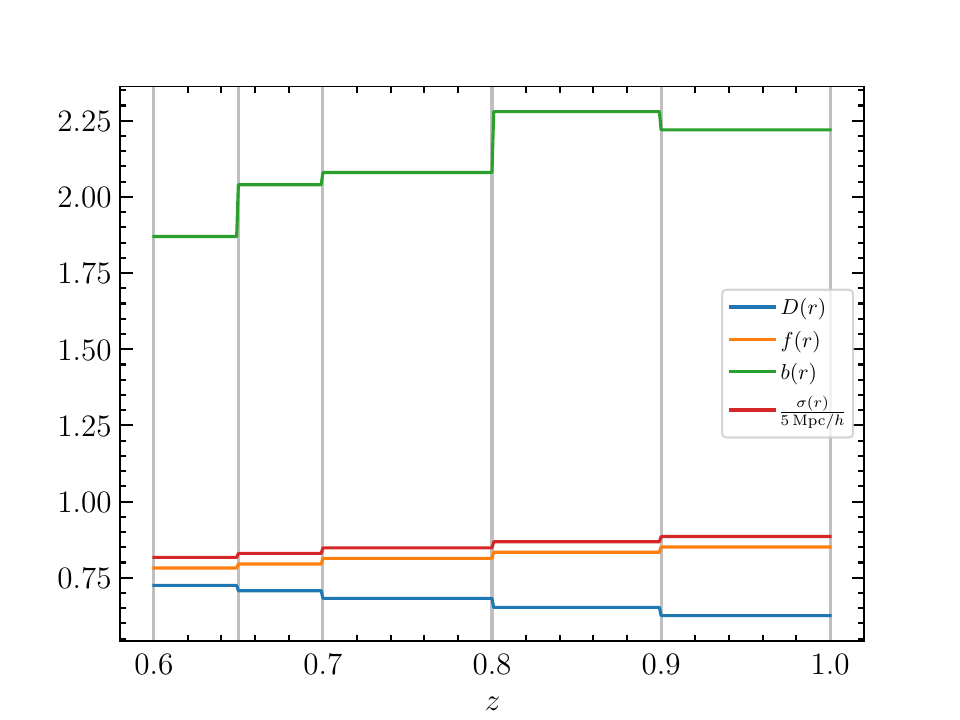}
  \caption{
    eBOSS DR16 LRG growth factor, growth rate, bias, and FoG evolution.
  }
  \label{fig:ez_complete_bias}
\end{figure}
\begin{figure}
  \centering
  \incgraph[0.39]{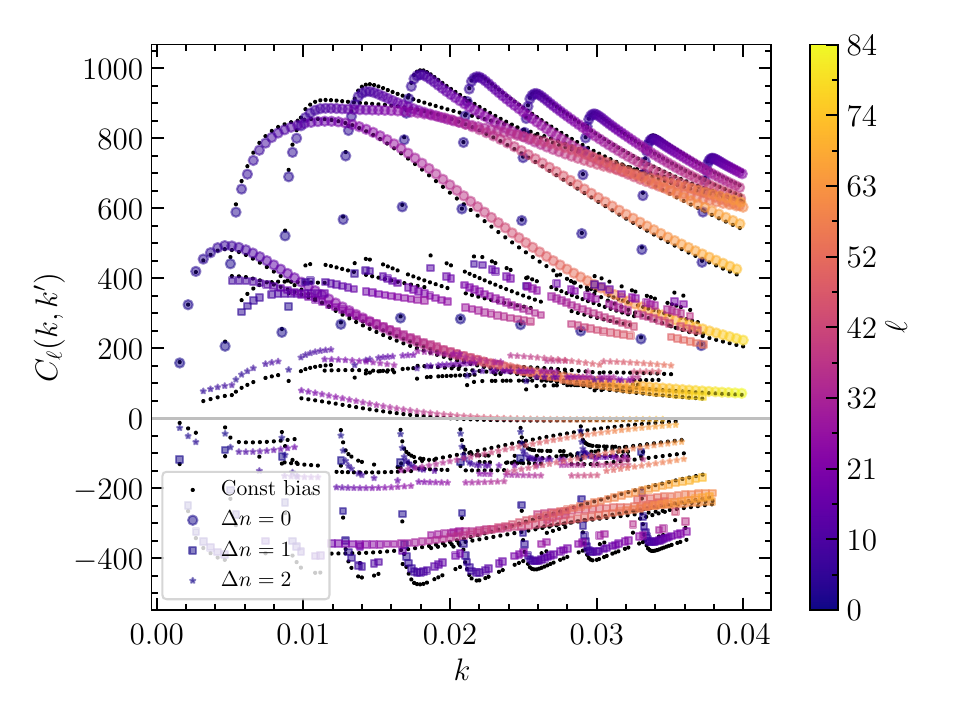}
  \incgraph[0.39]{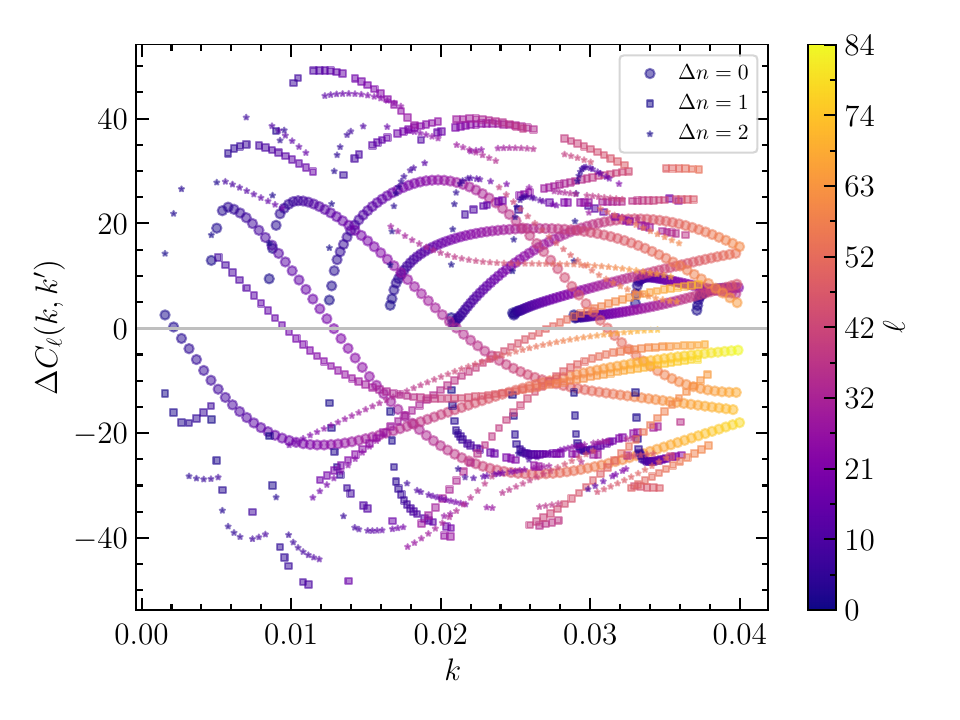}
  \caption{The SFB power spectrum is sensitive to redshift evolution, as exemplified here by 
    comparing the SFB power spectrum with the biases for the 5 redshift slices
    set equal (black points) with the step-function bias (larger colored
    points). Both are the theoretical curves.
  }
  \label{fig:ez_complete_redshift_evolution}
\end{figure}
\begin{figure*}
  \centering
  \incgraph[0.28]{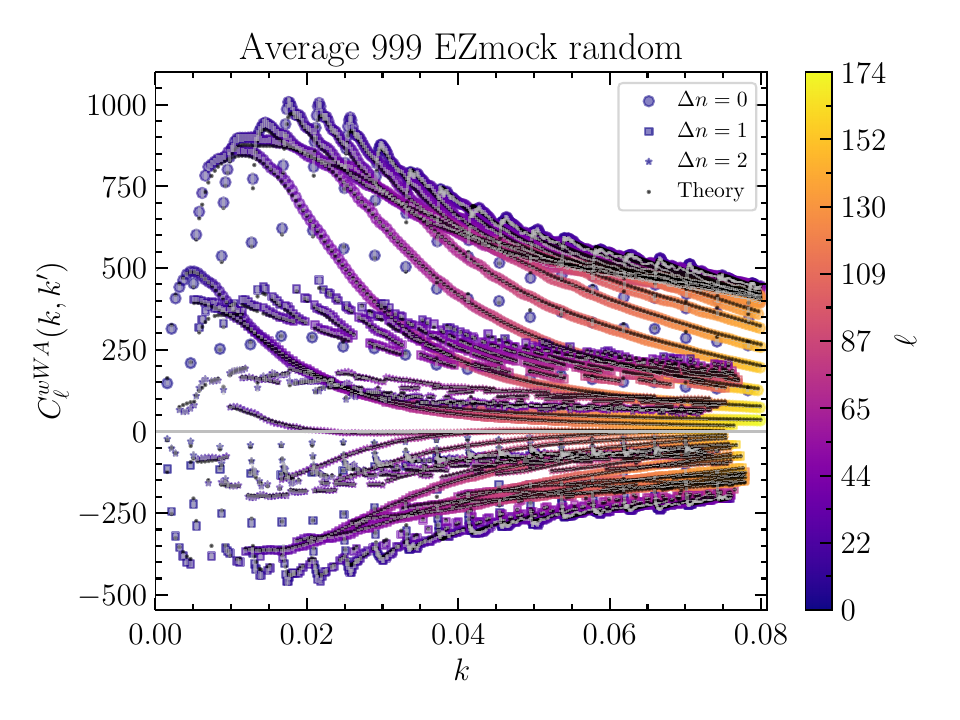}
  \incgraph[0.28]{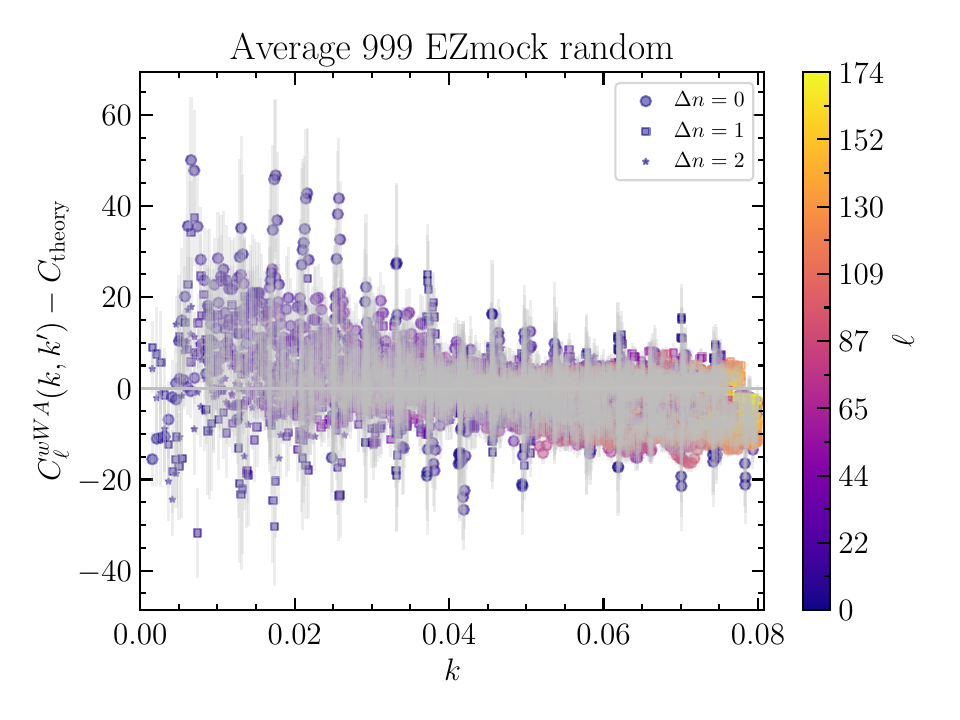}
  \incgraph[0.28]{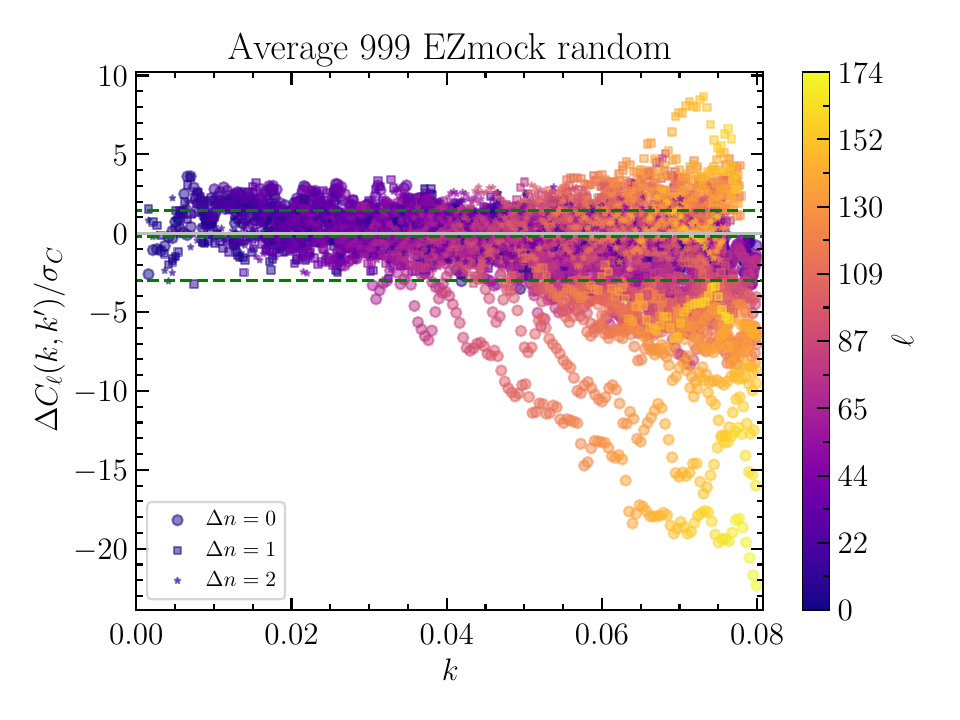}
  \incgraph[0.14]{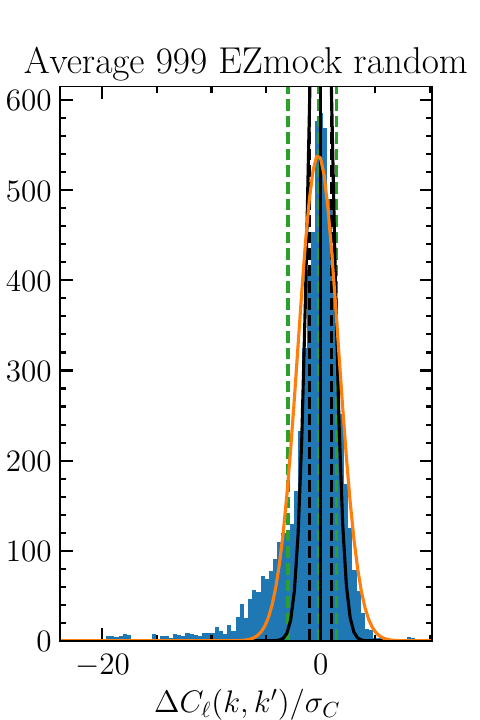}
  \caption{
    This figure is similar to \cref{fig:ez_complete_zslices}, except that now
    the full redshift range over all five slices of the \emph{complete} LRG NGC
    EZmocks is used.
    Without any further fitting of bias and cosmological parameters compared to
    \cref{fig:ez_complete_zslices}, we get generally good agreement between the
    mocks' average and theory, with a significant discrepancy only on smaller
    scales, as seen in the center right panel.
  }
  \label{fig:ez_complete}
\end{figure*}
To model the evolution of structure and galaxy bias with redshift, the EZmock
catalogs for the LRG sample are generated in 5 redshift slices
\citep{Zhao+:2021MNRAS.503.1149Z}. We show SFB
measurements from each of these slices in \cref{fig:ez_complete_zslices}.
In order to avoid leakage due to RSD from neighboring slices we
chose redshift slices that are thinner by $\Delta z=0.05$ on each side
than the original slices used to generate the mocks.

For each redshift slice, the left column of \cref{fig:ez_complete_zslices}
shows the average over 999 \emph{complete} EZmock simulations and a theory
curve where the linear galaxy bias was adjusted to fit the measurement in that
slice. The center left column shows the residuals with the error bars estimated
from the 999 EZmocks. The center right column shows the residuals relative to
the standard deviation of that mode, ignoring mode-couplings, and dashed
horizontal lines indicating the 16th, 50th, and 84th percentiles. The left-most
column shows a histogram of those relative residuals as well as a Gaussian fit
in orange, and a standard Gaussian in black.

Slices 1 and 2 cover a smaller redshift range and, thus, have fewer modes (and
no $\Delta n=2$ modes). Modes are also coupled due to the limited sky coverage,
so that the number of
degrees of freedom is fairly small, and slice 2, in particular, is quite noisy.
Slice 3 contains the most galaxies, and the 16th, 50th, and 84th percentiles
are nearly exactly coincident with the 1-$\sigma$ lines.

The linear galaxy biases thus obtained are plotted as step-functions in
\cref{fig:ez_complete_bias}.

The sensitivity of the SFB power spectrum to the evolution of the linear galaxy
bias $b(z)$ is shown in \cref{fig:ez_complete_redshift_evolution}, where we
compare theory curves for constant bias and the step-function bias. The
difference is on the level of \SI{\sim5}{\percent}.

Using the bias model in \cref{fig:ez_complete_bias}, we now show the comparison
with the \emph{complete} EZmocks with all 5 redshift slices in
\cref{fig:ez_complete}. In the figure, we extend the range of modes up to
$k_\max=\SI{0.1}{\h\per\mega\parsec}$. However, we only show up to
$k=\SI{0.08}{\h\per\mega\parsec}$ because of the incomplete window convolution
above that.

As \cref{fig:ez_complete} shows, the model is unbiased on large scales.
However, small angular modes with high $\ell$ (yellow points in the figure) are
significantly biased relative to their covariance. This comes from the fact
that we measured the radial selection function from the data which is afflicted
by systematics that are not present in the \emph{complete} EZmocks. We remedy
this situation with the \emph{realistic} EZmocks in the next section.

\subsection{\emph{Realistic} EZmocks}
\begin{figure*}
  \centering
  \incgraph[0.28]{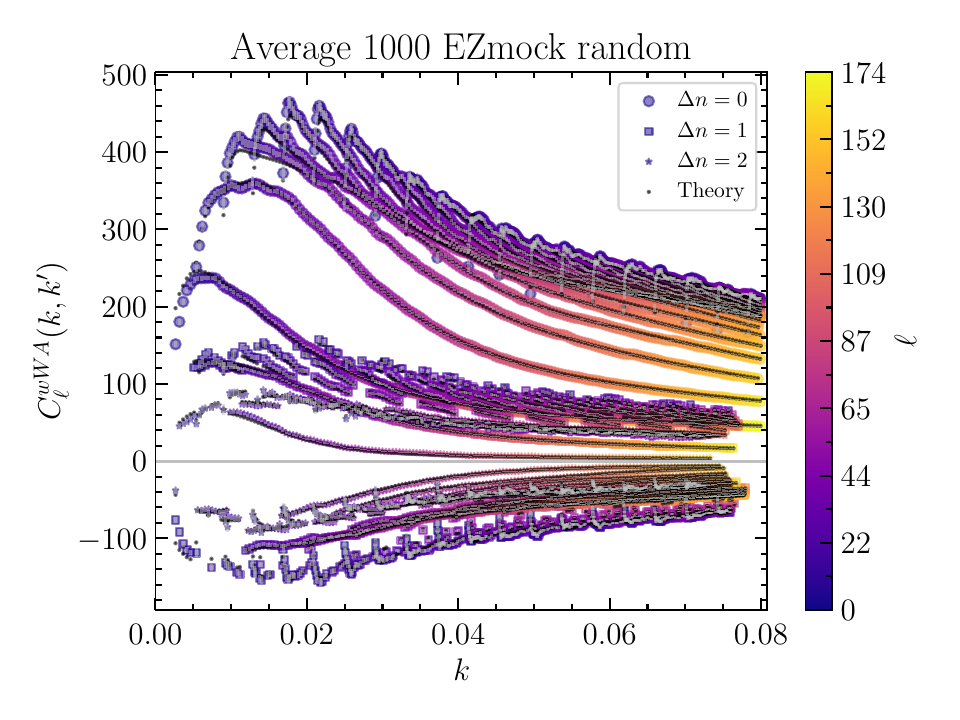}
  \incgraph[0.28]{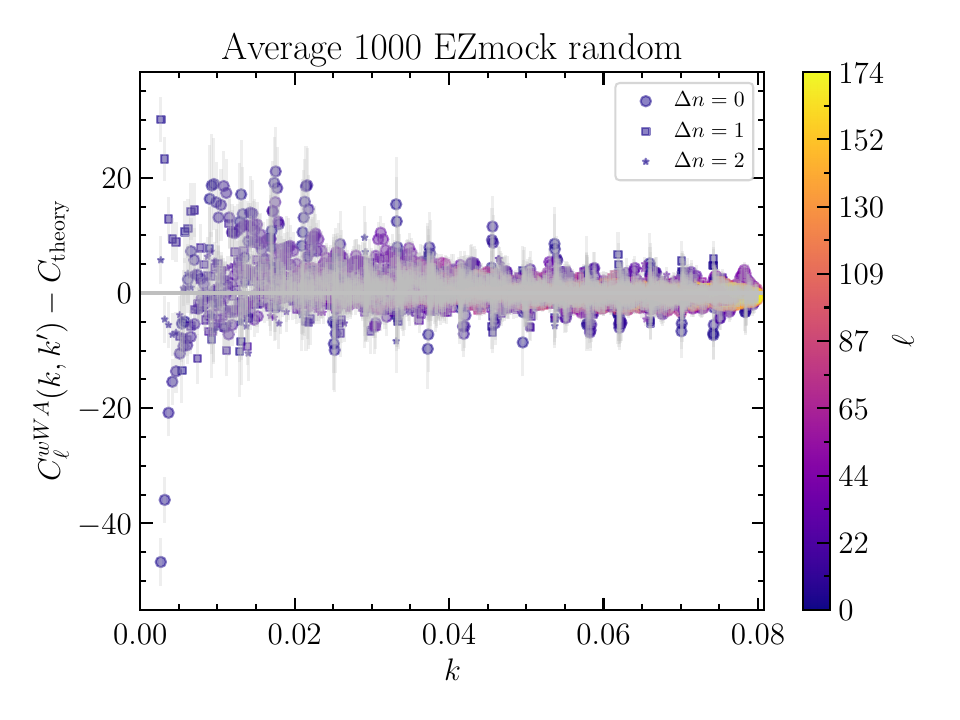}
  \incgraph[0.28]{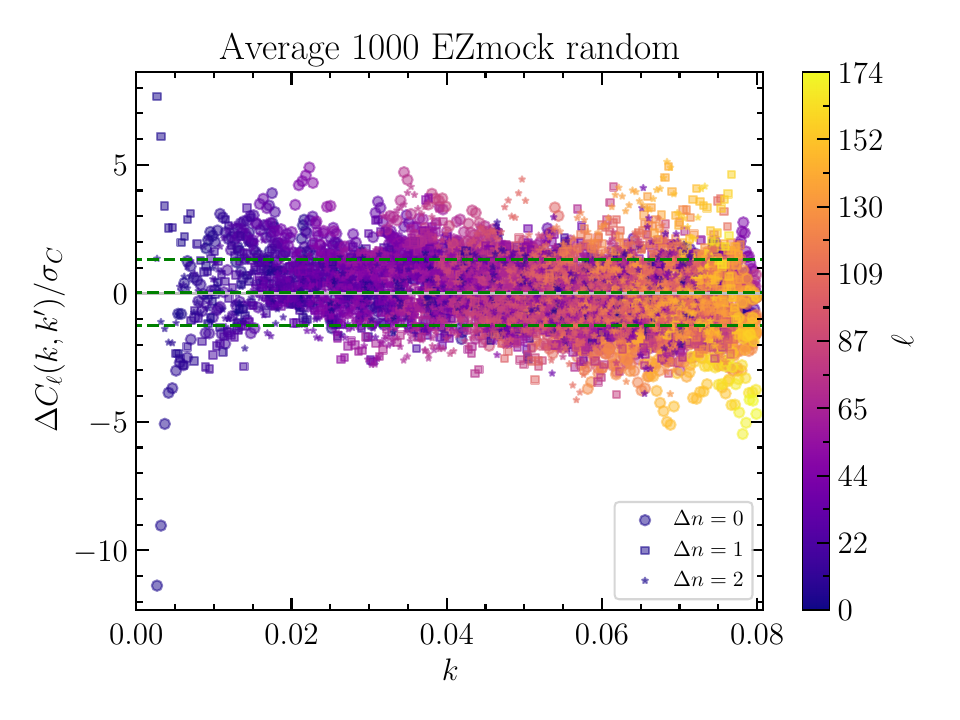}
  \incgraph[0.14]{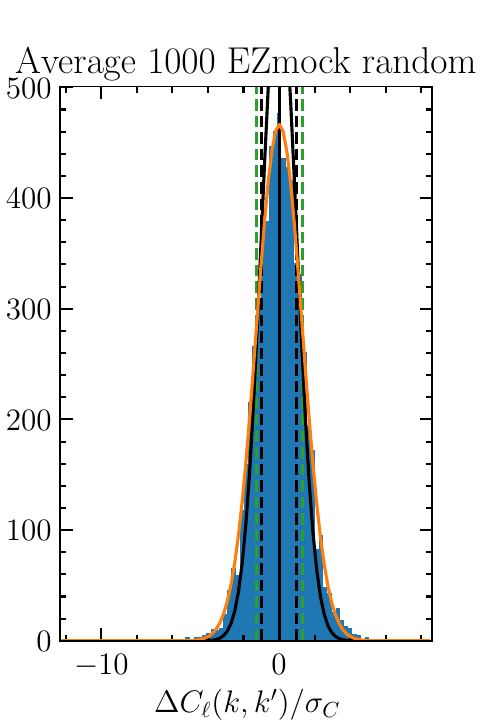}
  \incgraph[0.28]{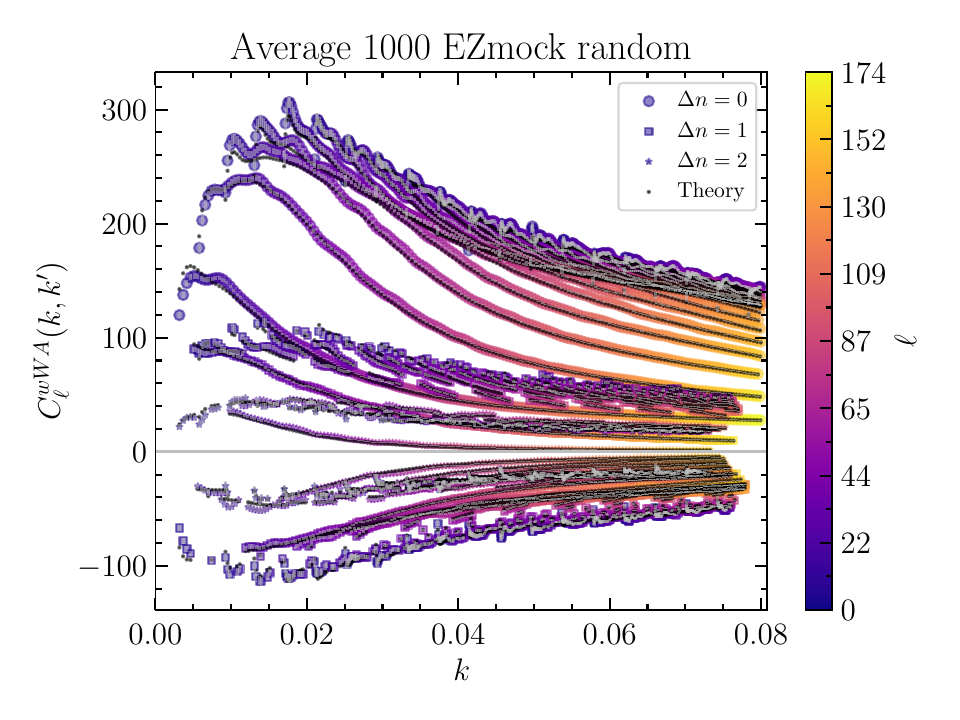}
  \incgraph[0.28]{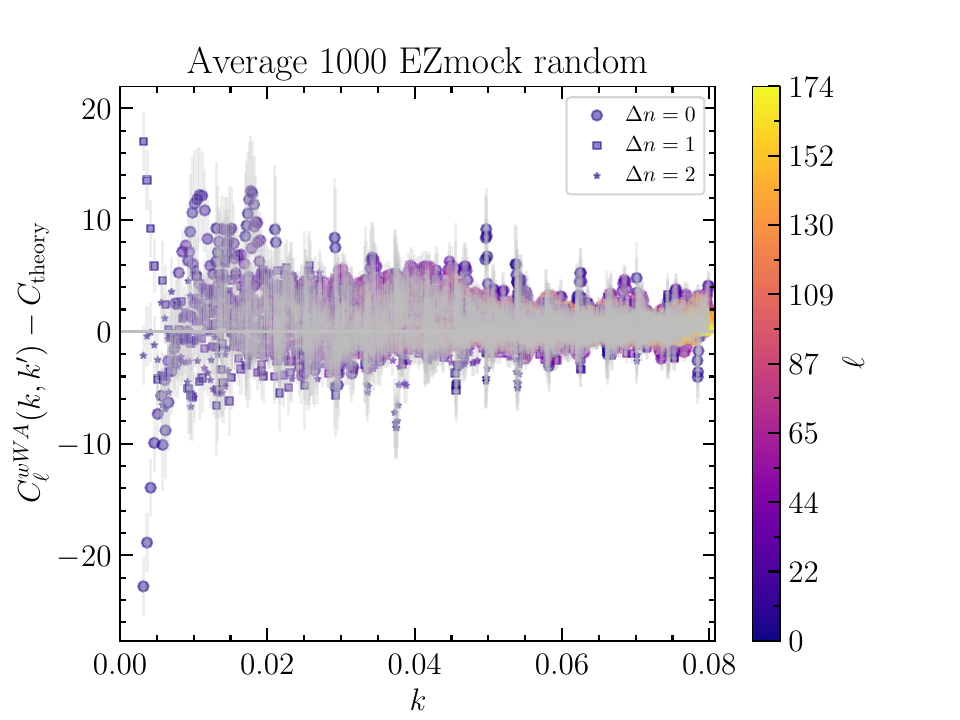}
  \incgraph[0.28]{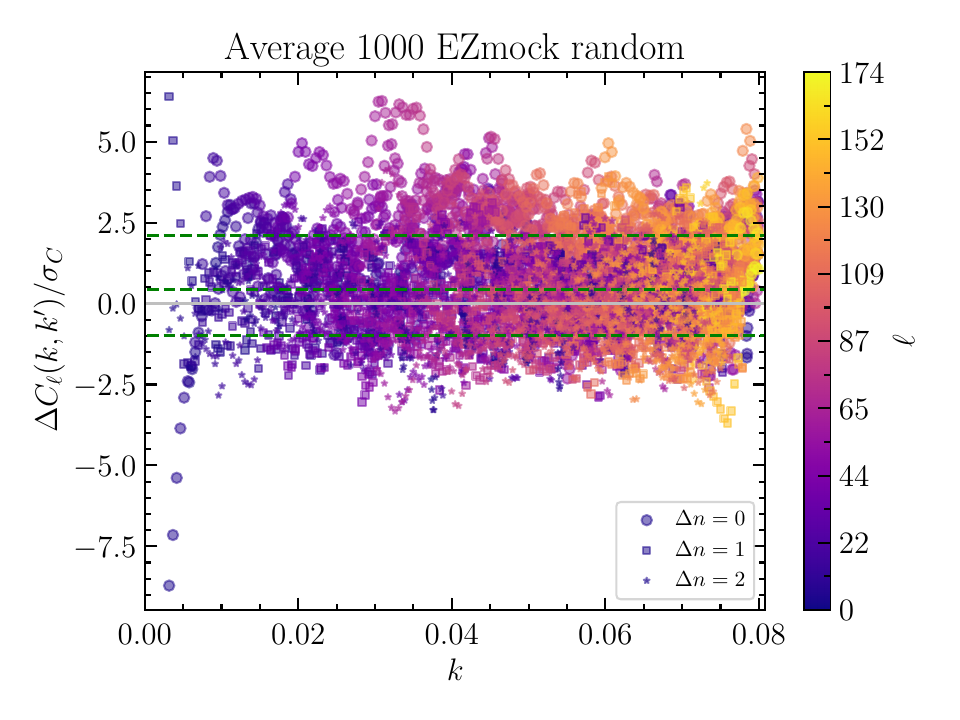}
  \incgraph[0.14]{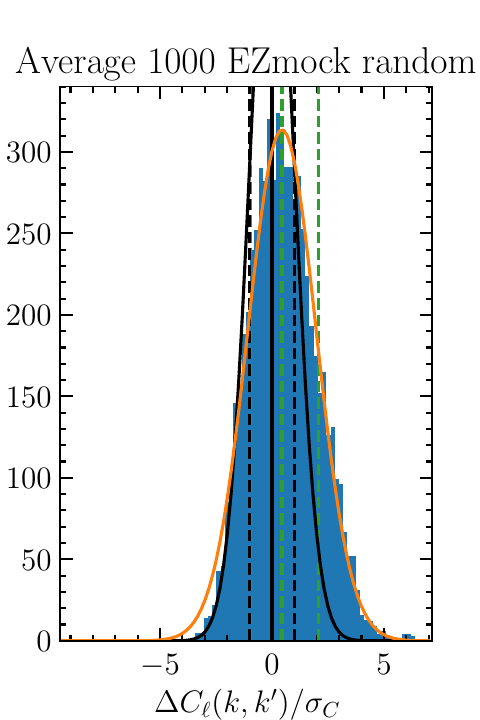}
  \caption{
    This figure is similar to \cref{fig:ez_complete}, except that here we used
    the \emph{realistic} EZmocks that include systematics, and we have also
    included FKP weights in the power spectrum estimation. Furthermore, the top
    row of panels shows the comparison between mocks' average and theory for
    the NGC, and the bottom row for the SGC. The same cosmological and bias
    parameters were used for NGC and SGC.
  }
  \label{fig:ez_realistic_fkp}
\end{figure*}
Several observational systematics enter the analysis of the eBOSS DR16 ELG
samples. Building ontop of the \emph{complete} EZmocks, the \emph{realistic}
EZmocks add the effects of photometric systematics, close-pair fiber
collisions, and redshift failures. Weights to correct for these systematics are
provided by the EZmocks.

In addition to the systematic weights we apply FKP weights according to
\cref{eq:fkp_weights}. We chose to apply our own FKP weights instead of those
provided by EZmocks, for no other than practical reasons, e.g., to apply them
to our lognormal simulations.

\cref{fig:ez_realistic_fkp} shows the average over the 1000 EZmocks for the NGC in
the top panels and for the SGC in the bottom panels of the figure. The
simulations and data analysis are fully consistent with each other, and the
agreement is within the statistical expectation for both NGC and SGC.

\section{MCMC fitting and Covariance matrix}
\label{sec:mcmc}
In this section we test our likelihood construction and MCMC fitting. The
covariance matrix is estimated from 1000 realistic EZmocks
\citep{Zhao+:2021MNRAS.503.1149Z}. We briefly describe how we invert such a
noisy covariance matrix using the eigenvector method outlined in
\citet{Wang+:2020JCAP...10..022W}, and then we apply that to a few realizations
of the EZmocks.

Assuming the SFB power spectrum modes are Gaussian distributed, we construct
the likellihood as follows.
\ba
-2\ln\mathcal{L}
&=
\mathrm{const}
+ (\hat C - C)^T M^{-1} (\hat C - C)\,,
\ea
where $\hat C$ is the vector of measured $C_{\ell nn'}$ values, $C$ is the SFB
power spectrum model and $M$ is the covariance matrix.

\subsection{On the inversion of noisy convariance matrices}
\begin{figure}
  \centering
  \incgraph{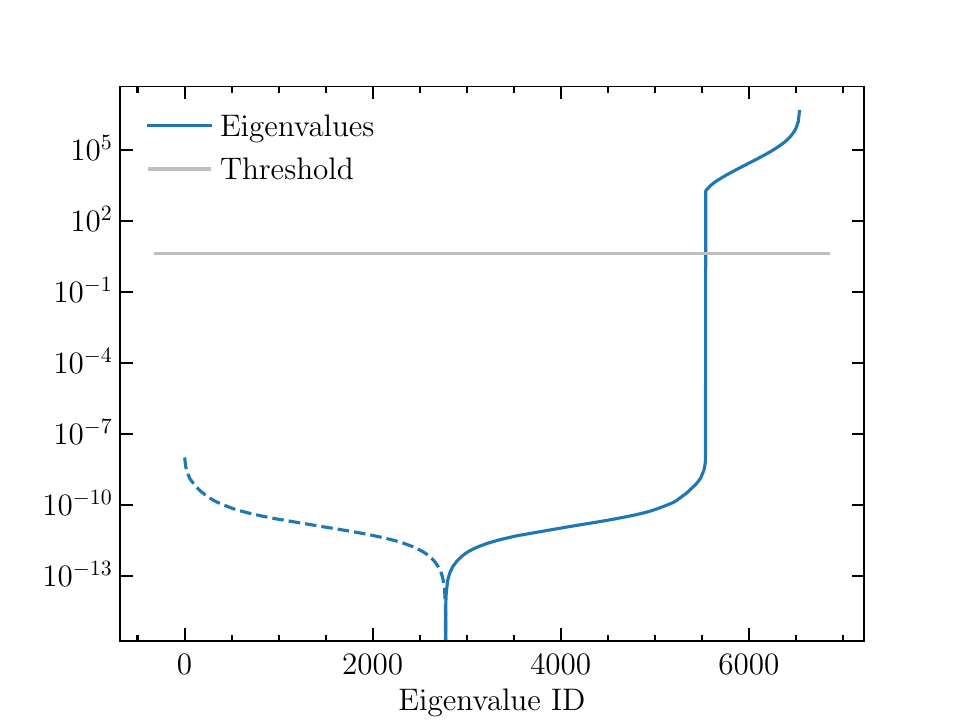}
  \caption{Eigenvalues of the full covariance matrix estimated from 1000
    EZmocks, sorted from smallest to largest, dashed means negative. The
    largest 1000 eigenvalues are well-constrained. The smaller eigenvales are
    consistent with numerical machine-level noise around a mean of zero. The
    grey line indicates the cutoff for which we use eigenvalues in the matrix
    inversion.
  }
  \label{fig:cov_eigvals}
\end{figure}
A covariance matrix estimated from simulations is afflicted by noise. In our
case, the number of modes exceeds the number of simulations. Thus, only 1000
eigenvalues of the covariance matrix are measured from the 1000 simulations,
and the rest are zero within machine precision. An example is shown in
\cref{fig:cov_eigvals}.

Therefore, the inverse of the measured covariance matrix will be dominated by
the noise of the unmeasured eigenvalues. To solve this we use the approach outlined in \citet{Wang+:2020JCAP...10..022W}
that ensures that the inverted matrix is dominated by well-constrained
eigenmodes. The assumption is that the eigenvectors with large eigenvalues are
the best determined. Thus, only the eigenvectors with the largest eigenvalues
are retained.

There are a few unfortunate consequences of this approach.
First, the modes
that have a small variance and, thus, would contribute most to the likelihood
are thrown away. However, here we are interested in the largest scales, and
these modes tend to have larger uncertainties due to the limited volume in the
survey. Thus, we expect these large modes to be retained in this approach.
Second, among the modes with large uncertainties, the noise in the covariance
estimate means that there is some stochasticity as to the exact modes that get
retained. Some modes will be over estimated, and those will definitely be
retained, which merely degrades the constraints. However, it is also possible
(e.g., if the numbers of modes is similar to the number of simulations) that some
covariance entries get underestimated, in which case we could get biased
constraints. We leave a more detailed investigation to a future paper.

The inversion works essentially as a lossy compression algorithm.
First, the variable condition number of an eigenmode $i$ is defined as
\ba
\varrho_i &= \frac{\lambda_\max}{\lambda_i}\,,
\ea
where $\lambda_1 \leq \ldots \leq \lambda_N=\lambda_\max$ are the eigenvalues
sorted from smallest to largest. We then keep only the largest eigenvectors
starting at the smallest $i_\min$ such that
$\varrho_{i_\min}\geq\varrho_\mathrm{threshold}=1000$. The compression matrix
$R$ is then constructed from these $J=N-i_\min+1$ eigenvectors $\bm{e}_j$,
\ba
R &= (\bm{e}_{i_\min}, \ldots, \bm{e}_N)^\dagger \in \mathbb{C}^{J \times N}\,,
\ea
which satisfies $RR^\dagger=1$. Then, the compressed data vector is
\ba
\Delta &= R\delta\,,
\ea
and the compressed covariance matrix is
\ba
C &= RMR^\dagger\,,
\ea
which is positive-definite if $M$ is positive-definite.
The likelihood analysis is then performed in compressed space with data vector
$\Delta$ and covariance matrix $C$.

\citet{Wang+:2020JCAP...10..022W} have shown that the matrix $R$ can be
recycled between iterations of the MCMC fitting procedure as long as the
parameter space that is being explored stays in a similar region.

\subsection{MCMC fitting}
\begin{figure}
  \centering
  \incgraph[0.32]{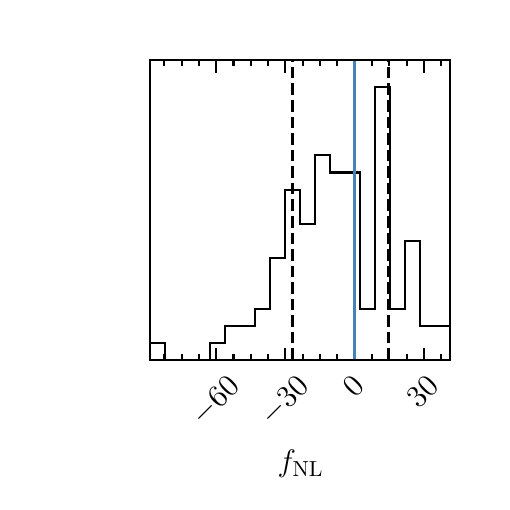}
  \caption{Histogram of best-fit $\fnl$ for 100 simulations. The blue line marks the
    fiducial $\fnl=0$, the dashed lines the 16th and 84th percentiles.
  }
  \label{fig:100sims_fnl}
\end{figure}
\begin{figure*}
  \centering
  \incgraph[1]{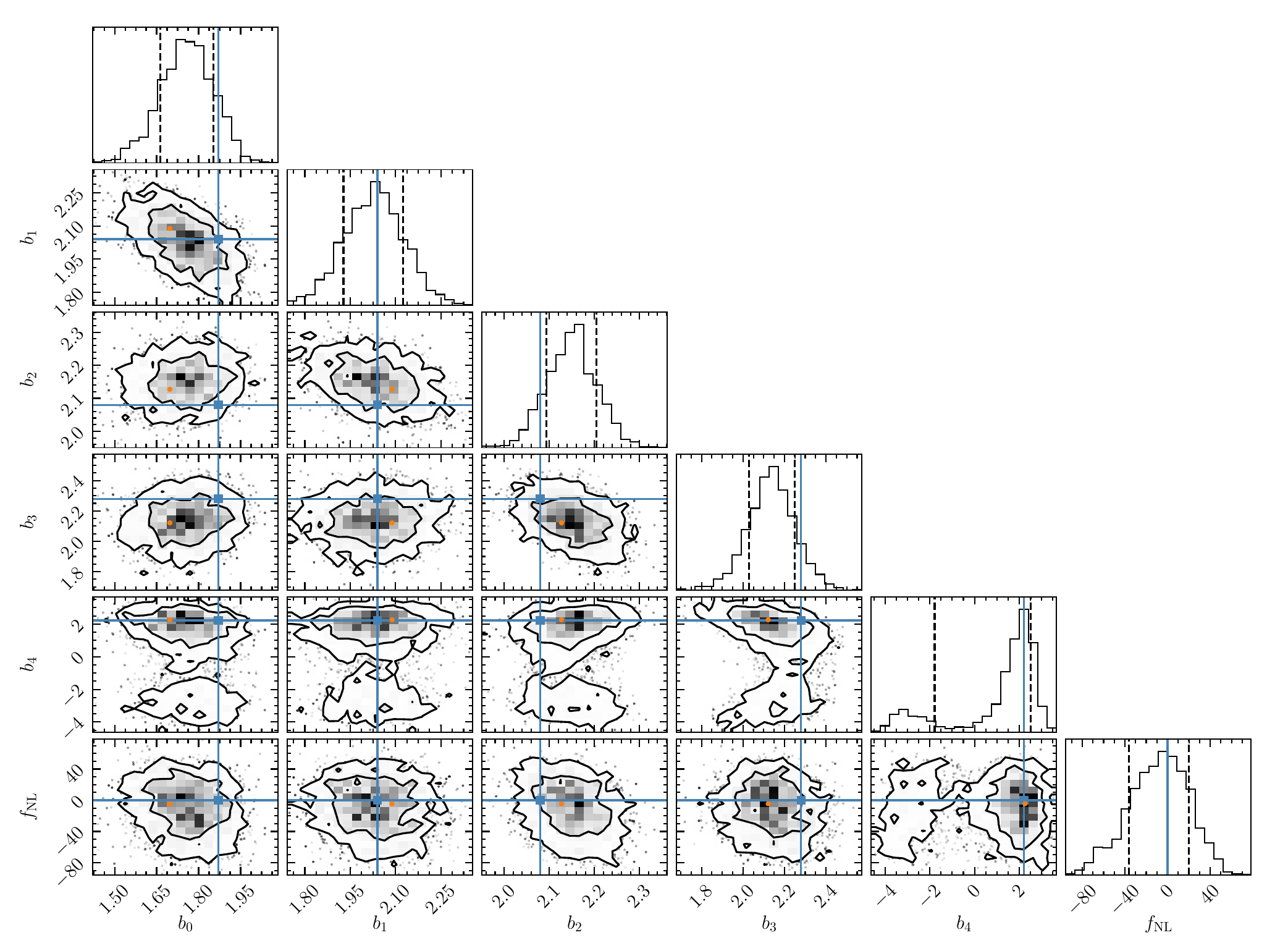}
  \caption{MCMC results for a single EZmock. The blue crosses mark the fiducial
    value as measured from all EZmocks, the orange point marks the point that had
    the highest likelihood in the chain. All parameters except the bias $b_4$
    for the highest redshift slice are reasonably well constrained and
    consistent with the fiducial.
  }
  \label{fig:mcmc_b_fnl}
\end{figure*}
\begin{figure*}
  \centering
  \incgraph[1]{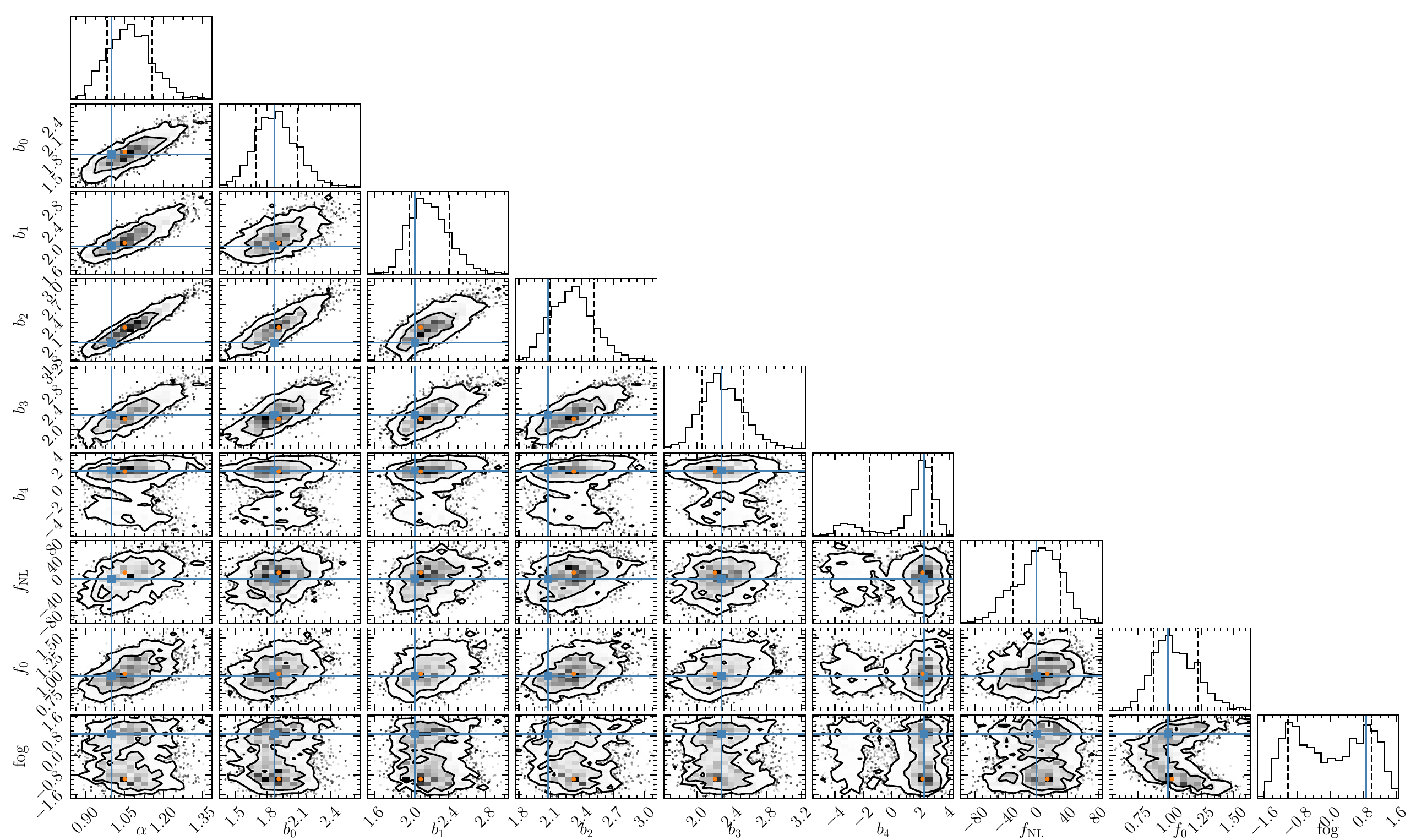}
  \caption{Similar to \cref{fig:mcmc_b_fnl}, except with additional parameters
  for RSD, FoG, and Alcock-Paczynski.}
  \label{fig:mcmc_all}
\end{figure*}
As a final step in this auto-correlation SFB validation paper, we use single
EZmock simulations as data. First, we will measure the bias parameters and
$\fnl$ from the average of the simulations, then we include RSD, FoG and
single-parameter Alcock-Paczynski.

For our MCMC we developed in Julia an adaptive Metropolis-Hastings sampler
\citep{Roberts+:doi:10.1198/jcgs.2009.06134}. \emph{Adaptive} here means that
the covariance matrix of the proposal distribution is updated every few steps.
Therefore, the chain is Markovian only asymptotically in the limit of a large
number of steps. We use the following procedure to get a robust estimate of the
errors. The number of steps is a multiple of $M=n(n+1)$, where $n$ is the
number of parameters. We first run the chain for $100M$ steps to get an
estimate for the maximum likelihood point. Then, we run it again for $100M$
points to get an estimate of the covariance matrix for the parameters. We then
run it twice more, each time updating both the starting point with the new
maximum lileklihood point and the initial covariance matrix. Finally, we run
the adaptive Metropolis-Hastings sampler for $700M$ steps, and these samples
will be the ones used as output of the MCMC.

\cref{fig:100sims_fnl} shows a histogram over 100 sims, fitting only $\fnl$ and
keeping all other parameters constant. We used only the NGC, and $\ell\geq4$,
and $k_\max=\SI{0.08}{\h\per\mega\parsec}$. While the noise is large, the set
of ensembles is consistent with $\fnl=0$, as we would expect given the input
simulations.

\cref{fig:mcmc_b_fnl} shows an MCMC run with the 5 biases (one for each
redshift slice) and $\fnl$ as parameters. The measured bias parameters are
essentially the same as when we fix $\fnl=0$, and they are consistent with the
average measured from all EZmocks in \cref{sec:ez_complete}, generally within
the \SI{68}{\percent} contours.

Finally, \cref{fig:mcmc_all} shows an MCMC run for the 5 biases, $\fnl$, Kaiser
parameter $f_0$, FoG $\sigma_u$, and Alcock-Paczynski parameter $\gamma$. The
FoG parameter only enters as its square, hence negative values are equally
allowed.

While the error contours are large, they are largely unbiased. Crucially, the
EZmocks were created with $\fnl=0$, and we successfully recover that from the
simulations.

\section{Conclusion}
\label{sec:conclusion}
In this paper we validate our pipeline for using the \emph{SuperFaB} estimator
for cosmological inference on the spherical Fourier-Bessel pseudo power
spectrum. We develop a reasonably fast SFB power spectrum calculator along the
lines of \citet{Khek+:2022arXiv221205760K}. We derive in detail the shot noise
and the local average effect with weights.

We start the validation with full-sky lognormal simulations in real and
redshift space \citep[similar to][]{Agrawal+:2017JCAP...10..003A}, and then
progressively add redshift-space distortions, realistic masks, radial
selection, and systematic weights using the EZmocks of
\citet{Zhao+:2021MNRAS.503.1149Z}. We also verify the use of potential and
velocity boundary conditions.

We model the EZmocks bias prescription by steps as shown in
\cref{fig:ez_complete_bias}. For the NGC and SGC separately, we verify that the
model including all systematics is consistent with the measured modes of the
EZmocks, \cref{fig:ez_realistic_fkp}.

Finally, we construct a Gaussian likelihood for the power spectrum, we measure
the covariance matrix from all \num{1000} EZmocks with a
large-eigenvalue-inversion, and we verify that the measured $\fnl$ from 100
EZmocks is consistent with zero, \cref{fig:100sims_fnl}, and that all the
parameters of the model are consistent with their fiducial values.

For the future we plan to apply the method on the data, improve the
covariance estimate, and develop the multi-tracer SFB analysis.

\acknowledgements
\textcopyright 2023. All rights reserved.
The authors would like to thank Zhao Cheng for helpful discussion on the EZmock catalogs.
The authors also thank Kendrick Smith for pointing out possible problems with the approach to the covariance matrix.
The authors are indebted to the anonymous referee who pointed out many places to make the paper significantly more readable.
Part of this work was done at Jet Propulsion Laboratory, California Institute of Technology, 
under a contract with the National Aeronautics and Space Administration. This work was 
supported by NASA grant 15-WFIRST15-0008 \textit{Cosmology with the High Latitude Survey} Roman Science Investigation Team (SIT).
Henry S. G. Gebhardt's research was supported by an appointment to the NASA Postdoctoral Program at the Jet Propulsion Laboratory, administered by Universities Space Research Association under contract with NASA.
Part of this work was done at the Aspen Center of Astrophysics.

\bibliography{../../doc/references}

\appendix

\section{Useful formulae}
\label{app:sfb_useful_formulae}
Spherical Bessel functions and spherical harmonics satisfy orthogonality
relations
\ba
\label{eq:jljlDelta}
\delta^D(k-k')
&= \frac{2kk'}{\pi}\int_0^\infty\dd{r}\,r^2\,j_\ell(kr)\,j_\ell(k'r)\,, \\
\label{eq:YlmYlmDelta}
\delta^K_{\ell\ell'}\delta^K_{mm'}
&= \int\dd{\Omega}_{\rhat}\,Y_{\ell m}(\rhat)\,Y^*_{\ell'm'}(\rhat)\,.
\ea

The Laplacian in spherical coordinates is
\begin{align}
    \nabla^2f &= \frac{1}{r^2}\,\frac{\partial}{\partial r}\left(r^2\,\frac{\partial f}{\partial r}\right)
    + \frac{1}{r^2\sin\theta}\,\frac{\partial}{\partial\theta}\left(\sin\theta\,\frac{\partial f}{\partial\theta}\right)
    \nonumber \\
    &\quad
    + \frac{1}{r^2\sin^2\theta}\,\frac{\partial^2f}{\partial\phi^2}\,.
    \label{eq:laplacian_spherical}
\end{align}

The SFB transform pair is
\ba
\delta(\vr) &= \int\dd k\,\sum_{\ell m}
\left[\sqrt{\frac{2}{\pi}}\,k\,j_\ell(kr)\,Y_{\ell m}(\theta,\phi)\right]
\delta_{\ell m}(k)\,,
\label{eq:sfb_fourier_pair_a}
\\
\delta_{\ell m}(k) &=
\int\dd^3r
\left[\sqrt{\frac{2}{\pi}}\,k\,j_\ell(kr)\,Y^*_{\ell m}(\rhat)\right]
\delta(\vr)\,.
\label{eq:sfb_fourier_pair_b}
\ea

\section{Shot noise and FKP Weighting}
\label{app:fkp_shotnoise}
Here we consider the effect of the weighting on the clustering signal and shot
noise. We start with the Poisson statistics of the sampled field
\citep{Peebles:1973ApJ...185..413P,Feldman+:1994ApJ...426...23F}
\ba
\<n(\vr)\,n(\vr')\>
&=
\bar n(\vr) \, \bar n(\vr') \left[1 + \xi(\vr,\vr')\right]
\vs&\quad
+ \bar n(\vr)\,\delta^D(\vr - \vr')\,,
\\
\alpha\<n(\vr)\,n_r(\vr')\>
&=
\bar n(\vr) \, \bar n(\vr')\,,
\\
\alpha^2\<n_r(\vr)\,n_r(\vr')\>
&=
\bar n(\vr) \, \bar n(\vr')
\vs&\quad
+ \alpha\,\bar n(\vr)\,\delta^D(\vr - \vr')\,,
\ea
where $n(\vr)$ is the number density of the data catalog, $n_r(\vr)$ that of
the random catalog, and we define $\nbar(\vr)=\<n(\vr)\>=\alpha\<n_r(\vr)\>$,
and $\alpha$ is the ratio of the number of points in the data catalog to
the number of points in the random catalog. With
\cref{eq:delta_fkp_weighted} the correlation function becomes
\ba
\<\delta^\obs(\vr)\,\delta^\obs(\vr')\>
&=
w(\vr)w(\vr')
W(\vr)\,W(\vr')\,\xi(\vr,\vr')
\vs&\quad
+ (1+\alpha)\,\frac{w^2(\vr)W(\vr)\,\delta^D(\vr'-\vr)}{\bar n}\,,
\label{eq:xi_obs}
\ea
where we used the window function $W(\vr)=\nbar(\vr)/\nbar$, which is how we
define it throughout the paper. \cref{eq:xi_obs} shows that there is a
difference between the window and the weighting. This difference appears in the
shot noise term, which must still include the square of the weighting function,
but contains the window only linearly.

The SFB transform of the shot noise term becomes (i.e.,
\cref{eq:sfb_discrete_fourier_pair_b})
\ba
N^\obs
&= 
\frac{1+\alpha}{\bar n}\,\mathrm{SFB}^2[w^2(\vr)W(\vr)\delta^D(\vr'-\vr)]
\\
&= \frac{1+\alpha}{\bar n}\,(w^2W)_{n\ell m}^{n'\ell'm'}\,,
\label{eq:shotnoise}
\ea
where our notation $(w^2W)$ signifies the window convolution matrix
\cref{eq:wmix}, but with $W \to w^2 W$.
The pseudo-power shot noise is
\ba
\label{eq:Nshot_lnn}
N_{\ell nn'}^{\obs}
&=
\frac{1+\alpha}{\nbar}
\,\frac{1}{\sqrt{4\pi}}\,
\int\dd r\,r^2
\,g_{n\ell}(r)
\,g_{n'\ell}(r)
\,(w^2W)_{00}(r)
\,.
\ea

\cref{eq:xi_obs} also shows that the coupling matrix is modified by
calculating it with the substitution $W(\vr)\to w(\vr)W(\vr)$.

In summary, the following changes are introduced by the weighting:
\begin{enumerate}
  \item In the estimator, each galaxy needs to be multiplied by the weight
    $w(\vr)$ at its location, see \cref{sec:sfb_estimator}. (If no random catalog is
    used, then this also means that $W_{\ell n}(\rhat)$ must be calculated with
    the substitution $W(\vr)\to w(\vr)W(\vr)$.)
  \item The coupling matrix is computed with the substitution $W(\vr)\to
    w(\vr)W(\vr)$.
  \item The shot noise is calculated with the substitution $W(\vr)\to
    w^2(\vr)W(\vr)$.
  \item The local average effect changes as a result of the weighting, and we
    defer the details to \cref{sec:sfb_local_average_effect_with_weighting}.
\end{enumerate}

\section{Local Average Effect with weighting}
\label{sec:sfb_local_average_effect_with_weighting}
In this section we recognize that the average number density $\nbar$ in
\cref{eq:delta_fkp_weighted} must in practice be measured from
the survey itself. This is often called the \emph{integral constraint}
\citep{Beutler+:2014MNRAS.443.1065B, deMattia+:2019JCAP...08..036D} or the
\emph{local average effect} \citep{dePutter+:2012JCAP...04..019D,
Wadekar+:2020PhRvD.102l3521W}.

Compared to \citet{GrasshornGebhardt+:2021PhRvD.104l3548G} we include a
weighting $w(\vr)$. We also consider an extension where $\nbar(z)$ is
independently determined at every redshift \citep{deMattia+:2019JCAP...08..036D}.

Measuring the average number density is accomplished by dividing the total
number of galaxies in the survey (or the total number in each redshift bin) by
the effective volume. However, the total number of galaxies in the survey is a
stochastic quantity such that the average number density is given by
\ba
\nbar &= \left(1 + \bar\delta\,\right) \nbar^\true\,,
\ea
where $\nbar^\true$ is the true density contrast if one were to measure on a
much larger survey volume. Defining the effective volume
\ba
\label{eq:Veff}
V_\mathrm{eff} &= \int \dd^3r\,W(\vr)\,,
\ea
the average density contrast within the survey volume is
\ba
\label{eq:deltabar}
\bar\delta &= \frac{1}{V_\mathrm{eff}}\int\dd^3\vr\,W(\vr)\,\delta(\vr)\,,
\ea
or, if we measure the number density as a function of redshift,
\ba
\label{eq:deltabarz}
\bar\delta(z) &= \frac{1}{4\pi f^\mathrm{eff}_\mathrm{sky}(z)}\int\dd^2\hat r\,W(\vr)\,\delta(\vr)\,,
\ea
where we define the effective sky fraction as
\ba
\label{eq:fsky_eff}
f^\mathrm{eff}_\mathrm{sky}(z) &= \frac{1}{4\pi} \int\dd^2\hat{r}\,W(\vr)
\,.
\ea
That is, we fold the radial selection $\phi(z)$ into the effective sky fraction.
Next, we define the triplet $\mu=(n_\mu,\ell_\mu,m_\mu)$. Then, the average
density field in SFB space becomes
\ba
\label{eq:deltabar_nlm}
\bar\delta_\mu
&=
\delta^K_{\ell_\mu 0} \delta^K_{m_\mu 0}
\begin{cases}
  V_\mathrm{eff}^{-1} \, d_\mu d^*_\nu \, W_{\nu\rho} \delta_\rho
  & \text{ for }\bar\delta=const\,,
  \\
  \widetilde W_{\mu\rho} \delta_\rho
  & \text{ for }\bar\delta=\bar\delta(z)\,,
\end{cases}
\ea
where we sum over repeated indices and we defined the SFB transform of a unit
uniform field,
\ba
d_\mu
&=
\sqrt{4\pi}\,\delta^K_{\ell_\mu0}\delta^K_{m_\mu0}
\int\dd r\,r^2\,g_{n_\mu0}(r)\,,
\ea
and
\ba
\label{eq:Wtilde}
\widetilde W_{\mu\rho}
&=
\int\dd^3r
\,g_\mu(r)
\,g_\rho(r)
\,Y^*_\mu(\rhat)
\,Y_\rho(\rhat)
\,\frac{W(\vr)}{f^\mathrm{eff}_\mathrm{sky}(z)}
\,,
\ea
which is the mixing matrix if the window is $W/f^\mathrm{eff}_\sky$
\citep[e.g., Eq.~(34) in][]{GrasshornGebhardt+:2021PhRvD.104l3548G}. If the
window is separable, then $W/f^\mathrm{eff}_\sky$ is purely a function of
$\rhat$, and $\widetilde W$ is proportional the coupling matrix due to the
angular window. The Kronecker-deltas in \cref{eq:deltabar_nlm} ensure that
$\widetilde W$ only needs to be calculated for $\ell_\mu=m_\mu=0$, and it
describes the coupling of the angular DC mode to the higher multipoles for a
partial sky.

With our model in \cref{eq:delta_fkp_weighted}, the measured
density contrast is \citep[similar to][]{Taruya+:2021PhRvD.103b3501T}
\ba
\label{eq:delta_estimated}
\delta^\obs(\vr) \equiv \delta^{wWA}(\vr)
&= w(\vr)\,W(\vr)\,\frac{\delta(\vr) - \bar\delta}{1+\bar\delta}\,,
\ea
where $\delta(\vr)$ is the true density contrast, the superscript `A' refers to
the local average effect, and the superscript `W' refers to the effect of the
window convolution.

The SFB transform of \cref{eq:delta_estimated} is
\ba
\label{eq:delta_W_sfb}
\delta^{wWA}_{\mu}
&\simeq
(wW)_{\mu\rho}
\,[\delta_\rho - \bar\delta_\rho]\,,
\ea
to first order in $\bar\delta$, and we defined the mixing matrix
$(wW)_{\mu\rho}=w_{\mu\sigma}W_{\sigma\rho}$.

The observed weighted correlation function is
\ba
\label{eq:deltaWAdeltaWA}
\<\delta^{wWA}_{\mu}\delta^{wWA,*}_{\nu}\>
&=
(wW)_{\mu\rho}
\Big[
\<\delta_{\rho}\delta^*_{\lambda}\>
- \<\delta_{\rho}\bar\delta^*_\lambda\>
\vs&\quad
- \< \bar\delta_\rho \delta^*_{\lambda} \>
+ \< \bar\delta_\rho \bar\delta^*_{\lambda} \>
\Big]
(wW)_{\lambda\nu}\,,
\ea
where we used that the mixing matrix is Hermitian,
$(wW)^*_{\kappa\lambda}=(wW)_{\lambda\kappa}$.

\cref{eq:deltaWAdeltaWA} contains four terms. The first term is the clustering
term,
\ba
\label{eq:P1}
P^1_{\mu\nu}
&=
(wW)_{\mu\rho}\<\delta_{\rho}\delta^{*}_{\lambda}\> (wW)_{\lambda\nu}
\,.
\ea
The other terms in \cref{eq:deltaWAdeltaWA} we call $P^2_{\mu\nu}$,
$P^3_{\mu\nu}$, and $P^4_{\mu\nu}$ with similar definitions as $P^1_{\mu\nu}$,
and they are due to the local average effect. The second and third terms are
related by taking the adjoint. That is,
\ba
P^3_{\mu\nu}
&=
(wW)_{\mu\rho}
\< \bar\delta_\rho \delta^*_{\lambda} \>
(wW)_{\lambda\nu}
=
P^{2,*}_{\nu\mu}
\,.
\label{eq:P3}
\ea

The second and third terms are the computationally most expensive to calculate.
The reason is that the correlation between the two isotropic fields
$\delta(\vr)$ and $\bar\delta(r')$ comes from the anisotropic region on the
partial sky defined by the window function. Thus, the correlation
$\<\delta(\vr)\bar\delta(r')\>$ is anisotropic, and this couples
$\ell=0$ modes with higher-$\ell$ modes. Hence, treating
$\<\delta_\rho \bar\delta^*_\lambda\>$ as a pseudo-power spectrum with
$(\ell_\rho,m_\rho)=(\ell_\lambda,m_\lambda)$ is inadequate, in general.
However, we use the heuristic that only modes up to
$\ell_\rho\lesssim1/(2f_\sky)$ couple to $\ell_\lambda=0$, since that is
approximately the width of the window function in spherical harmonic space.

We treat the clustering and shot noise terms separately so that $P^i=C^i+N^i$. We
define the observed weighted power spectrum and shot noise as
\ba
\label{eq:CwW}
C^{wW}_{\mu\nu}
&=
(wW)_{\mu\sigma}C_{\sigma\rho} (wW)_{\rho\nu}
\,,\\
\label{eq:NwW}
N^{wW}_{\sigma\rho}
&=
\frac{1}{\nbar}\, (w^2W)_{\sigma\rho}
=
\frac{1}{\nbar}\, w_{\sigma\alpha} W_{\alpha\beta} w_{\beta\rho}
\,,
\ea
where $C_{\sigma\rho}$ enforces the isotropy condition $\ell_\sigma=\ell_\rho$
and $m_\sigma=m_\rho$, and the shot noise is from \cref{eq:shotnoise}.
Then,
\ba
\label{eq:C1}
P^1_{\mu\nu}
&=
C^{wW}_{\sigma\rho} + N^{wW}_{\mu\nu}\,.
\ea

To calculate the local average effect terms we treat them in two different ways,
depending on whether we consider a constant $\bar\delta$ or $\bar\delta(z)$.
In either case, the result can be separated into clustering and shot noise
terms. Furthermore, we will only consider the pseudo-SFB power spectrum.

\subsection{Constant average density contrast}
Considering a constant $\bar\delta$, the result will be written as the matrix
equations
\ba
\label{eq:CwWA}
C^{wWA} &= C^{wW} - \widetilde T C + \tr(C^WD)\,D^{wW}
\,,\\
\label{eq:NwWA}
N^{wWA} &= N^{wW} - [2\nbar^{-1} - \tr(N^WD)]\,D^{wW}
\,,
\ea
where $\widetilde T$ is given in \cref{eq:T23}. The traces are given in
\cref{eq:trace_CWD,eq:trace_NWD}, and $D^{wW}$ can be calculated efficiently,
as shown in \citet{GrasshornGebhardt+:2021PhRvD.104l3548G}.

Next, we calculate the second term in \cref{eq:deltaWAdeltaWA} assuming
constant $\bar\delta$. Using \cref{eq:deltabar_nlm}, we write
\ba
P^2_{\mu\nu}
&=
w_{\mu\sigma} W_{\sigma\rho}\<\delta_{\rho}\bar\delta^*_\lambda\> W_{\lambda\kappa} w_{\kappa\nu}
\\
&=
V_\mathrm{eff}^{-1}w_{\mu\sigma}
[ C^W_{\sigma\iota} + N^W_{\sigma\iota} ]
d_\iota d^*_\lambda W_{\lambda\kappa} w_{\kappa\nu}
\vs
&=
V_\mathrm{eff}^{-1}w_{\mu\sigma}
[ C^W_{\sigma\iota} + N^W_{\sigma\iota} ]
d_\iota d^{wW,*}_\nu
\vs
&=
V_\mathrm{eff}^{-1}w_{\mu\sigma}
C^W_{\sigma\iota}
d_\iota d^{wW,*}_\nu
+
\nbar^{-1}
D^{wW}_{\mu\nu}
\vs
&=
V_\mathrm{eff}^{-1}
(wW)_{\mu\rho}
[ C_{\rho\omega}
W_{\omega\iota}
d_\iota d^*_\lambda ]
(wW)_{\lambda\nu}
+
\nbar^{-1}
D^{wW}_{\mu\nu}
\,,
\ea
where we used that $N^W_{\sigma\iota}=\nbar^{-1} W_{\sigma\iota}$, and
$(wW)_{\mu\rho}$ is the mixing matrix for $w(\vr)W(\vr)$, and we defined
\ba
\label{eq:DwWmunu}
D^{wW}_{\mu\nu}
&=
(wW)_{\mu\alpha} \, D_{\alpha\beta} \, (wW)_{\beta\nu}\,,
\ea
along with
\ba
\label{eq:Dmunu}
D_{\alpha\beta} &= d_\alpha \, d^*_\beta\,.
\ea
The term in brackets is
\ba
[ C_{\rho\omega}
W_{\omega\iota}
d_\iota d^*_\lambda ]
&=
\sum_{n_\omega}
C_{\ell_\rho n_\rho n_\omega}
W_{n_\omega \ell_\rho m_\rho}
\,d_\lambda
\,,
\ea
where we used
\ba
d_{n \ell m}^{W}
&=
\sum_{n'}
W_{n \ell m}^{n' 00}
d_{n' 00}
\\
&=
\int\dd r\,r^2\,g_{n\ell}(r)
\,W_{\ell m}(r)
\int\dd r'\,r'^2
\sum_{n'} g_{n'0}(r) \,g_{n'0}(r')
\vs
&=
W_{n\ell m}\,,
\label{eq:dWnlm=Wnlm}
\ea
and the last step follows from the orthogonality of the basis functions in
harmonic space, $\sum_n g_{n0}(r) g_{n0}(r')=r^{-2}\delta^D(r-r')$.
Thus, we get
\ba
P^2_{\mu\nu}
&=
C^{2,wW}_{\mu\nu} + \nbar^{-1} D^{wW}_{\mu\nu}
\,,
\ea
where
\ba
C^{2,wW}_{\mu\nu}
&=
(wW)_{\mu\rho}
\left[
V_\mathrm{eff}^{-1}
\sum_{n_\omega}
C_{\ell_\rho n_\rho n_\omega}
W_{n_\omega \ell_\rho m_\rho}
\,d_\lambda
\right]
(wW)_{\lambda\nu}
\\
&=
V_\mathrm{eff}^{-1}
\sum_{n_\rho \ell_\rho m_\rho}
(wW)_{n_\mu \ell_\mu m_\mu}^{n_\rho \ell_\rho m_\rho}
\sum_{n_\omega}
W_{n_\omega \ell_\rho m_\rho}
C_{\ell_\rho n_\rho n_\omega}
\vs&\quad\times
(wW)^*_{n_\nu \ell_\nu m_\nu}
\,,
\ea
where the last line follows again from \cref{eq:dWnlm=Wnlm} and the Hermitian
property of $(wW)_{\lambda\nu}$.
For the pseudo-power spectrum we set $(\ell_\mu,m_\mu)=(\ell_\nu,m_\nu)$,
and average over $m_\mu$,
\ba
C^{2,wW}_{\ell_\mu n_\mu n_\nu}
&=
\sum_{\ell_\rho n_\rho n_\omega}
T_{\ell_\mu n_\mu n_\nu}^{\ell_\rho n_\rho n_\omega}
\,C_{\ell_\rho n_\rho n_\omega}\,,
\label{eq:C2}
\ea
where
\begin{widetext}
\ba
\label{eq:T2}
T_{\ell_\mu n_\mu n_\nu}^{\ell_\rho n_\rho n_\omega}
&=
\frac{1}{V_\mathrm{eff}\(2\ell_\mu+1\)}
\sum_{m_\mu}
(wW)^*_{n_\nu \ell_\mu m_\mu}
\sum_{m_\rho}
(wW)_{n_\mu \ell_\mu m_\mu}^{n_\rho \ell_\rho m_\rho}
W_{n_\omega \ell_\rho m_\rho}
\,.
\ea
\end{widetext}
Further, we use \cref{eq:P3} to calculate both terms two and three in
\cref{eq:deltaWAdeltaWA},
\ba
\label{eq:T23}
\widetilde T_{\ell_\mu n_\mu n_\nu}^{\ell_\rho n_\rho n_\omega}
&=
T_{\ell_\mu n_\mu n_\nu}^{\ell_\rho n_\rho n_\omega}
+
T_{\ell_\mu n_\nu n_\mu}^{\ell_\rho n_\rho n_\omega,*}
\,.
\ea
The local average effect primarily affects the large scale modes. We recommend
calculating $\widetilde T$ fully up to $\ell_\max\sim 1/f_\sky$, and setting
$\widetilde T=0$ on smaller scales.

For the last term in \cref{eq:deltaWAdeltaWA} we use \cref{eq:deltabar_nlm} to get
\ba
P^4_{\mu\nu}
&=
w_{\mu\sigma}W_{\sigma\rho}\<\bar\delta_{\rho}\bar\delta^{*}_{\lambda}\> W_{\lambda\kappa} w_{\kappa\nu}
\\
&=
w_{\mu\sigma}W_{\sigma\rho}D_{\rho\epsilon}
[ C^W_{\epsilon\alpha} + N^W_{\epsilon\alpha} ]
D^*_{\alpha\lambda} W_{\lambda\kappa} w_{\kappa\nu}\,,
\label{eq:deltabar_sq}
\ea
where we have avoided needing to calculate the ill-defined
$N_{\epsilon\alpha}$. However, the window-convolved $N^W_{\epsilon\alpha}$ is
well-defined, as shown by \cref{eq:NwW} because the weighting matrix $w$ is
generally invertible.

To simplify further, we specialize to the case of a constant $\bar\delta$.
\cref{eq:Dmunu} then allows writing \cref{eq:deltabar_sq} as
\ba
P^4_{\mu\nu}
&=
\left[ C^W_{\epsilon\alpha} + N^W_{\epsilon\alpha} \right]
D_{\alpha\epsilon}
\,D^{wW}_{\mu\nu}\,,
\ea
where we used the definition \cref{eq:DwWmunu}.
Further, we find
\ba
\label{eq:trace_CWD}
[C^W_{\epsilon\alpha}D_{\alpha\epsilon}]
&=
\sum_{\ell nn'} (2\ell+1) \, C_{\ell nn'} D^W_{\ell nn'}\,,
\\
\label{eq:trace_NWD}
[N^W_{\epsilon\alpha}D_{\alpha\epsilon}]
&=
\nbar^{-1} \sum_n V_\mathrm{eff}^{-1} d_{n00} d^W_{n00}\,.
\ea
Then, the power spectrum with local average effect can be calcualted with
\cref{eq:CwWA,eq:NwWA}.

\subsection{Redshift-dependent average density contrast}
For the radial local average effect, we use separate approaches for the
clustering and shot noise terms, because each are afflicted by separate
computational concerns. For the shot noise it is non-trivial to calculate the
window-corrected shot noise $N=W^{-1}/\nbar$, and using the approach that works
for the shot noise would require going to very high $k_\max$ for the clustering
term.

We start with the fourth term in \cref{eq:deltaWAdeltaWA}.
Using the second line in \cref{eq:deltabar_nlm}, we get
\ba
P^4_{\mu\nu}
&=
(wW)_{\mu\rho}
\< \bar\delta_\rho \bar\delta^*_{\lambda} \>
(wW)_{\lambda\nu}
\\
&=
(wW)_{\mu\rho}
\,\delta^K_{\ell_\rho 0}
\delta^K_{m_\rho 0}
\widetilde W_{\rho\sigma}
\< \delta_\sigma \,\delta^*_\tau \>
\widetilde W_{\tau\lambda}
\delta^K_{\ell_\lambda 0}
\delta^K_{m_\lambda 0}
\,(wW)_{\lambda\nu}
\,.
\ea
Thus, the fourth clustering and shot noise terms are
\ba
\label{eq:C4munu_z}
C^4_{\mu\nu}
&=
(wW)_{\mu\rho}
\,\delta^K_{\ell_\rho 0}
\delta^K_{m_\rho 0}
\,C^{\widetilde W}_{\rho\lambda}
\,\delta^K_{\ell_\lambda 0}
\delta^K_{m_\lambda 0}
\,(wW)_{\lambda\nu}
\,,\\
\label{eq:N4munu_z}
N^4_{\mu\nu}
&=
(wW)_{\mu\rho}
\,\delta^K_{\ell_\rho 0}
\delta^K_{m_\rho 0}
\,N^{\widetilde W}_{\rho\lambda}
\,\delta^K_{\ell_\lambda 0}
\delta^K_{m_\lambda 0}
\,(wW)_{\lambda\nu}
\,.
\ea

To calculate the shot noise term, consider that \cref{eq:Wtilde} implies that
\ba
\widetilde W_{\rho\sigma}
&=
\big(f^{\mathrm{eff},-1}_\sky\big)_{\rho\kappa}
\,W_{\kappa\sigma}\,,
\ea
where $\big(f^{\mathrm{eff},-1}_\sky\big)_{\rho\kappa}$ is the mixing matrix for
$f^{\mathrm{eff},-1}_\sky(z)$, and it is proportional to
$\delta^K_{\ell_\rho\ell_\kappa}\,\delta^K_{m_\rho m_\kappa}$.
Therefore, \cref{eq:NwW} implies
\ba
N^{\widetilde W}_{\rho \lambda}
&=
\nbar^{-1}
\big(f^{\mathrm{eff},-1}_\sky\big)_{\rho \kappa}
\widetilde W_{\kappa \lambda}
=
\nbar^{-1}
\,\widetilde{\widetilde W}_{\rho \lambda}
\,,
\ea
where we defined
\ba
\label{eq:Wtildetilde}
\widetilde{\widetilde W}_{\rho\lambda}
&=
\int\dd^3r
\,g_\rho(r)
\,g_\lambda(r)
\,Y^*_\rho(\rhat)
\,Y_\lambda(\rhat)
\,\frac{W(\vr)}{f^\mathrm{eff,2}_\mathrm{sky}(z)}
\,,
\ea
in analogy with \cref{eq:Wtilde}. However, we will only need the
$\ell_\rho=\ell_\lambda=0$ terms,
\ba
\widetilde{\widetilde W}_{n_\rho 0 0}^{n_\lambda 0 0}
&=
\int\dd r\,r^2
\,g_\rho(r)
\,g_\lambda(r)
\,\frac{1}{f^\mathrm{eff}_\mathrm{sky}(z)}
=
\big(f^\mathrm{eff,-1}_\mathrm{sky}\big)_{n_\rho 0 0}^{n_\lambda 0 0}
\,,
\ea
with the definition \cref{eq:fsky_eff}.
Since $\big(f^\mathrm{eff,-1}_\mathrm{sky}\big)_{\rho\lambda} \propto
\,\delta^K_{\ell_\rho \ell_\lambda}\delta^K_{m_\rho m_\lambda}$, setting
$\ell_\rho=0$ will also set $\ell_\lambda=0$. Thus, the shot noise term
\cref{eq:N4munu_z} becomes
\ba
N^4_{\ell_\mu n_\mu n_\nu}
&=
\nbar^{-1}
\sum_{\ell_\rho n_\rho}
\delta^K_{\ell_\rho 0}
\,\frac{1}{2 \ell_\mu + 1}
\sum_{m_\mu m_\rho}
(wW)_{n_\mu \ell_\mu m_\mu}^{n_\rho \ell_\rho m_\rho}
\vs&\quad\times
\sum_{n_\lambda \ell_\lambda m_\lambda}
\big(f^\mathrm{eff,-1}_\mathrm{sky}\big)_{n_\rho 0 0}^{n_\lambda \ell_\lambda m_\lambda}
\,(wW)_{n_\lambda \ell_\lambda m_\lambda}^{n_\nu \ell_\mu m_\mu}
\\
&=
\nbar^{-1}
\sum_{n_\rho}
\,\frac{1}{2 \ell_\mu + 1}
\sum_{m_\mu}
(wW)_{n_\mu \ell_\mu m_\mu}^{n_\rho 0 0}
\,(w\widetilde W)_{n_\rho 0 0}^{n_\nu \ell_\mu m_\mu}
\,,
\ea
which is what we use to calculate the $N^4$ term.
Alternatively, \cref{eq:N4munu_z} can be written
\ba
N^4_{\ell_\mu n_\mu n_\nu}
&=
\sum_{\ell_\rho n_\rho n_\lambda}
\!\!\big(\mathcal{M}^{wW}\big)_{\ell_\mu n_\mu n_\nu}^{\ell_\rho n_\rho n_\lambda}
\,\delta^K_{\ell_\rho 0}
\,\big(N^{\widetilde W}\big)_{n_\rho 0 0}^{n_\lambda 0 0}
\,.
\ea
Thus, $N^4$ is the projection of $1/f^\mathrm{eff}_\sky(z)$ into the weighted
partial sky.
which is more readily interpreted as a term projected onto the weighted partial
sky. However, it requires going to high frequencies if, e.g., the selection
function has a strong redshift-dependence.
\begin{widetext}
\ba
N^2_{\ell_\mu n_\mu n_\nu}
&=
\frac{1}{\nbar}\,
\frac{1}{2\ell_\mu+1} \sum_{m_\mu}
\sum_{n_\sigma n_\lambda}
(wW)_{n_\mu \ell_\mu m_\mu}^{n_\sigma 0 0}
(f_\sky^{-1})_{n_\sigma 0 0}^{n_\lambda 0 0}
\,(wW)_{n_\lambda 0 0}^{n_\nu \ell_\mu m_\mu}
\,.
\ea

The clustering term \cref{eq:C4munu_z} pseudo-power is
\ba
C^4_{\ell_\mu n_\mu n_\nu}
&=
\sum_{\ell_\rho n_\rho n_\lambda}
\left[
\frac{1}{2\ell_\mu + 1}
\sum_{m_\mu}
(wW)_{n_\mu \ell_\mu m_\mu}^{n_\rho 0 0}
\,(wW)_{n_\lambda 0 0}^{n_\nu \ell_\mu m_\mu}
\right]
\delta^K_{\ell_\rho 0}
\,C^{\widetilde W}_{0 n_\rho n_\lambda}
\\
&=
\sum_{\ell_\rho n_\rho n_\lambda}
\big(\mathcal{M}^{wW}\big)_{\ell_\mu n_\mu n_\nu}^{\ell_\rho n_\rho n_\lambda}
\,\delta^K_{\ell_\rho 0}
\sum_{\ell_\sigma n_\sigma n_\tau}
\big(\mathcal{M}^{\widetilde W}\big)_{\ell_\rho n_\rho n_\lambda}^{\ell_\sigma n_\sigma n_\tau}
\,C_{\ell_\sigma n_\sigma n_\tau}
\,,
\ea
\end{widetext}
where we introduced the pseudo-power coupling matrices $\mathcal{M}^X$ that
replaces the window with $X$.
Symbolically, we may write this as a matrix equation,
\ba
C^4
&=
\mathcal{M}^{wW}
\,\delta^K_{\ell 0}
\,\mathcal{M}^{\widetilde W}
\,C
\,.
\ea
The matrix $\mathcal{M}^{\widetilde W}$ first couples modes due to the partial
sky coverage at each redshift. Then, the Kronecker-delta selects the angular DC
mode for the local average effect. Finally, the coupling matrix
$\mathcal{M}^{wW}$ projects the power into the observed partial sky with
weighting.

The Kronecker-delta may be written in matrix form as
\ba
(\delta^K)_{\ell_\alpha n_\alpha n_\beta}^{\ell_\rho n_\rho n_\lambda}
&=
\delta^K_{\ell_\rho 0}
[ \delta^K_{\ell_\rho \ell_\alpha}
\delta^K_{n_\rho n_\alpha}
\delta^K_{n_\lambda n_\beta}]
\ea
That is, it is the unit matrix, with all elements set to zero where $\ell_\rho\neq0$.

For a full-sky survey $W(\vr)=\phi(r)$, we get $f^\mathrm{eff}_\mathrm{sky}(z)
= \phi(r)$, and so $ \widetilde W_{\mu\rho} = \delta^K_{\mu\rho} $ is the
identity. Hence, $\mathcal{M}^{\widetilde W}$ is also the identity. Further,
$\mathcal{M}^{wW}\propto\,\delta^K_{\ell_\mu \ell_\rho}$, for an FKP-style
weighting that only depends on the selection function $\phi(r)$. Thus,
$C^4\propto\,\delta^K_{\ell0}$, as required.

The simplest case, $f_\mathrm{sky}(r)=w(\vr)=W(\vr)=1$, yields
\ba
C^4_{\ell_\mu n_\mu n_\nu}
&=
\delta^K_{\ell_\mu 0}
\,C_{0 n_\mu n_\nu}
\,.
\ea

\subsection{\texorpdfstring{$\bar\delta(z)$}{deltabar(z)} second and third terms}
For the second and third terms in \cref{eq:deltaWAdeltaWA} in the case of a
radial loacl average effect $\bar\delta(z)$, we get using
\cref{eq:deltabar_nlm}
\ba
P^2_{\mu\nu}
&=
(wW)_{\mu\rho} \<\delta_{\rho}\bar\delta^*_\lambda\> (wW)_{\lambda\nu}
\\
&=
(wW)_{\mu\rho}
\<\delta_{\rho} \delta^*_\kappa\>
\widetilde W_{\kappa\lambda}
\delta^K_{\ell_\lambda 0}
\delta^K_{m_\lambda 0}
(wW)_{\lambda\nu}
\,.
\ea
Thus, the clustering and shot noise terms are
\ba
C^2_{\mu\nu}
&=
(wW)_{\mu\rho}
\,C_{\rho \kappa}
\,\widetilde W_{\kappa\lambda}
\delta^K_{\ell_\lambda 0}
\delta^K_{m_\lambda 0}
(wW)_{\lambda\nu}
\,,\\
N^2_{\mu\nu}
&=
(wW)_{\mu\rho}
\,N_{\rho \kappa}
\,\widetilde W_{\kappa\lambda}
\delta^K_{\ell_\lambda 0}
\delta^K_{m_\lambda 0}
(wW)_{\lambda\nu}
\,.
\ea

Since $N=\nbar^{-1} W^{-1}$ [from \cref{eq:NwW}], and $\widetilde
W=(f^\mathrm{eff,-1}_\sky) W = W (f^\mathrm{eff,-1}_\sky)$, the pseudo shot
noise is equal to the fourth term,
\ba
N^2 = N^3 = N^4\,.
\ea

The clustering pseudo-power $C^2$ becomes
\begin{widetext}
\ba
C^2_{\ell_\mu n_\mu n_\nu}
&=
\frac{1}{2 \ell_\mu + 1} \sum_{m_\mu}
(wW)_{n_\mu \ell_\mu m_\mu}^{n_\rho \ell_\rho m_\rho}
\,C_{\ell_\rho n_\rho n_\kappa}
\,\widetilde W_{n_\kappa \ell_\rho m_\rho}^{n_\lambda 0 0}
\,(wW)_{n_\lambda 0 0}^{n_\nu \ell_\mu m_\mu}
\\
&=
\sum_{\ell_\rho n_\rho n_\kappa}
\left[
  \frac{1}{2 \ell_\mu + 1} \sum_{m_\mu m_\rho}
  (wW)_{n_\mu \ell_\mu m_\mu}^{n_\rho \ell_\rho m_\rho}
  \sum_{n_\lambda}
  \widetilde W_{n_\kappa \ell_\rho m_\rho}^{n_\lambda 0 0}
  \,(wW)_{n_\lambda 0 0}^{n_\nu \ell_\mu m_\mu}
\right]
C_{\ell_\rho n_\rho n_\kappa}
\,.
\ea
Including the third term $C^3$ via \cref{eq:P3}, we get
\ba
C^{2+3}_{\ell_\mu n_\mu n_\nu}
&=
C^2_{\ell_\mu n_\mu n_\nu}
+ \<n_\mu \leftrightarrow n_\nu\>^*
=
\sum_{\ell_\rho n_\rho n_\kappa}
T_{\ell_\mu n_\mu n_\nu}^{\ell_\rho n_\rho n_\kappa}
\,C_{\ell_\rho n_\rho n_\kappa}
\,,
\ea
where
\ba
T_{\ell_\mu n_\mu n_\nu}^{\ell_\rho n_\rho n_\kappa}
&=
\frac{1}{2 \ell_\mu + 1}
\sum_{n_\lambda m_\rho}
\widetilde W_{n_\kappa \ell_\rho m_\rho}^{n_\lambda 0 0}
\sum_{m_\mu}
\left[
  (wW)_{n_\mu \ell_\mu m_\mu}^{n_\rho \ell_\rho m_\rho}
  \,(wW)_{n_\lambda 0 0}^{n_\nu \ell_\mu m_\mu}
  +
  (wW)_{n_\nu \ell_\mu m_\mu}^{n_\rho \ell_\rho m_\rho}
  \,(wW)_{n_\lambda 0 0}^{n_\mu \ell_\mu m_\mu}
\right],
\ea
\end{widetext}
where we exploit that the elements of $T$ must be real.
We also exploit the symmetry $C_{lnn'}=C_{ln'n}$. This adds the term with
$n_\kappa$ and $n_\rho$ interchanged whenever $n_\kappa$ and $n_\rho$ are
different.

For a full-sky survey, $W(\vr)=\phi(r)$, we have
$f^\mathrm{eff}_\sky(z)=\phi(r)$ and $\widetilde
W_{\mu\rho}=\delta^K_{\mu\rho}$, and $(wW)_{\mu\rho}\propto\,\delta^K_{\ell_\mu
\ell_\rho}\delta^K_{m_\mu m_\rho}$. Therefore,
\ba
T_{\ell_\mu n_\mu n_\nu}^{\ell_\rho n_\rho n_\kappa}
&=
\delta^K_{\ell_\mu 0}
\left[
  \big(\mathcal{M}^{wW}\big)_{\ell_\mu n_\mu n_\nu}^{\ell_\rho n_\rho n_\kappa}
  +
  \big(\mathcal{M}^{wW}\big)_{\ell_\mu n_\nu n_\mu}^{\ell_\rho n_\rho n_\kappa}
\right]
\delta^K_{\ell_\rho 0}
\,.
\ea

This basically suppresses modes with
\ba
\ell \lesssim \frac{1}{f_\sky}\,.
\ea
However, to calculate these to high precision, we need to calculate the mixing
matrices up to
\ba
\ell_\max \lesssim \frac{2}{f_\sky}\,,
\ea
or even up to $\frac{3}{f_\sky}$.

\section{Radial spherical Fourier-Bessel modes with velocity boundary conditions}
\label{app:gnl_velocity}
In this appendix we derive the radial basis functions of the Laplacian with
Neumann boundary conditions at $r_\min$ and $r_\max$. Specifically, we require
the derivatives of the basis functions to vanish on the boundary.
\citet{Fisher+:1995MNRAS.272..885F} refered to these as \emph{velocity}
boundaries, since it implies that the velocities vanish on the boundary.

The radial part of the Laplacian eigenequation in spherical polar coordinates
\cref{eq:laplacian_spherical} is
\ba
0
&=
\frac{\dd}{\dd r}\left(r^2\,\frac{\dd g_{\ell}(kr)}{\dd r}\right)
+ \big[(kr)^2 - \ell(\ell+1)\big]\,g_{\ell}(kr)\,,
\label{eq:helmholtz}
\ea
where function $g$ depends on $\ell$. Our first aim is to derive the discrete
spectrum of $k$ modes for a given $\ell$. We then use that to derive the form
of the $g_\ell$. Following \citet{Fisher+:1995MNRAS.272..885F}, we demand that
the orthonormality relation \cref{eq:gnl_orthonormality}
\ba
\int_{r_\min}^{r_\max}\dd r\,r^2\,g_{n\ell}(r)\,g_{n'\ell}(r)
&=
\delta^K_{nn'}
\ea
is satisfied, where we defined $g_{n\ell}(r)=g_\ell(k_{n\ell}r)$. However, we
modify the approach in \citet{Fisher+:1995MNRAS.272..885F} to integrate from
$r_\min$ to $r_\max$, which will in general add spherical Bessels of the second
kind to the solution. \cref{eq:helmholtz} multiplied by $g_{\ell}(kr)$ yields
\ba
&\int_{r_\min}^{r_\max}\dd r\,
\frac{\dd}{\dd r}\left(r^2\,\frac{\dd g_{\ell}(kr)}{\dd r}\right)g_{\ell}(k'r)
\vs
&=
\int_{r_\min}^{r_\max}\dd r\,
\big[\ell(\ell+1) - (kr)^2\big]\,g_{\ell}(kr)\,g_{\ell}(k'r)\,.
\ea
Subtract from this equation the same equation with $k$ and $k'$ interchanged,
\ba
&\big[k'^2 - k^2\big]\int_{r_\min}^{r_\max}\dd r\,r^2
\,g_\ell(kr)\,g_{\ell}(k'r)
\vs
&=
\int_{r_\min}^{r_\max}\dd r\,
\bigg\{
\frac{\dd}{\dd r}\left(r^2\,\frac{\dd g_\ell(kr)}{\dd r}\right)g_{\ell}(k'r)
\vs&\quad
- \frac{\dd}{\dd r}\left(r^2\,\frac{\dd g_{\ell}(k'r)}{\dd r}\right)g_{\ell}(kr)\bigg\}\,.
\label{eq:integrate_glgl_step1}
\ea
Partial integration of the terms on the right hand side (r.h.s.) gives
\ba
&\int\dd r\,\frac{\dd}{\dd r}\!\(kr^2 g'_\ell(kr)\)g_\ell(k'r)
\vs
&=
\left.kr^2g'_\ell(kr)g_\ell(k'r)\right|_{r_\min}^{r_\max}
- kk'\int\dd r\,r^2\,g'_\ell(kr)g'_\ell(k'r)\,,
\ea
where primes denote derivatives with respect to the argument.
Then, \cref{eq:integrate_glgl_step1} becomes
\ba
&\big[k'^2 - k^2\big]\int_{r_\min}^{r_\max}\dd r\,r^2
\,g_\ell(kr)\,g_{\ell}(k'r)
\vs
&=
\left.kr^2g'_\ell(kr)g_\ell(k'r)\right|_{r_\min}^{r_\max}
-\left.k'r^2g'_\ell(k'r)g_\ell(kr)\right|_{r_\min}^{r_\max}\,.
\label{eq:glgl_kk_integral}
\ea
The r.h.s.\  will vanish for any $k$ whenever
\ba
0
&=
Akr_\max^2g'_\ell(kr_\max)
- Br_\max^2g_\ell(kr_\max)
\vs&\quad
- akr_\min^2g'_\ell(kr_\min)
+ br_\min^2g_\ell(kr_\min)\,,
\label{eq:glgl_orthogonality_condition}
\ea
for any constants $a$, $b$, $A$, and $B$ (given by $k'$).

To choose $a$, $b$, $A$, and $B$, we need to choose boundary conditions.
The Neumann boundary conditions are
\ba
g'_\ell(kr_\max) &= 0\,,\\
g'_\ell(kr_\min) &= 0\,.
\ea
These will ensure that a spherical boundary in real space remains a spherical
boundary in redshift space as long as the observer is at $r=0$. Thus, the
theoretical redshift-space modeling will be greatly simplified.

Writing the basis as a combination of spherical Bessel functions of the first
and second kinds,
\ba
\label{eq:gl_definition}
g_\ell(kr)=c_{n\ell}\,j_\ell(kr) + d_{n\ell}\, y_\ell(kr)
\,,
\ea
where we anticipate that the function $g_\ell$ will also depend on $n$.
Then,
\ba
\label{eq:knl_rmin_condition_v1}
0
&=
c_{n\ell}\,j'_\ell(k_{n\ell}\,r_\min) + d_{n\ell}\,y'_\ell(k_{n\ell}\,r_\min)
\,,\\
\label{eq:knl_rmax_condition_v1}
0
&=
c_{n\ell}\,j'_\ell(k_{n\ell}\,r_\max) + d_{n\ell}\,y'_\ell(k_{n\ell}\,r_\max)\,.
\ea
This is equivalent to choosing $B=b=0$ in
\cref{eq:glgl_orthogonality_condition}. Thus, the conditions
\cref{eq:knl_rmin_condition_v1,eq:knl_rmax_condition_v1} lead to a set of
orthogonal basis functions $g_{n\ell}(r)=g_\ell(k_{n\ell}r)$.

Both $j_\ell$ and $y_\ell$ satisfy the two recurrence relations
\ba
\label{eq:jl_recursion_lp1}
j'_\ell(kr) &= -j_{\ell+1}(kr) + \frac{\ell}{kr}\,j_\ell(kr)\,,
\\
\label{eq:jl_recursion_lm1}
j'_\ell(kr) &= j_{\ell-1}(kr) - \frac{\ell+1}{kr}\,j_\ell(kr)\,.
\ea
Solving both \cref{eq:knl_rmin_condition_v1,eq:knl_rmax_condition_v1} for $d_{n\ell}/c_{n\ell}$,
\begin{widetext}
\ba
\label{eq:knl_condition}
\frac{-kr_\min\,j_{\ell+1}(kr_\min) + \ell\,j_\ell(kr_\min)}
{-kr_\min\,y_{\ell+1}(kr_\min) + \ell\,y_\ell(kr_\min)}
&=
\frac{-kr_\max\,j_{\ell+1}(kr_\max) + \ell\,j_\ell(kr_\max)}
{-kr_\max\,y_{\ell+1}(kr_\max) + \ell\,y_\ell(kr_\max)}
\,,
\ea
\end{widetext}
which needs to be satisfied for $k_{n\ell}$. When $r_\min=0$, then
$d_{n\ell}=0$ so that the $y_\ell$, which diverge as their argument vanishes,
do not contribute.

The solution $k_{n\ell}=0$ is valid for $\ell=0$ only, because otherwise the
normalization for $g_{n\ell}$ does not exist, see \cref{eq:gl_definition} and
the next section.

\subsection{Normalization}
The normalization of $g_\ell$ is obtained by dividing
\cref{eq:glgl_kk_integral} by $k'^2-k^2$, and taking the limit $k'\to
k=k_{n\ell}$,
\ba
1
&=
\int_{r_\min}^{r_\max}\dd r\,r^2
\,g^2_\ell(kr)
=
\lim_{k'\to k}
\frac{-\left.k'r^2g'_\ell(k'r)g_\ell(kr)\right|_{r_\min}^{r_\max}}
    {(k' - k)(k' + k)}
    \,.
    \label{eq:gl_normlization_definition}
\ea
Thus, the key is to calculate
\ba
\lim_{k'\to k_{n\ell}}\frac{g'_\ell(k'R)}{k'-k_{n\ell}}
&=
R\,g''_\ell(k_{n\ell}R)\,,
\ea
where $R$ is either $r_\min$ or $r_\max$, and we expanded $g'_\ell$ in a Taylor
series around $k'\sim k_{n\ell}$. Hence,
\ba
1
&= 
-\frac12\left.r^3g_\ell(k_{n\ell}r)\,g''_\ell(k_{n\ell}r)\right|_{r_\min}^{r_\max}
\label{eq:gl_normalization}
\,.
\ea
We can evaluate by using
\ba
g''_\ell(kr)
&=
\frac{1}{(kr)^2}\left[
  \(\ell (\ell - 1) - (kr)^2\)g_\ell(kr) + 2 kr\,g_{\ell+1}(kr)
\right],
\ea
which follows from \cref{eq:jl_recursion_lp1,eq:jl_recursion_lm1}.

For the special case $k_{n\ell}=\ell=0$, the normalization is given by
\ba
\int_{r_\min}^{r_\max}\dd r\,r^2 &= \frac13\(r_\max^3 - r_\min^3\).
\ea

Choosing $k_{n\ell}$, $c_{n\ell}$, and $d_{n\ell}$ that satisfy
\cref{eq:knl_condition,eq:knl_rmax_condition_v1,eq:gl_normalization}
guarantees the orthonormality of the $g_\ell$,
\ba
\int_{r_\min}^{r_\max}\dd r\,r^2\,g_\ell(k_{n\ell}\,r)\,g_{\ell}(k_{n'\ell}\,r)
&=
\delta^K_{nn'}
\,.
\ea
Note that the condition $\ell=\ell'$ is \emph{not} enforced by the $g_\ell$.
Instead, $\ell=\ell'$ comes from the spherical harmonics, i.e., \cref{eq:YlmYlmDelta}.

\end{document}